\documentstyle[aps,epsfig,multicol]{revtex}
\def\top#1{\vskip #1\begin{picture}(290,80)(80,500)\thinlines \put(
65,500){\line( 1, 0){255}}\put(320,500){\line( 0, 1){5}}\end{picture}}
\def\bottom#1{\vskip #1\begin{picture}(290,80)(80,500)\thinlines \put(
330,500){\line( 1, 0){255}}\put(330,500){\line( 0, -1){5}}\end{picture}}
\begin{document}
\draft
\title{Theories of Low-Energy Quasi-Particle States in Disordered
  $d$-Wave Superconductors}
\author{Alexander Altland${}^a$, B.D. Simons${}^b$, and M.R.
  Zirnbauer${}^c$}
\address{${}^a$Theoretische Physik III, Ruhr-Universit\"at-Bochum,
  44780 Bochum, Germany\\ ${}^b$Cavendish Laboratory, Madingley Road,
  Cambridge CB3 0HE, UK\\ ${}^c$Institut f\"ur Theoretische Physik,
  Universit\"at zu K\"oln, Z\"ulpicher Strasse 77, 50937 K\"oln,
  Germany}
\maketitle                                 
\begin{abstract}
  The physics of low-energy quasi-particle excitations in disordered
  $d$-wave superconductors is a subject of ongoing intensive research.
  Over the last decade, a variety of conceptually and methodologically
  different approaches to the problem have been developed.
  Unfortunately, many of these theories contradict each other, and the
  current literature displays a lack of consensus on even the most
  basic physical observables.  Adopting a symmetry-oriented approach,
  the present paper attempts to identify the origin of the
  disagreement between various previous approaches, and to develop a
  coherent theoretical description of the different low-energy regimes
  realized in weakly disordered $d$-wave superconductors.  We show
  that, depending on the presence or absence of time-reversal
  invariance {\it and} the microscopic nature of the impurities, the
  system falls into one of four different symmetry classes.  By
  employing a field-theoretical formalism, we derive effective
  descriptions of these universal regimes as descendants of a common
  parent field theory of Wess-Zumino-Novikov-Witten type.  As well as
  describing the properties of each universal regime, we analyse a
  number of physically relevant crossover scenarios, and discuss
  reasons for the disagreement between previous results.  We also
  touch upon other aspects of the phenomenology of the $d$-wave
  superconductor such as quasi-particle localization properties, the
  spin quantum Hall effect, and the quasi-particle physics of the
  disordered vortex lattice.
\end{abstract}
\pacs{74.20.-z, 74.25.Fy, 71.23.-k, 71.23.An, 72.15.Rn}
\tableofcontents
\begin{multicols}{2}
\narrowtext 
\section{Introduction}
\label{sec:intro}

In recent years it has become clear that the superconducting phase of
the hole-doped cuprate superconductors is of unconventional, $d$-wave,
symmetry.  Motivated by this observation, the properties of
quasi-particle states in $d$-wave superconductors have come under
intense scrutiny \cite{Gorkov,Lee,hwe,zhh,ntw,sfbn,mp,Balatsky94,%
  Balatsky96,Joynt,Balents,smf,sf,vsf,dl,mag,Pepin,Fukui,fk,ahm}. A
particular subject that turned out to be not only experimentally
relevant but also theoretically intricate is the influence of {\it
  static disorder} on the large-scale features of the $d$-wave phase.
Leaving aside the pair-breaking effect of the disorder which strongly
influences the high-temperature properties, attempts to resolve the
impact of impurity scattering on low temperature spectral and
transport properties of a disordered $d$-wave superconductor have
ignited controversy in the recent literature.
  
The key feature which distinguishes $d$-wave from the more
conventional $s$-wave superconductors, and makes the disorder problem
harder to solve, is the presence of four isolated ``nodes'' on the
Fermi surface in the vicinity of which low-lying Dirac-type
\cite{fradkin} quasi-particle excitations exist.  Beginning with the
work of Gorkov and Kalugin \cite{Gorkov}, and later Lee \cite{Lee},
the earliest considerations, based on approximate self-consistent
treatments, concluded that an arbitrarily weak impurity potential
induces a finite density of states (DoS) at the Fermi surface, and
leads to weak localization of all quasi-particle states in the
two-dimensional (2d) system. Although the first result found support,
at least superficially, in an analysis by Ziegler et al.~\cite{zhh},
these conclusions were found to be in contradiction with those of
other authors.

In particular, following the early considerations of
Refs.~\cite{Gorkov,Lee}, Nersesyan, Tsvelik and Wenger \cite{ntw}
(NTW) developed a complementary approach in which the quasi-particle
Green functions of a disordered $d$-wave superconductor were mapped on
a conformally invariant fermion-replica field theory.  Based on this
mapping, NTW proposed that the 2d system is characterized by critical
properties of the quasi-particle states at the Fermi level.  Their
model predicted that the DoS vanishes as $\nu(E) \sim |E|^\alpha$ with
some disorder-dependent exponent $\alpha$.

To further add to this controversy, Senthil, Fisher, Balents, and
Nayak \cite{sfbn} (SFBN) proposed a description by an alternative
fermion-replica field theory, namely the principal chiral non-linear
sigma model over the symplectic group ${\rm Sp}(2r)$ (with $r = 0$).
Void of any signatures of criticality, the large-scale behaviour of
the two-dimensional model was argued to describe a phase of localized
states, the ``spin insulator'', thereby finding qualitative agreement
with the early considerations and contradicting the findings of NTW.
In contradiction to {\it both} NTW and the early works, the
quasi-particle DoS at the Fermi level was found to vanish with a
universal exponent of unity.

Since this work, further analysis of the field-theoretic scheme
generated fresh, and seemingly contradictory proposals.  Fendley and
Konik \cite{fk} have argued that, when the scattering between pairs of
opposite nodes is neglected, the low-energy properties of the weakly
disordered system are described by a {\it critical} field theory that
has ${\rm Sp}(2r)$ (with $r=0$) for its target space and includes a
Wess-Zumino term.  The same theory, now based on a peculiar
implementation of a time-reversal symmetry breaking lattice operator,
had been suggested by Fukui \cite{Fukui}.  Albeit belonging to a
different universality class, both proposals \cite{Fukui} and
\cite{fk} share key aspects with the theory of NTW, including the
existence of extended quasi-particle states. 

How, if indeed at all, can the largely contradictory approaches be
reconciled with each other?  And is it possible to unify elements of
these approaches into one coherent theoretical formulation?  This is
the first complex of questions we are going to address in this paper.

A second and more pragmatic point we are going to address concerns the
microscopic nature of the impurity scattering.  It will turn out that
much of the controversy in the field-theoretical literature described
above can be attributed to the sensitivity of the $d$-wave
quasi-particle physics to the range of the scattering potential. In a
number of recent publications \cite{kulic1,kulic2,kulic3}, it has been
argued that both experimental and theoretical evidence hints at a
strong enhancement of forward scattering in $d$-wave superconductors.
Without going into details, we here merely state the essence of the
argument, viz.~that a significant amount of inter-node scattering
would have a strong pair-breaking effect, thereby being in conflict
with the very {\it formation} of stable $d$-wave order.  Actually
figuring out the meaning of the attribute ``significant'' is a
delicate issue we will not even attempt to address. Notwithstanding
this uncertainty, one may ask the principal question of what kind of
low-energy theory governs the behaviour of quasi-particle Hamiltonians
that are strongly anisotropic (albeit certainly not {\it exclusively}
forward scattering) on the bare microscopic level.  In particular, the
question arises whether, as in disordered metallic systems, the
impurity form factor merely leads to a renormalization of the
scattering time or whether it might have a more substantial effect.
As we are going to argue below, the latter is the case here: the
extreme cases of pure forward scattering and isotropic scattering fall
into distinct universality classes with qualitatively different
properties.  And, although the isotropic limit is ultimately
attractive (in the renormalization group sense), one might speculate
that under realistic conditions (finite temperatures, experimental
resolutions, etc.) the forward-scattering limit could well be the
relevant one.

A second aspect that makes the limit of weak forward scattering an
interesting object of study concerns the physics of the mixed state:
as with conventional type-II superconductors, the application of a
magnetic field of intermediate strength $(H_{c1} < H < H_{c2})$ forces
vortices into the $d$-wave superconductor.  The impact of vortex
formation on the low-energy density of quasi-particle states of {\it
  pure} $d$-wave superconductors has been the subject of an ongoing
debate \cite{gs,anderson,ft,melnikov,mhs,wm}.  Recently, significant
progress was made by Franz and Tesanovic \cite{ft}, who exploited a
novel singular gauge transformation, mapping the problem again on the
problem of Dirac fermions.  This transformation, which is analogous to
that applied in the theory of the fractional quantum Hall effect, is
applicable even in the presence of randomness.  As a result, one
obtains an effective low-energy model, essentially Dirac fermions with
random scalar and vector potential, that can be applied to explore the
quasi-particle DoS of the {\it disordered} vortex lattice.  We will
return to this subject below.

Although, for a non-expert, the detailed field-theoretical analysis
contained in this paper is often somewhat involved, we believe that
the main conclusions of our survey of the dirty ``$d$-wave system''
--- for lack of a better terminology we will often refer to the
quasi-particles of a disordered $d$-wave superconductor by that word
--- should be widely accessible.  We have, therefore, chosen to
summarize the complete phase diagram here in the introduction. At the
same time, this allows us to organize and place into context many of
the existing field-theoretical works on the subject.

\subsection{Intrinsic and Extrinsic Symmetries}

While, at first sight, the various points brought up above seem
unrelated, the microscopic analysis below will show that they do, in
fact, all find their origin in a common microscopic mechanism.  Just
as for the clean $d$-wave Gorkov Hamiltonian \cite{gorkov2,fn_gorkov},
the Hilbert space for the low-energy Hamiltonian of the pure
forward-scattering system foliates into four sectors not connected by
the disorder potential.  These nodal sectors can be grouped into pairs
related by parity.  The low-energy physics of the individual sectors
turns out to be more intricate than that of the complete system, which
is comprised of four Dirac nodes coupled by impurity scattering.  In
general it will, of course, be the behaviour of the full system that
matters.  There are, however, a number of situations in which
signatures of the decoupled system remain visible.  Exploring these
scenarios, which are accompanied by a drastic change in the
phenomenology of the system, is one of the major issues to be
addressed in this paper.

The origin of the relative non-triviality of the decoupled system can
be understood qualitatively by noticing that the structure of any
theory describing the low-lying quasi-particle excitations of a
disordered system is essentially fixed by two types of symmetries:
``intrinsic'' symmetries, such as the behaviour of the Hamiltonian
under spin-rotation, time-reversal, particle-hole transformations
etc., and ``extrinsic'' symmetries such as translational or rotational
invariance or, for that matter, parity.  Specifically, the unperturbed
$d$-wave superconductor is a particle-hole symmetric, spin-rotation
and time-reversal invariant system, thereby falling into the symmetry
class $C$I in the classification of Ref.~\cite{Altland}.  The
implementation of these intrinsic symmetries into an effective
low-energy theory of a superconductor system was formulated some time
ago by Oppermann \cite{Oppermann}, and was recently rederived in the
context of $d$-wave superconductivity by SFBN \cite{sfbn}.

\begin{figure}[hbt]
  \centerline{\epsfxsize=3.5in
    \epsfbox{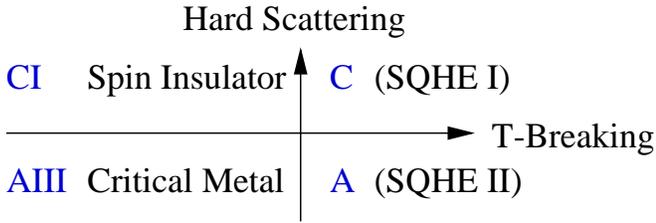}}
  \caption{Symmetry classes realized in dirty $d$-wave superconductors
    with conserved spin. The two types of perturbation causing
    crossover between classes are indicated.}
  \label{fig:sym}
\end{figure}

The behaviour of the $d$-wave system under extrinsic symmetry
operations is more remarkable.  In fact, as we will see, it is
disregard for these symmetries that has driven astray some of the
field theories based on intrinsic symmetry: not all terms allowed by
the intrinsic symmetries respect the extrinsic symmetries.  The point
is that individual nodes are non-invariant under ``parity'', by which
we mean any operation that reverses the orientation of two-dimensional
space \cite{fn_parity}.  Of course, the full four-node system emerging
from a manifestly invariant Gorkov Hamiltonian is parity invariant,
but individual nodes are not.  Remarkably, this parity non-invariance
and the intrinsic symmetries conspire to let the system of isolated
nodes fall into a symmetry class that differs from the Wigner-Dyson
classes {\it and} the classes commonly attributed to bulk
superconductors.  The complete phase diagram of the $d$-wave system in
fact separates into a total of four distinct regions (see
Fig.~\ref{fig:sym}) distinguished by the presence/absence of
time-reversal invariance and by the correlation radius of the disorder
potential. Taking each region in turn, we here summarize the
phenomenology and the basic properties of the respective low-energy
theories \cite{class_warning}.

\subsection{Symmetry Class $A$III}

So, beginning with the time-reversal invariant system with only
forward scattering, one finds that the corresponding symmetry class is
not $C$I, but rather $A$III, i.e.~a symmetry class that is typically
realized in disordered relativistic (chiral) fermion systems.  Indeed,
it is straightforward to verify that the low-energy physics of
individual nodes is described by a model of Dirac fermions in the
presence of some quenched random gauge field.  The implications of
this correspondence for the disordered system were first noticed and
analysed in the seminal work of NTW \cite{ntw}, who pointed out that
the low-energy physics of the individual nodes is critical.  NTW
argued that a general mechanism uncovered long ago by Witten
\cite{Witten84} applies in particular to the $d$-wave system: in a
Lagrangian formulation based on non-Abelian bosonization, non-trivial
transformation behaviour with respect to parity is reflected by the
presence of a Wess-Zumino-Novikov-Witten (WZW) term in the action
functional.  Specifically, NTW mapped the $d$-wave system on a field
theory with ${\rm U}(1) \times {\rm SU}(r) \times {\rm SU}(N)$
symmetry ($r$ being the number of replicas).  In the limit of isolated
nodes $(N = 1)$, the theory has the field manifold ${\rm U}(1) \times
{\rm SU}(r)$, with the ${\rm SU}(r)$ sector being governed by a
so-called level-$1$ WZW action.  The latter theory enjoys the
existence of an infinite-dimensional (current algebra) symmetry, which
makes it solvable to a large extent.  Several of its features,
including the existence of algebraically decaying correlations
(absence of localization!), and the scaling of the DoS in the vicinity
of the Fermi energy can be derived rigorously.

Within the framework of the supersymmetric formulation, the properties
of the forward-scattering system are found to be described by two
copies of a WZW model of level $1$ and type $A|A$, i.e.~a WZW model
whose fields take values in a supermanifold that, for reasons
discussed in body of the paper, is labeled by $A|A$.  Each copy
derives from a pair of nodes on the Fermi surface related by parity or
inversion symmetry.  Formally, the total zero-energy effective action
is given by $S = W[M_1,\gamma] + W[M_2^{-1} , \gamma^{-1}] + g^\prime
\int d^2r \, {\cal O}_{00}^\prime$ where the dimensionless parameter
$\gamma$ (defined microscopically later) specifies the degree of
anisotropy of the Dirac cones, and
\begin{eqnarray*}
  &&W[M,\gamma] \equiv {i\over 12\pi} \Gamma[M] + g \int d^2r\, {\cal
    O}_{00} \\ && -{1\over 8\pi} \int d^2r \, {\rm STr} \left(
    \gamma^{-1} \partial_1 M^{-1}\partial_1 M + \gamma \partial_2
    M^{-1}\partial_2 M\right)
\end{eqnarray*}
is the anisotropic WZW action.  The fields $M_1$ and $M_2$ take values
in (a maximal Riemannian subspace of) the supergroup ${\rm GL}(2|2)$,
$\Gamma[M]$ represents the WZW term, and ${\cal O}_{00}$, ${\cal
  O}_{00}^\prime$ are low-angle scattering operators whose specific
structure will be described in Section \ref{sec:boso}. Note that,
since the WZW term is odd under $M \to M^{-1}$ (i.e.~$\Gamma[M^{-1}] =
- \Gamma[M]$), the WZW terms for the two sectors of nodes carry {\it
  opposite} sign, which reflects the fact that they derive from Dirac
operators carrying opposite orientations. Further, the fact that ${\rm
  GL}(2|2)$ and not ${\rm GL}(1|1)$ -- the supersymmetric analog of a
fermion replica theory with group manifold ${\rm U}(r)$ -- appears as
the relevant degree of freedom, has to do with the behaviour of the
system under time reversal. This aspect, unimportant for the physics
of the system of isolated nodes, becomes vitally relevant once
inter-node scattering is switched on.

Perhaps the most characteristic feature of models of Dirac fermions
subject to Abelian gauge randomness is their stability under
renormalization.  Indeed, the low-energy theory defined by the action
$W$ is a fixed point of the renormalization group for each value of
the dimensionless disorder strength $g$, as long as $g$ does not
exceed some critical value.  The absence of running couplings bears
consequences for various physical observables.  One finds that, for
low energies, the DoS vanishes as $ \nu(E) \propto |E|^\alpha$, with
the $g$-dependent family of exponents
\begin{eqnarray*}
  \alpha = {1 - 2g / \pi \over 1 + 2g / \pi} \;,
\end{eqnarray*}
in accord with the analysis of NTW.  We also recapitulate that the
eigenstates of the random gauge Dirac system are extended \cite{mcw}.

While exhibiting critical spectral and transport properties, the
effective action for $d$-wave superconductors in class $A$III is
unstable against the influence of short-range isotropic scattering, a
matter not correctly handled by NTW.  Impurity scattering involving
large momentum transfer couples the nodes and leads to a ``locking''
of the fields at the different nodes.  This process bears two
consequences for the structure of the low-energy theory: first, and in
contradiction to the naive expectation, the field configurations
surviving the locking of the formerly independent ${\rm GL}(2|2)$
fields take values in a different manifold, viz.~${\rm OSp}(2|2)$.
Second, the locking $M_1 = M_2$ implies that in the full action the
WZW terms cancel, which makes the restoration of parity invariance in
the coupled system explicit.  (In parentheses we note that the absence
of a WZW term is a sufficient but {\it not} a necessary condition for
parity invariance of physical observables: for a single WZW theory,
the sign change in $\Gamma[M]$ caused by a parity operation can be
compensated by a field transformation $M = {\rm e}^X \to M^{-1} = {\rm
  e}^{-X}$, i.e.~the field $X$ is a pseudo-scalar.  However, in a
system comprised of {\it two} coupled sub-theories carrying opposite
orientations, this compensating transformation is inhibited on either
sub-theory.  Moreover, there exists no room for a WZW term in the
locked full theory, as the sum of coupled Dirac theories for the
$d$-wave system is anomaly-free.)  Since it was the WZW term that
rendered the NTW theory critical, it is evident that the low-energy
behaviour of the full theory will be of different type.

\subsection{Symmetry Class $C$I}

A fermion-replica theory of the full system, a non-linear sigma model
on the field manifold ${\rm Sp}(2r)$, was first derived and analysed
by SFBN.  Essentially, this model is a $d$-wave version of the ${\rm
  Sp}(2r)$ non-linear sigma model formulation of time-reversal
invariant superconducting glasses developed earlier by Oppermann
\cite{Oppermann}.  Void of any elements supporting critical behaviour,
the predictions of that theory differ greatly from NTW: all
quasi-particle states are localized, a fact that led SFBN to coin the
term ``spin insulator''; the DoS still vanishes at zero energy but in
a manner altogether different from the scaling obtained by NTW.  All
in all, the low-energy theory of the composite four-node system seems
to have little overlap with the critical theory of the decoupled
forward-scattering system.

Once the fields belonging to the different nodal sectors become
locked, the total effective action from above takes the form $S[M] =
W[M,\gamma] + W[M^{-1},\gamma^{-1}]$, i.e.
\begin{eqnarray*}
  S[M]= - {\gamma+\gamma^{-1} \over 8\pi} \int d^2r \, {\rm STr}
  \left( \partial_\mu M^{-1} \partial_\mu M\right) \;,
\end{eqnarray*}
where the field $M$ now belongs to the group manifold ${\rm OSp}
(2|2)$.  Technically this is a principal chiral non-linear sigma model
of type $D|C$.  It represents a supersymmetric extension of the action
derived by SFBN \cite{sfbn}, and leads to the prediction of
quasi-particle localization of all states in two dimensions --- the
spin insulator phase --- and a DoS that vanishes at zero energy.

It seems unlikely that symmetry class $A$III is realized in the
physical $d$-wave system. As mentioned earlier, the bare microscopic
theory will probably contain a large forward-scattering amplitude,
plus a certain amount of inter-node scattering rendering it a member
of symmetry class $C$I. This hard-scattering admixture is a relevant
perturbation which means that the theory will ultimately flow towards
the strong-coupling fixed point in class $C$I.  On the other hand,
realistic experimental conditions are likely to prevent the system
from exploring all limiting regions of the phase diagram. This
motivates us to ask how fast the flow towards the strong-coupling
regime takes place and what properties the crossover theories,
situated somewhere in between the independent-node and the strongly
coupled theory, possess.  We will return to this issue below.

\subsection{Symmetry Class $C$}

An alternative mechanism of allowing elements of the decoupled theory
to resurface is by breaking parity {\it explicitly}. At least
theoretically, this is realized in the so-called spin quantum-Hall
(SQH) system \cite{smf,Laughlin}, where the nodal degeneracy is lifted
via the addition of an $i d_{xy}$ component to the order parameter.
The resulting low-energy phase shares much of its phenomenology ---
the formation of edge states, quantized values of the transverse
component of the conductivity tensor, etc. --- with the integer
quantum Hall effect and has, therefore, been christened the spin
quantum Hall state.  (The prefix ``spin'' indicates that in the
superconductor problem, the behaviour of spin currents or thermal
currents is considered.  Because of the non-conservation of
quasi-particle charge, the study of charge transport coefficients
becomes meaningless.) Originally, it was suggested \cite{Laughlin}
that the formation of a secondary $id$-order parameter component
giving the $d+id$ state might be quasi-spontaneously driven by a weak
external magnetic field.  Recent work \cite{lhw} has cast some doubt
on the applicability of this scenario and, therefore, on the
experimental relevance of the spin quantum Hall effect.
Theoretically, however, the $d+id$ system remains an interesting
object of study: first, it represents a rare example of a system with
an exactly solvable quantum-Hall phase transition driven by disorder
\cite{glr}.  Second, unlike with $d$-wave superconductors in the
previously discussed symmetry class $C$I, aspects of the physics of
individual nodes survive the thermodynamic limit.

In fact, the $id$-component does more than lift the node degeneracy,
implying that it does not lead to a resurrection of the WZW model: the
inclusion of an $id_{xy}$ component not only breaks parity but also
time-reversal invariance, and hence reduces the symmetry of the system
from $C$I down to $C$.  Within a fermion-replica approach, the class
$C$ system is described by a field theory taking values on the
manifold ${\rm Sp}(2r)/{\rm U}(r)$.  This field theory does not
support a WZW term.  It does, however, admit for the existence of a
closely related parity-breaking term in the Lagrangian, viz.~a
topological theta term.  Based on phenomenological considerations, it
has in fact been argued \cite{smf} that a non-linear sigma model over
${\rm Sp}(2r)/{\rm U}(r)$ with a theta term should be the relevant
theory of the disordered $d+id$ system.

Exploring how such a theory may emerge as a descendant of the class
$C$I theory of the non-degenerate node system will be a further topic
to be addressed below. The explicit form of the resulting
supersymmetric soft-mode action for the system of class $C$ reads as
\begin{equation}
  - {1\over 8}\int d^2r\, {\rm STr} \left( \sigma_{11}^0 \partial_\mu
    Q \partial_\mu Q + \sigma_{12}^0 \epsilon^{\mu\nu} Q\partial_\mu Q
    \partial_\nu Q \right) \;,
\label{nlsm_DIIICI}
\end{equation}
where the WZW term has transformed into a topological theta term.  In
contrast to the conventional integer quantum Hall effect exhibited by
normal systems, the fields of the saddle-point manifold $Q$ belong to
the coset space ${\rm OSp}(2|2)/ {\rm GL}(1|1)$, with the relevant
Riemannian symmetric superspace inside that manifold being of type
$D{\rm III}|C{\rm I}$.

Unlike its normal relative, the critical theory for the spin quantum
Hall effect for class $C$ is readily amenable to theoretical
investigation.  Indeed the properties of this phase have been explored
in a number of recent numerical and analytical works.  Surprisingly, a
mapping \cite{smf} of the class $C$ theory onto a network model
\cite{Kagalovsky} shows the transition to be in the same universality
class as classical percolation \cite{glr}.  This correspondence, which
applies only for the statistics of various low-order moments, relies
on a remarkable cancellation of interference channels, and allows
several of the critical exponents to be determined exactly
\cite{glr,cardysqh}.  Very recently, a proposal for the general
critical theory of the spin quantum Hall transition has been made by
Bernard and LeClair \cite{bele}, although its validity still remains a
matter of some controversy.

\subsection{Symmetry Class $A$}

To exhaust the number of different low-energy scenarios realized in
the $d$-wave system, one may consider a situation where a perturbation
breaking time-reversal symmetry is superimposed on an impurity
potential causing only forward scattering.  Below we will identify the
resulting system as a member of symmetry class $A$, i.e. the standard
Wigner-Dyson class of unitary symmetry.  We find that, as with its
class $C$ counterpart, the class $A$ system also supports a (spin)
quantum Hall transition.  Somewhat surprisingly, however, that
transition turns out to fall into the universality class of the
standard integer quantum Hall transition.

Formally, the low-energy theory is again described by an action
functional including a theta term, similar to the one above.  The main
differences with the previously discussed case are that (a) one
obtains two copies $S[Q,\gamma]$ of the action, one for each nodal
sector, and (b) that the fields $Q_i$ $(i=1,2)$ belong to the coset
space ${\rm GL}(2|2)/ ({\rm GL}(1|1)\times {\rm GL}(1|1))$.  This
means that the field theory is of type $A{\rm III}|A{\rm III}$ which
is the supersymmetric extension of Pruisken's theory for the
conventional integer quantum Hall effect.

Finally, we remark on a second and, arguably, more physically
motivated scenario in which the classes $A$ and $C$ are engaged ---
the mixed phase of a type-II superconductor.  On symmetry grounds
alone, one would expect the Hamiltonian of the vortex phase of the
dirty $d$-wave system to belong to one of these classes.  Although the
nature of the quasi-particle states in the vortex phase of the clean
$d$-wave system has been addressed in a number of (seemingly
controversial) publications \cite{gs,anderson,ft,melnikov,mhs,wm,ye},
a field-theoretic analysis of this disordered system has not been
given in the literature.  This motivates us to focus our discussion of
the symmetry classes $C$ and $A$ on the mixed state, not on the spin
quantum Hall system.

Furthermore, for brevity, we have chosen not to comment upon the
influence of the breaking of spin rotational invariance either by
magnetic impurities or spin-orbit scatterers.  Such systems, which lie
outside the scope of present work, have been addressed in the recent
literature and we refer to Refs.~\cite{sfD,rg,bsz} for a discussion of
the phenomenology of these systems.

Altogether, the key characteristics of the field theories for the
$d$-wave system are summarized in the following table:
\end{multicols}
\widetext
\top{-2.8cm}
 \begin{center}
   \begin{tabular}{{|c||c|c|l|l|l|l|}}\hline
     Class & ${\rm T}$& hard scattering& Goldstone modes &
     saddle-point manifold & NL$\sigma$M& fermion-replica analog\\ 
     \hline $A$III& + & $-$ & ${\rm GL}(2|2) \times {\rm GL}(2|2)$ &
     ${\rm GL}(2|2)$ &$A|A$& ${\rm U}(2r)$\\ $A$& $-$ & $-$ & ${\rm
       GL}(2|2)$ & ${\rm GL}(2|2)/({\rm GL}(1|1) \times {\rm
       GL}(1|1))$& $A$III$|A$III &${\rm U}(2r)/({\rm U}(r) \times {\rm
       U}(r))$ \\ $C$I& + & + & ${\rm OSp}(2|2)\times {\rm OSp}(2|2)$
     & ${\rm OSp}(2|2)$ &$D|C$ & ${\rm Sp}(2r)$\\ $C$& $-$ & + & ${\rm
       OSp}(2|2)$ & ${\rm OSp}(2|2)/{\rm GL}(1|1)$ &$D$III$|C$I &${\rm
       Sp}(2r)/{\rm U}(r)$\\ \hline
   \end{tabular}\\
 \end{center}
\bottom{-2.7cm}
\begin{multicols}{2}
\narrowtext\noindent
Note that the contents of this table are in complete agreement with
Ref.~\cite{rss}, where each of the ten symmetry classes, as enumerated
by the large families of Cartan's list of irreducible symmetric
spaces, was put in correspondence with the saddle-point manifold of a
non-linear sigma model.  This coincidence is by no means accidental.
Although Ref.~\cite{rss} was formulated in the random-matrix setting,
the translation scheme employed there is of universal validity and
unerringly identifies the {\it internal} (field-theoretical) global
symmetries for each symmetry class.  The present treatment has the
virtue of being well adapted to the specific case of $d$-wave
superconductors, and shows quite explicitly how the hierarchy of
Goldstone modes emerges in this context.

\subsection{Self-consistent Theories}

Before leaving this introductory section, let us return to the
controversy between the predictions made by the various
field-theoretical approaches, and the self-consistent diagrammatic
schemes developed in the literature.  Readers who find the following,
purely qualitative, remarks cryptic are referred to Section
\ref{sec:comparison} for a more substantial discussion.  The last few
years have seen a build-up of confusion concerning the behaviour of
the density of states at very small energies: straightforward
diagrammatic analysis predicts a finite DoS at zero energy, whereas,
in the field theory approaches, unbounded fluctuations of Goldstone
modes lead to a vanishing DoS.  An intuitive argument in favor of a
finite DoS is that any average over random potential fluctuations
should introduce some imaginary self energy $\Sigma$ in the Green
function of the Dirac operator.  This self energy would lead to a
smearing of spectral structures over some energy window of width given
by $\Sigma$ and would, therefore, be in conflict with the formation of
a cusp in the energy-dependent DoS.  This argument, however, is overly
simple; in particular it does not account for the full complexity of
the scattering problem in random Dirac (or gapless superconductor)
systems.  Let us try to discuss this point on a somewhat more
technical level:

The various classes of diagrams appearing in perturbative approaches
can be grouped, roughly speaking, into two categories: diagrams of the
self-energy type (commonly evaluated in SCBA or similar
approximations), and various types of ladder diagrams (alias diffusion
modes).  While the former are usually associated with some smearing of
average single-particle quantities by disorder, the latter describe
mesoscopic fluctuations which become operative at long range.  More
precisely, ladder diagrams describe the long-range quantum
interference between retarded and advanced single-particle Green
functions.  For normal-conducting systems, the distinction between
these diagrammatic elements is perfectly canonical: single-particle
Green functions are renormalized by self-energy diagrams whereas
higher-order Green functions with poles on both sides of the real
energy axis may be influenced by diffusion modes.

For superconductors however, and for gapless superconductors in
particular, this distinction breaks down.  In previous diagrammatic
approaches, the DoS of dirty $d$-wave superconductors was computed
with (self-consistent) account for self-energy diagrams.  This led to
some smearing and, therefore, to a finite DoS.  However, as was shown
in Ref.~\cite{Altland} for the prototypical case of
superconductor/normal quantum dots, the single-particle DoS of a
gapless superconductor may be affected by diffusion-like modes.  This
phenomenon can be understood on various levels of sophistication.  It
can be explained by qualitative semi-classical
reasoning \cite{Altland} or, more technically, by the fact that unlike
in normal systems the single-particle Green function of a
superconductor possesses poles on both sides of the real energy axis.
While for a bulk $s$-wave superconductor, all poles are shifted far
into the complex plane by the bulk order parameter (a manifestation of
Anderson's Theorem \cite{AndersonT}), the situation in gapless
systems is different.  Here, multiple impurity scattering suppresses
the self energy.  Technically, the suppression is caused by ladder
diagrams that sum up to yield diffusion modes.  These modes enter {\it
  already at the single-particle level} \cite{Altland}, and are
missing from previous diagrammatic works on $d$-wave
superconductivity.

In the limit of zero energy, the diffusion modes become massless (or
of infinite range) implying that a purely perturbative approach is met
with a problem.  (Technically, in two dimensions, the contribution of
each diffusion mode scales as $g^{-1} \ln(E/E_0)$, where $E_0$ is some
cutoff energy and $g$ a measure of the dimensionless (spin)
conductance of the system. For $E$ sufficiently small, the expansion
becomes uncontrolled.)  At this point it is good to remember that all
the concepts discussed above have a field-theoretical analog, with an
optional non-perturbative extension.  In particular, the diffusion
modes of the diagrammatic approach have the significance of
perturbative excitations of the Goldstone modes discussed above ---
very much like a spin wave can be interpreted as a perturbative
manifestation of the Goldstone mode related to the intrinsic
spin-rotation symmetry of a ferromagnet.  The fluctuations of these
modes, restricted only by finite energies, become stronger and
stronger as the energy approaches zero. For sufficiently small
energies, (a) perturbative (diagrammatic) schemes for treating these
modes are ruled out, and (b) the non-perturbative analyses discussed
below produce a vanishing DoS.  (In passing we note that such
mechanisms are not unfamiliar in mesoscopic physics.  For example, for
energy differences $\omega = E_1-E_2$ larger than the mean level
spacing $\Delta$, spectral correlations of generic normal disordered
systems can be treated by diffusion-mode diagrammatic approaches.  For
$\omega / \Delta < 1$ perturbation theory breaks down, a phenomenon
that manifests itself in the appearance of unphysical divergences.  In
this regime, a non-perturbative integration over the appropriate
Goldstone mode \cite{efetov} produces the correct result.)
Summarizing, we find that (i) the issue of the DoS is crucially
related to long-range modes of quantum interference operating on the
single-particle level, (ii) to obtain the correct, vanishing,
behaviour of the DoS, a non-perturbative treatment of these modes in
inevitable but that (iii) {\it signatures} for the incompleteness of
previous perturbative analyses can be found within the diagrammatic
framework.

Now, it is important to realize that the arguments above by no means
amount to a categoric prediction of a vanishing zero-energy DoS in
bulk $d$-wave superconductors.  The point is that the weak disorder
field-theoretical approach is based on certain model assumptions which
leave room for complementary scenarios.  Therefore, to complete our
introductory discussion, let us briefly mention a number of competing
theories of the disordered $d$-wave system which lie outside the
field-theoretic scheme outlined above.  Typically such theories rely
on peculiarities of the $d$-wave system.  For example, Balatsky and
Salkola \cite{Balatsky96} have proposed a mechanism whereby the weak
coupling of ``marginally-bound'' impurity states at the Fermi level
leads to the absence of quasi-particle localization.

When subject to a {\em single} impurity in the unitarity limit, a
quasi-particle state is created at zero energy
\cite{Nagaosa95,Krivenko95}.  In a recent work, P\'epin and Lee
\cite{Pepin} have proposed that, when subjected to many such
impurities, these quasi-particle states broaden into a narrow
delocalized band around zero energy.  On this basis, these authors
have proposed that the properties of the dirty $d$-wave system at very
low energies are characterized by extended quasi-particle states.
Based on the RG analysis contained in this paper, it is our belief
that, even for scatterers in the unitarity limit, there must ultimately
be a cross-over at low-energy scales to the physics advocated by SFBN.

Finally, some authors \cite{ahm} have inquired into the importance of
maintaining self-consistency of the $d$-wave order parameter in a
disordered background.  As others before, in this work we will take a
pragmatic view and will not attempt a synthesis of self-consistent
aspects of the theory.

A second and probably more severe limitation regards the role of
interactions.  As with more conventional disordered normal and
superconductor systems, the low-energy physics of quasi-particles in
$d$-wave superconductors, too, will be affected by various mechanisms
of inter-particle interactions: zero-bias anomalies, interaction
contributions to transport coefficients, interaction induced
renormalization of the propagation in the Cooper channel, and more.
Beyond that, one should expect that {\it collective} excitations, both
of purely electronic and of phononic type, contribute to (thermal)
transport coefficients, and further modify the low-energy behaviour of
quasi-particle states.  None of these effects will be considered in
the present paper.  (For a discussion of interaction effects in
$d$-wave superconductors, see Ref.~\cite{vsf}.)  Needless to say, this
may impose severe limitations on the relevance of the present work for
the physics of real $d$-wave superconductors.

To what extent do the predictions of the weakly disordered theory bear
comparison with experiment?  As mentioned above, superconductors of
unconventional $d$-wave symmetry are realized in the high-temperature
cuprate superconductors.  Here the influence of both magnetic and
non-magnetic impurities has been investigated intensively.  In
$d$-wave systems, both types of disorder are pair-breaking and lead to
a rapid depression of $T_c$ with doping. However, at very low doping
concentrations, $T_c$ remains a significant fraction of its optimal
value. In this case the influence of disorder on the low-energy
quasi-particle properties have been studied extensively.

In particular, recent angle resolved photoemission experiments
\cite{Kaminski99} confirm both the integrity of the large-scale nodal
structure of the spectrum and confirm the existence of long-lived
quasi-particle states near the Fermi level. The same measurements show
a large anisotropy of the Dirac nodes with, for example,
Bi${}_2$Sr${}_2$CaCu${}_2$O${}_8$ exhibiting a ratio of $t / \Delta
\equiv \gamma = 20$.

At very low temperatures, phase coherence properties of the
quasi-particle states should be accessible via spin transport.  The
latter have been investigated indirectly through measurements of the
thermal conductivity at very low temperatures: relating the thermal
conductance $\kappa(T)$ to the quasi-particle spin conductivity
$\sigma_s$ through the Einstein relation, one obtains the
Wiedemann-Franz ratio $\kappa(T) / T \sigma_s = (4\pi^2/3) k_B^2$.
Taking the bare value for the spin conductance from theory, one
obtains \cite{sfbn,dl}
\begin{eqnarray*}
  {\kappa (T\ll 1/\tau) \over T} = 4 {\pi^2\over 3} k_B^2 \sigma_s \to
  {k_B^2 \over 3} \left( \gamma + \gamma^{-1} \right) \;.
\end{eqnarray*}
Remarkably, this result is found to be in close agreement with
experiment \cite{Chiao99}. In BiSCCO the extrapolated value of the
longitudinal part of the thermal conductivity remains constant over a
wide range of impurity concentrations. This result has been triumphed
(see, e.g., Ref.~\cite{Millis}) as experimental support for the
self-consistent theory of Lee \cite{Lee}.

However, it should be noted that this result does not sit comfortably
with the class $C$I theory, which predicts that the spin conductance
should exhibit weak localization corrections which renormalize
$\sigma_s$ down from its universal bare value. Whether the seeming
coincidence of the measured conductance and the bare value signifies
strongly enhanced forward scattering, or whether, as seems more
likely, it can be explained in terms of a short phase coherence
length, is as yet unclear.  On this point, it is interesting to note
that low-temperature measurements of the quasi-particle conductivity
in the underdoped cuprate YBa${}_2$Cu${}_4$O${}_8$ \cite{Hussey}
indicate that the quasi-particle states are localized.  Although it
would be useful to explore the thermal Hall transport \cite{Ong},
experiments have been so far unable to access the same very low
temperature regime.

The paper is organized as follows: in Section \ref{sec:basics} we will
formulate some basic concepts required for the construction of a
low-energy theory of the $d$-wave system.  This will include a
symmetry analysis of the disordered Hamiltonian, its representation in
terms of a (supersymmetric) generating functional, and a
renormalization group analysis of the relevancy of the various
channels of disorder.  In Section \ref{sec:proceed} we ask how an
effective low-energy theory can be distilled from the microscopic
functional integral.  We will find that standard schemes, such as a
straightforward gradient expansion, are met with severe difficulties.
In Section \ref{sec:boso}, non-Abelian bosonization will be applied as
an alternative method to derive the low-energy theory.  In a number of
subsections the predictions of this theory for the four symmetry
variants introduced above will be discussed.  In Section
\ref{sec:comparison} we compare these results with the phenomenology
predicted by other approaches, such as self-consistent $T$-matrix
approaches. To complement the theoretical analysis of the low-energy
properties of the d-wave system, in
Section~\ref{sec:numer-analys-quasi} we present a limited review of
the progress made in the numerical investigation of the density of
states and show how these results relate to the different regimes
discussed in the text. Finally, we close with a brief discussion in
Section~\ref{sec:discussion}.

\section{Dirty $d$-wave (basics)}
\label{sec:basics}

In this section we discuss the essential concepts needed for the
construction of a low-energy effective theory describing the
quasi-particles of a disordered spin-singlet superconductor with
$d_{x^2-y^2}$ symmetry.  We start out by reviewing some elementary
features of the Gorkov Hamiltonian for that system.  Much attention
will be paid to the generic symmetries, which turn out to be quite
subtle.  We then express the Green functions of the Hamiltonian as a
field integral.  The result is a supersymmetric generating functional,
which will serve as the basic platform for the low-energy effective
field theories developed in later sections.  Similar considerations
for a class of related two-dimensional systems, having a low-energy
limit modeled by a disordered Dirac equation, were first made by
Fradkin \cite{fradkin}.

\subsection{The Hamiltonian}
\label{sec:d_wave_ham}

Basic to the theory of quasi-particle transport in spin-singlet
superconductors are the Gorkov equations \cite{gorkov2}.  They follow
from a quasi-particle Hamiltonian which in second quantized language,
and on a lattice of sites $i$, is written as
\end{multicols}
\widetext
\top{-2.8cm}
\begin{eqnarray}
  H = \sum_{\langle ij\rangle} (c_{i\uparrow}^\dagger,
  c_{i\downarrow}^{\vphantom{\dagger}})
  \left(\begin{array}{cc} t_{ij}-\mu\delta_{ij} &\Delta_{ij}\\
      \Delta_{ij}& -t_{ij}+\mu\delta_{ij}\end{array}\right)
  \left(\begin{array}{c}c_{j\uparrow}\\
      c_{j\downarrow}^\dagger\end{array}\right),
\label{hamil}
\end{eqnarray}
\bottom{-2.7cm}
\begin{multicols}{2}
\narrowtext\noindent
where $t_{ij}$ are the hopping matrix elements, and $\Delta_{ij}$ is
the lattice $d$-wave order parameter.  Time-reversal and spin-rotation
invariance constrain these matrices to be real symmetric: $t_{ij} =
t_{ji}=\bar t_{ij}$ and $\Delta_{ij} = \Delta_{ji} = \bar\Delta_{ij}$.
Their precise form will be specified shortly.  The operators
$c_{i\sigma}^\dagger$ create spin-1/2 fermions, $\mu$ is the chemical
potential, and the sum $\sum_{\langle ij\rangle}$ extends over nearest
neighbours on a two-dimensional square lattice with spacing $a$.  By
making a particle-hole transformation on down spins,
\begin{eqnarray*}
  d_{\uparrow} \equiv c_\uparrow \;, \qquad 
  d_\downarrow \equiv c^\dagger_\downarrow \;,
\end{eqnarray*}
the Hamiltonian is cast in the concise form
\begin{eqnarray*}
  H=\sum_{\langle ij\rangle} d^\dagger_i ((t_{ij}-\mu\delta_{ij}) 
  \sigma_3 + \Delta_{ij} \sigma_1)d_j \;,
\end{eqnarray*}
where $\sigma_1,\sigma_2,\sigma_3$ are the Pauli matrices.  By way of
warning, we remark that this transformation makes it seem as though
charge were conserved and spin was not.  In reality, the opposite is
true: quasi-particle charge is not conserved, whereas spin is. To
facilitate future manipulations, we perform a $\pi/2$-rotation
$d\mapsto \exp(i\pi \sigma_1/4)d$, to transform to
\begin{equation}
  \label{hatH}
  H = \sum_{\langle ij\rangle} d^\dagger_i 
  ((t_{ij}-\mu\delta_{ij}) \sigma_2 +\Delta_{ij} \sigma_1)d_j \;.
\end{equation}

\begin{figure}[hbt]
  \centerline{\epsfxsize=3.0in \epsfbox{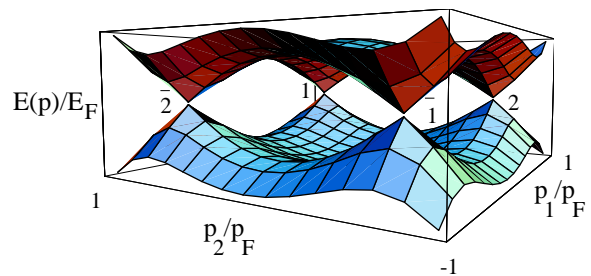}}
  \vskip 0.2truein
  \caption{The quasi-particle spectrum of a clean $d$-wave superconductor 
    showing the existence of four Dirac nodes.}
  \label{fig:dirac}
\end{figure}

The matrix elements $t_{ij}$ and $\Delta_{ij}$ consist of clean and
dirty parts.  The clean part of the Hamiltonian, $H_0$, which we focus
on first, assumes its simplest form in momentum space:
\begin{eqnarray*}
  H_0 = \sum_{k}d^\dagger_k \left[(t(k)-\mu)\sigma_2 + 
  \Delta(k) \sigma_1\right]d_k \;.
\end{eqnarray*}
The kinetic energy and the $d$-wave order parameter are taken to be
\begin{eqnarray*}
  t(k) &=& t (\cos (k_x a)  + \cos (k_y a) ),\\
  \Delta(k) &=& \Delta (\cos (k_x a)  - \cos (k_y a) ).
\end{eqnarray*}
It is the momentum dependence of the order parameter $\Delta(k)$ that
distinguishes $d$-wave superconductors from their relatives with
$s$-wave pairing.  The dispersion relation of $H_0$,
\begin{eqnarray*}
  \epsilon(k) = \pm \sqrt{(t(k)-\mu)^2 + \Delta(k)^2} \;,
\end{eqnarray*}
is displayed in Fig.~\ref{fig:dirac} for the special value $\mu = 0$
(the half filled band). 

The most important feature of the function $\epsilon(k)$ is that it
vanishes (for $\mu = 0$) at the four points 
\begin{eqnarray*}
  (k_x,k_y) = a^{-1} (\pm\pi/2, \pm \pi/2) \;.
\end{eqnarray*}
When the chemical potential is shifted away from half filling, these
nodal points move to different locations in the Brillouin zone.  No
qualitative change is caused by that, and we shall therefore stick to
the case $\mu = 0$.  After the introduction of small momentum offsets
by setting $(k_x,k_y) = (\pm\pi/(2a)+ \delta k_x, \pm \pi/(2a) +
\delta k_y)$, the dispersion relation in the vicinity of the nodal
points is approximated as
\begin{eqnarray*}
  \epsilon(k+\delta k) \simeq \pm a \sqrt{t^2  k_2^2 + \Delta^2 k_1^2}\;,
\end{eqnarray*}
where $k_{{}^1_2} \equiv \delta k_x \mp \delta k_y$.  Thus the
dispersion relation near the nodal points is linear in $k$, and the
low-energy physics will therefore be of Dirac type.  Anticipating that
we will be primarily interested in the nodal regions, we introduce
four species of fermion operators,
\begin{eqnarray*}
  d^n_{\delta k}\equiv d_{k^n + \delta k} \;,
\end{eqnarray*}
where $n \in \{1,\bar 1, 2,\bar 2\}$ labels the nodes, with the 
assignments to node momenta $k^n$ being
\begin{eqnarray*}
  1&\to& a^{-1}(-\pi/2,\pi/2)\;,\\ 2&\to& a^{-1}(\pi/2,\pi/2)\;, \\
  \bar 2&\to& a^{-1}(-\pi/2,-\pi/2)\;,\\ \bar 1&\to& a^{-1} (\pi/2,-\pi/2)\;.
\end{eqnarray*}
If we restrict attention to the low-energy sector, the Hamiltonian 
of the clean system decomposes into four nodal sub-Hamiltonians:
\begin{eqnarray*}
  H_0 = \sum_{n,k} d^{n\dagger}_k H_0^n(k) d^n_k \;.
\end{eqnarray*}
With the definition of the characteristic speed $v$ and the 
anisotropy parameter $\gamma$ by
\begin{eqnarray*}
  v = ta \;, \quad \gamma = {t\over \Delta} \;,
\end{eqnarray*}
the continuum real space representations of the operators $H_0^a$ are
\begin{eqnarray*}
  H^1_{0} &=& -H^{\bar{1}}_{0} = 
  -iv(\gamma^{-1}\sigma_1\partial_1+\sigma_2\partial_2)\;,\\
  H^2_{0} &=& -H^{\bar{2}}_{0} = 
  -iv(\gamma^{-1}\sigma_1\partial_2+\sigma_2\partial_1)\;.
\end{eqnarray*}
Here $\partial_i \equiv \partial_{x_i}$ ($i=1,2$) and $x_{{}^1_2}
\equiv x \mp y$.
%For later convenience, we introduce the velocities $v_1 = v\gamma^{-1}
%= \Delta a$ and $v_2 = v = ta$.

The structure of $H_0$ makes the connection between clean $d$-wave
superconductors and the two-dimensional Dirac Hamiltonian manifest.
An equivalent but more convenient representation of $H_0$ reads
\begin{eqnarray*}
  H_0=\left(\matrix{&H_0^{12}\cr H_0^{21}&}\right) \;,
\end{eqnarray*}
where the four nodal sub-Hamiltonians $H_0^a$ have been assembled into
a single block $H_0^{12}$, and $H_0^{21} = H_0^{12\dagger}$.  The
explicit form of $H_0^{12}$ in the four-component nodal space is
\begin{equation}
  \label{H_0}
  H_0^{12} = v\left(\matrix{
      -i\partial^{(1)} & & &\cr
      &i\partial^{(1)} & & \cr
      &&-i\bar \partial^{(2)} &\cr
      &&&i\bar \partial^{(2)}}\right)
  \quad\matrix{1\cr\bar1\cr 2\cr\bar2}
\end{equation}
with
\begin{eqnarray*}
  \partial^{(1)} &=&  \gamma^{-1} \partial_1 -i \partial_2 \;,\\
  \bar \partial^{(2)} &=& \gamma^{-1} \partial_2 -i \partial_1 \;.
\end{eqnarray*}
The overbar over $\bar \partial^{(2)}$ does not denote complex
conjugation, but is motivated by the observation that, for $\gamma
= 1$, $\bar\partial^{(2)}$ vanishes on holomorphic functions $f(x_1 + 
i x_2)$, while $\partial^{(1)}$ annihilates antiholomorphic functions
$f(x_1 - ix_2)$.

We now introduce disorder. In general, both the normal part of the
Hamiltonian and the order parameter will contain a random component,
which we denote by $V_{\rm imp}$ and $\Delta_{\rm imp}$, respectively.
Specifically, we assume that
\begin{itemize}

\item 
$V_{\rm imp}$ and $\Delta_{\rm imp}$ are uncorrelated Gaussian 
distributed variables with

\item zero mean: $\langle V_{\rm imp}\rangle = \langle \Delta_{\rm imp}
\rangle = 0$,

\item and the same variance:
\begin{eqnarray*}
  \langle X_{{\rm imp}}({\bf r}) X_{{\rm imp}}({\bf r}') \rangle = 
  {g\over 2}  f\left({|{\bf r}-{\bf r}'|/\xi}\right),
\end{eqnarray*}
\end{itemize}
where $X = V$ or $X = \Delta$.  The parameter $\xi$ is a correlation
length which we assume to be much larger than the lattice constant
$a$. (This assumption is motivated by the fact that impurity
scattering in real $d$-wave superconductors seems to be predominantly
forward \cite{kulic1,kulic2,kulic3}. To reproduce that feature, we
have to take $V_{\rm imp}$ and $\Delta_{\rm imp}$ to be slowly
varying.)  The strength of the disorder is measured by the constant
$g$, and $f$ is a correlation function normalized to unity: $ \int
d^2r f(|{\bf r}|/\xi) = 1$.

Later, we will argue that taking $V_{\rm imp}$ and $\Delta_{\rm imp}$
from the same distribution is simply a matter of technical
convenience. Allowing the distribution of disorder in the two channels
to be different will not change the universality class, and therefore
not alter the qualitative behaviour.

\begin{figure}[hbt]
  \centerline{\epsfxsize=2.5in
    \epsfbox{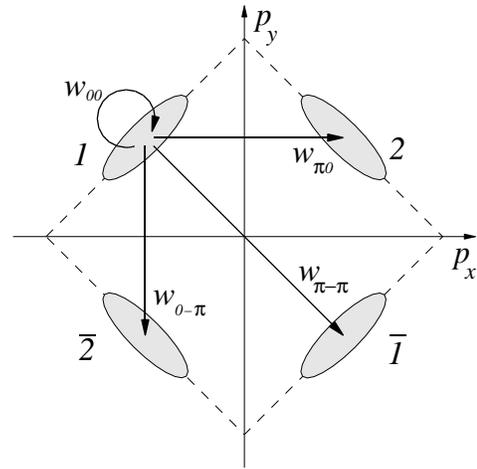}}
  \vspace{0.5cm}
  \caption{Impurity scattering between the nodes of a $d$-wave 
    superconductor.  The ellipsoidal lobes indicate the momentum
    support of the four low-energy sectors $1,\bar 1,2,\bar 2$. By way
    of example, the four different scattering channels $w_{pq}$ with
    mean momentum transfer $(p,q) = a^{-1}[(0,0), (\pi,0), (0,-\pi),
    (\pi,-\pi)]$ acting on node $1$ are indicated.}
  \label{fig:d_wave_scatt}
\end{figure}

The stochastic part of the quasi-particle Hamiltonian then has the form
\begin{eqnarray*}
  H_{\rm imp} \equiv V_{\rm imp} \sigma_2+ \Delta_{\rm imp} \sigma_1 =
  \left(\matrix{&H_{\rm imp}^{12}\cr H_{\rm imp}^{12\dagger}&}\right) \;,
\end{eqnarray*}
where $H_{\rm imp}^{12} = \Delta_{\rm imp} - i V_{\rm imp}$.
Projection of this operator on the space of four low-energy sectors
gives 
\begin{equation}
  H_{\rm imp}^{12}=\left(\begin{array}{llll}
      w_{00}&w_{-\pi\pi}&w_{-\pi0}&w_{0\pi}\\
      w_{\pi-\pi}&w_{00}&w_{0-\pi}&w_{\pi0}\\
      w_{\pi0}&w_{0\pi}& w_{00}&w_{\pi\pi}\\ 
      w_{0-\pi}&w_{-\pi0}& w_{-\pi-\pi}&w_{00}\\
    \end{array}\right)
  \qquad\matrix{1\vphantom{\big{.}}\cr\bar1 \vphantom{\big.}\cr
    2\vphantom{\big{.}}\cr \bar2\vphantom{\big.}} \;,
  \label{H_imp}
\end{equation}
where the matrix elements are random functions of position given by
\begin{eqnarray*}
  w_{p q} = {\rm e}^{i(p x + q y)} (\Delta_{\rm imp} - i V_{\rm imp}).
\end{eqnarray*}
They describe stochastic scattering between nodes that differ by a
momentum $(p,q)$ (c.f. Fig.~\ref{fig:d_wave_scatt}).   Writing
\begin{eqnarray*}
  g_{pq}\equiv \int d^2 r \langle \bar w_{pq}(0) w_{pq}({\bf r}) \rangle
\end{eqnarray*}
for the coupling between the nodes, and assuming the Gaussian
disorder specified above, we have
\begin{eqnarray*}
  g_{pq} = g_{-p-q} = g \int dx dy\, {\rm e}^{i(px + qy)} f(|{\bf r}|/\xi).
\end{eqnarray*}
Note two consequences which now follow from the inequality $\xi \gg
a$: (i) the variances $g_{pq}$ for large momentum transfer $(p,q)$ are
suppressed as compared to the variance $g$ in the soft channel, and
(ii) the various scattering channels are approximately statistically
independent.

The sum of the pure (\ref{H_0}) and disordered (\ref{H_imp}) operators
constitutes the full Gorkov Hamiltonian:
\begin{eqnarray}
  \label{H_canonical}
  H =\left(\matrix{&H^{12}\cr H^{12\dagger}&}\right),\qquad 
  H^{12} \equiv H_0^{12} + H_{\rm imp}^{12}\;,
\end{eqnarray}
which forms the basis for our analysis below.  Before turning to the
generic symmetries of this Hamiltonian, we introduce a perturbation
which is both interesting and physically relevant.

In the present paper we do not touch upon the subject of perturbations
involving the electron spin, but we will repeatedly comment on the
role of time-reversal invariance and the effects of breaking it.
Time-reversal symmetry is broken in the mixed state of type-II
superconductors, where a magnetic field penetrates in the form of
vortices.  Exactly how vortices influence the low-energy density of
quasi-particle states of a pure $d$-wave superconductor has been the
subject of an ongoing debate \cite{gs,anderson,ft,melnikov,mhs,wm}.
Progress was recently made by Franz and Tesanovic (FT) \cite{ft}, who
suggested to perform a singular gauge transformation similar to what
is done in the Chern-Simons gauge theory of the fractional quantum
Hall effect.  Because each vortex carries half a magnetic flux
quantum, a unit cell consisting of two vortices is used.  Dividing the
vortex lattice into $A$ and $B$ sublattices, FT make a singular gauge
transformation (centered around vortices) in the particle sector for
the $A$ sublattice, and in the hole sector for the $B$ sublattice.
The purpose of the transformation is to cancel the magnetic field on
average, thereby allowing to describe the quasi-particle states as
Bloch waves.  Although FT focus on the case of a perfectly regular
vortex lattice, the same transformation can be applied when the vortex
positions are given a random component.  After linearization and
projection on a single node (and after rotation by ${\rm e}^{i\pi
  \sigma_1/4}$), the gauge-transformed Hamiltonian for, say, node 1
is:
\begin{eqnarray*}
  H &=& H_0^1 + {m v \over 2} \Big( (v_2^A + v_2^B) \sigma_0 \\ &+&
  \sigma_2 (v_2^A - v_2^B) + \gamma^{-1} \sigma_1 (v_1^A - v_1^B)
  \Big) \;.
\end{eqnarray*}
Here ${\bf v}^{A,B}$ are the supercurrent velocities for the $A$ and
$B$ sublattices.  We see that the differences of the velocities add to
$V_{\rm imp}$ and $\Delta_{\rm imp}$, while their sum enters as a
scalar potential (proportional to the unit matrix $\sigma_0 = 1_2$ in
particle-hole space).  The latter fact was first understood by Volovik
\cite{volovik}, and is the message to carry away from the above
discussion: the main consequence of time-reversal symmetry breaking by
vortices is the addition of a scalar potential to the single-node
Dirac Hamiltonian.

Anticipating the discussion of the next subsection, we remark that the
different treatment of the particle and hole sectors by FT breaks the
symplectic symmetry (\ref{C_sym_coupled}) of the Gorkov Hamiltonian.
For our purposes, this causes no problems.  The only agent we will
need is the scalar potential proportional to $v_2^A + v_2^B$, and this
can be produced also from a gauge transformation that treats particles
and holes on an equal footing.

\subsection{Generic symmetries}
\label{sec:symmetries}

An essential pre-requisite to the construction of the low-energy
effective theory for dirty $d$-wave superconductors, just as for
disordered metals, is a solid understanding of the fundamental
symmetries of the quasi-particle Hamiltonian.  In recent years it has
become clear that, because of quantum interference channels that owe
their existence to the particle-hole degree of freedom of the Gorkov
equations, the low-energy quasi-particles of disordered
superconductors transcend the established framework of the ``threefold
way'' \cite{dyson}, namely the Wigner-Dyson classification scheme by
unitary, orthogonal, or symplectic symmetry.  Dirty superconductors
generically fall into one of four non-standard symmetry classes
\cite{Altland}.  Borrowing the notation from Cartan's table of
symmetric spaces, these have been termed $C$, $C{\rm I}$, $D$, and
$D{\rm III}$.  Specifically, spin-singlet superconductors (with
conserved quasi-particle spin) belong to the classes $C$ or $C{\rm
  I}$, depending on whether time-reversal symmetry is broken or not.

While these are the generic symmetry classes of dirty superconductors,
other classes appear, and complicate the symmetry pattern, if
system-specific conservation laws of exact or approximate nature are
present.  As we have reviewed, the low-energy quasi-particles of
$d$-wave superconductors organize into four nodal sectors.  In the
physically relevant case of ``soft'' disorder, the scattering between
nodes is suppressed as compared to the scattering within nodes. Of
course, on very large time scales any small amount of scattering
couples the nodes, leading to generic low-energy behaviour.  On
intermediate time scales, however, the nodes can be regarded as
isolated.  What symmetry class the disordered quasi-particles confined
to a single node belong to, is not immediate from general principles.
By inspection, we will find that the class is $A{\rm III}$ if time
reversal invariance is present, and $A$ if not.  While the latter is
just the Cartan label for the ``unitary'' Wigner-Dyson class, an
example for the former are Dirac fermions in a random ${\rm U}(N)$
gauge field (see e.g.~\cite{v1,v2}).  

Having sketched the general picture, we now turn to the details.  We
begin by recalling that the Gorkov Hamiltonian for a disordered
spin-singlet superconductor satisfies the relation
\begin{equation}
  \label{C_sym_coupled}
  {\rm C} : \quad H = - \sigma_2 H^T \sigma_2 \;,
\end{equation}
which we refer to as ``particle-hole'' ({\sc ph}) symmetry.  Its
physical origin is conservation of spin in conjunction with Fermi
statistics, and it is violated if and only if spin ceases to be a
constant of the motion.  For systems where time-reversal invariance
holds, which implies the absence of external fields and complex order
parameter components such as $id_{xy}$, the Hamiltonian matrix can be
chosen to be real symmetric: $H = \bar H = H^T$.  Because of the
rotation by ${\rm e}^{i\pi\sigma_1/4}$ made in going to equation
(\ref{hatH}), time-reversal symmetry here does not take its canonical 
form but is expressed by
\begin{equation}
  \label{T_sym_coupled}
  {\rm T} : \quad H = \sigma_1 H^T \sigma_1 \;.  
\end{equation}
When both ${\rm C}$ and ${\rm T}$ are valid symmetries, the
quasi-particles of the superconductor are said to be in class $C{\rm
  I}$.  When ${\rm T}$ is broken but ${\rm C}$ is still present, the
symmetry class changes to $C$.

The presence of both ${\rm C}$ and ${\rm T}$ constrains $H$ to be of
the form
\begin{eqnarray}
  H = \left( \matrix{&Z\cr \bar Z &} \right), \qquad Z^T = Z \;.
  \label{H_CI}
\end{eqnarray}
Comparing this with the previous section we notice that the
Hamiltonian (\ref{H_canonical}) is indeed off-diagonal but $H^{12}$,
unlike $Z$, is not symmetric.  Looking even further back we see that,
{\it before} linearization and projection on the four nodes, the
off-diagonal block $H^{12}_{ij}=\Delta_{ij}-it_{ij} = H^{12}_{ji}$
still obeyed the constraint of being a symmetric matrix.  What is the
origin of this apparent discrepancy?

The answer is this.  In writing $d^n_{\delta k}\equiv d_{k^n + \delta
  k}$ and making the assignment to nodes, we used {\it different}
conventions in the ``particle'' ($d_\uparrow$) and ``hole''
($d_\downarrow$) sectors, so as to arrange for $H^{12}$ in (\ref{H_0})
to have entries only on the diagonal.  If we had taken the conventions
to be the same, we would have obtained $H^{12} = Z_0 + Z_{\rm imp}$
with
\begin{equation}
  Z_0 = v\left(
    \matrix{0&i\partial^{(1)} & & \cr -i\partial^{(1)} &0 & & \cr
      &&0&i\bar \partial^{(2)} \cr &&-i\bar \partial^{(2)}&0} \right)
  \label{Z_0}
\end{equation}
and
\begin{eqnarray}
  \label{scatt_matrix}
  Z_{\rm imp}=\left(\begin{array}{llll}
      w_{\pi-\pi}&w_{00}& w_{0-\pi}&w_{\pi0}\\
      w_{00}&w_{-\pi\pi}& w_{-\pi0}&w_{0\pi}\\
      w_{0-\pi}&w_{-\pi0}& w_{-\pi-\pi}& w_{00}\\
      w_{\pi0}&w_{0\pi}& w_{00}&w_{\pi\pi}\\
    \end{array}\right) \;,
\end{eqnarray}
both of which are symmetric, in agreement with (\ref{H_CI}).  As
follows from the expression for $Z_0$, here one has to imagine that
the pure Dirac Hamiltonian flips the particle-hole spinor between the
particle state of one node, say $1$, and the hole state of its
conjugate, $\bar 1$.  This interpretation, while perfectly valid and
compatible with momentum conservation, is {\it not} the one commonly
adopted in the literature.  There, one follows the convention of
interchanging the two hole states, so that the Dirac Hamiltonian
simply flips between the particle and hole states of one and the same
node.  To facilitate comparison with the literature, we tacitly made
this change of basis in Section \ref{sec:d_wave_ham}.  The
transformation between the two bases is effected by a matrix we denote
by $\tau_1$:
\begin{equation}
  \label{node_repres}
  \tau_1 = \left(\matrix{0&1&&\cr 1&0&&\cr &&0&1\cr &&1&0}\right)
  \qquad\matrix{1\vphantom{\big{.}}\cr\bar1 \vphantom{\big.}\cr
    2\vphantom{\big{.}}\cr \bar2\vphantom{\big.}} \;.
\end{equation}
By multiplying the particle-hole spinor with this permutation
matrix in the hole sector (but not in the particle sector), the
symmetry relations (\ref{C_sym_coupled}) and (\ref{T_sym_coupled}) are
transformed into
\begin{eqnarray}
  {\rm C} &:& \quad H = - (\sigma_2 \otimes \tau_1) H^T (\sigma_2
  \otimes \tau_1) \;, \nonumber \\ {\rm T} &:& \quad H = \hphantom{-}
  (\sigma_1 \otimes \tau_1) H^T (\sigma_1 \otimes \tau_1) \;.
  \label{CT_sym_iso}
\end{eqnarray}
As is easily verified, these relations are obeyed by the Gorkov
Hamiltonian of equation (\ref{H_canonical}).

In summary, we distinguish between two choices of basis, and hence
between two ways of writing the Hamiltonian.  Both have their
respective advantages and disadvantages, and depending on the given
context we will use one or the other, whichever is better suited.  We
write 
\begin{eqnarray*}
H_{\rm e} = U H U^\dagger
\end{eqnarray*}
for the Hamiltonian in the canonical representation (\ref{H_CI}).

There now exist two different scenarios.  If the inter-node scattering
is assumed to be strong, a low-energy quasi-particle prepared in a
given initial configuration will quickly attain a state where its
wavefunction is uniformly spread over the space of four nodes.  In
that case, the presence of the nodal exchange operator $\tau_1$ in
(\ref{CT_sym_iso}) is of no consequence, and we are back to the
generic situation where the symmetry class is $C$ or $C{\rm I}$
depending on whether time-reversal invariance is broken or not.

On the other hand, if the inter-node scattering in $H_{\rm imp}$ is
negligibly weak, the four low-energy sectors decouple, and then the
two individual operations in (\ref{CT_sym_iso}), both of which relate
nodes to their conjugates, become ineffective.  What remains effective
is an operation relating each isolated node to itself.  By applying
the symmetries ${\rm C}$ and ${\rm T}$ in sequence,
\begin{eqnarray*}
  H \stackrel{{\rm C}}{\longrightarrow} - (\sigma_2 \otimes \tau_1)
  H^T (\sigma_2 \otimes \tau_1) \stackrel{{\rm T}}{\longrightarrow}
  - \sigma_3 H \sigma_3 \;,
\end{eqnarray*}
we see two facts: (i) the product operation CT does not mix the
low-energy sectors and (ii) if both ${\rm C}$ and ${\rm T}$ are good
symmetries, the nodal sub-Hamiltonians are odd under conjugation by
$\sigma_3$.  (The latter follows more directly from the basic equation
(\ref{hatH})).  Given the presence of disorder, no further symmetries
are expected.  Thus, the quasi-particle Hamiltonian for an isolated
node is constrained only by
\begin{eqnarray*}
  H = - \sigma_3 H \sigma_3 \;.
\end{eqnarray*}
This relation is the defining equation of the ``chiral'' symmetry
class $A{\rm III}$.

Any perturbation that breaks time-reversal invariance destroys ${\rm
  T}$ and hence ${\rm CT}$.  Under such conditions, the nodal
sub-Hamiltonians do not have any symmetry other than Hermiticity.  We
say that Hamiltonians of the last type belong to class $A$, the
standard Wigner-Dyson class with unitary symmetry.

This concludes our analysis of symmetries.  We have seen that,
depending on the hard or soft nature of the scattering potential and
the presence or absence of time-reversal invariance, the Gorkov
Hamiltonian for dirty $d$-wave superconductors belongs to one of four
symmetry classes: $C{\rm I}$, $C$, $A{\rm III}$, or $A$.  These
symmetries have physical consequences (such as singular vertex
corrections to the density of states, the thermal conductivity etc.),
which are best evaluated in a field-theoretical formalism.

\subsection{Field-integral formulation}
\label{sec:field_integral}

Our goal is to compute the disorder average of the Green function,
$\langle G(E) \rangle = \langle (E-H)^{-1} \rangle$, and for this
purpose we employ the machinery of supersymmetry \cite{efetov}.  In
that method, Green functions are generated from a Gaussian functional
integral,
\begin{eqnarray*}
  {\cal Z}[j,k] = \int {\cal D} \phi {\cal D}\psi \, 
  {\rm e}^{i\int \phi^T (E-H)\psi+\int (\phi^T j + k^T \psi)},
\end{eqnarray*}
where ${\rm Im\;}E > 0$ is assumed and $H$, defined in
(\ref{H_canonical}), is the Gorkov Hamiltonian projected on the four
nodal regions of the $d$-wave superconductor.
\begin{eqnarray*}
  \psi\equiv \left(\matrix{S\cr \chi} \right)\;, \qquad
  \phi\equiv \left(\matrix{\bar S\cr \kappa} \right) \;,
\end{eqnarray*}
are fields that have $2\times 2\times 4$ components each, where
$S,\bar S$ ($\chi,\kappa$) denote complex (Grassmann) fields with
$2\times 4$ components in the tensor product of {\sc ph} and node
space.  While convergence of the integral requires the commuting
components of the fields $\phi$ and $\psi$ to be related by complex
conjugation, the anticommuting components of the two fields are
independent. As usual, Green function matrix elements are obtained by
differentiating twice with respect to the sources $j$ and $k$.

A noteworthy feature due to the symmetry ${\rm C}$ in
(\ref{CT_sym_iso}) is that retarded (${\rm Im\;}E>0$) and advanced
(${\rm Im\;}E<0$) Green functions are related by
\begin{eqnarray*}
  G(E)=-(\sigma_2\otimes\tau_1)G^T(-E)(\sigma_2\otimes\tau_1)\;.
\end{eqnarray*}
This has the consequence that the functional ${\cal Z} [j, k]$
generating a single Green function can also be used to compute the
disorder average of the {\it two-particle} Green function $\langle
G(0^+)G(0^-)\rangle$ at zero energy ($E = 0^\pm \equiv 0\pm i\delta$).
As the emphasis here will be on the basic structure of the theory
(rather than on specific observables), we temporarily suppress the
source content and focus on the functional ${\cal Z}[0]$.

In the next step, we are going to adapt the functional integral to the
particular symmetries of the Gorkov Hamiltonian.  In normal
conductors, time-reversal symmetry is known to give rise, via the
so-called Cooperon mode, to quantum interference corrections to
diffusion that are {\it infrared singular} in dimension $d \le 2$.
Similar modes appear in the present case \cite{Altland}, as a result
of the discrete symmetries ${\rm C}$ and ${\rm T}$.  In the impurity
diagram technique, these modes emerge as divergent series of ladder
graphs, or maximally crossed diagrams.  For a semiclassical
interpretation of these modes we refer to \cite{Altland}.  The
field-theoretic approach elevates their status to those of Goldstone
modes due to the breaking of certain global continous symmetries,
which in turn derive from ${\rm C}$ and ${\rm T}$.

In view of this, we now exercise special care to translate the
fundamental symmetries of the Hamiltonian into symmetries of the
functional integral.  For that purpose we find it convenient to
transform to the representation $U H U^\dagger = H_{\rm e} = \sigma_1
{\rm Re}Z - \sigma_2{\rm Im}Z$ introduced in the previous subsection.
Using it, we write
\begin{eqnarray*}
  {\cal Z}[0] = \int\exp\left( iE(\tilde\beta^T\gamma+\beta^T\tilde
    \gamma) -i\tilde\beta^T Z \tilde\gamma - i \beta^T \bar Z \gamma
  \right) \;,
\end{eqnarray*}
where the unitary transformation $U$ was absorbed into the measure of
the functional integral (which is implicit, as is integration of the
exponent over position space).  $\beta, \tilde\beta$ ($\gamma,\tilde
\gamma$) are the ${\sc ph}$ components of $\phi$ ($\psi$).

Now recall that the operators $Z,\bar Z$ are {\it symmetric}.  In
order to firmly install this property in the functional integral, we
symmetrize and ``double'' \cite{efetov} the field space:
\begin{eqnarray*}
  \beta^T \bar Z \gamma = {\textstyle{1\over 2}}\beta^T\bar Z\gamma 
  + {\textstyle{1 \over 2}} \gamma^T \sigma_3^{\sc bf} \bar Z \beta
  \equiv \Psi^s \bar Z \Psi \;.
\end{eqnarray*}
Here the matrix $\sigma_3^{\sc bf}$ acts in the boson-fermion space of
commuting and anticommuting field components, and the spinor $\Psi$
and its transpose $\Psi^s$ are defined by
\begin{equation}
  \Psi = {1 \over \sqrt{2}} \pmatrix{ \gamma\cr \beta\cr} \;,
  \quad \Psi^s = {1 \over \sqrt{2}} \left( \beta^T \;,\; 
    \gamma^T \sigma_3^{\sc bf} \right) \;.
  \label{pspinor}
\end{equation}
We refer to the two-dimensional space comprising the upper ($\gamma$)
and lower ($\beta$) components of the spinor $\Psi$ as ``charge
conjugation'' ({\sc cc}) space.  Note that the scalar product
determined by the transposition rule $\Psi \mapsto \Psi^s$ is
symmetric: $\Psi^s \Psi' = \Psi'^s \Psi$.  Doing the same
symmetrization for the other terms in the exponent, we obtain
\begin{eqnarray}
  \label{fun_int_basic}
  &&{\cal Z}[0] = \int {\cal D}\Psi {\cal D}\tilde\Psi \, {\rm e}^{-\int 
    {\cal L} d^2r}, \quad {\cal L} = {\cal L}_D + {\cal L}_E \;,\\
  &&{\cal L}_D = i \tilde\Psi^s Z \tilde\Psi + i\Psi^s \bar Z \Psi,   
  \quad {\cal L}_E = -2iE \tilde\Psi^s \Psi \;. \nonumber
\end{eqnarray}
A node-resolved representation of the Lagrangian is gotten by 
making the decompositions
\begin{eqnarray*}
  \Psi \equiv \left(\matrix{\Psi_1\cr \Psi_{\bar 1}\cr
      \bar \Psi_2 \cr \bar \Psi_{\bar 2}}\right),\qquad 
  \tilde\Psi \equiv \left(\matrix{\bar \Psi_1\cr \bar \Psi_{\bar 1}\cr
      \Psi_2 \cr  \Psi_{\bar 2}}\right) \;.
\end{eqnarray*}
The overbar here does {\it not} mean complex conjugation.  Following
the conventions of conformal field theory, we use it to denote fields
whose correlation functions are antiholomorphic in the limit of zero
disorder.  In the final step, we substitute the nodal decomposition into
(\ref{fun_int_basic}), which gives
\begin{eqnarray}
  \label{L_basic}
  &&{\cal L}_D\equiv{\cal L}_D^1+{\cal L}_D^2+{\cal L}_D^{12},\nonumber\\
  &&{\cal L}_D^1 = 
  2 \bar \Psi_1^s (v\partial^{(1)}+iw_{00})\bar \Psi_{\bar 1} +
  2 \Psi_{\bar 1}^s (v\bar \partial^{(1)}+i\bar w_{00})\Psi_1 \nonumber\\
  &&\hspace{2.0cm} +i w_{\pi-\pi} \bar\Psi_1^s \bar\Psi_1^{\vphantom{s}} 
  +iw_{-\pi\pi}\bar\Psi_{\bar 1}^s\bar\Psi_{\bar 1}^{\vphantom{s}}\nonumber\\
  &&\hspace{2.0cm}+i\bar w_{\pi-\pi}\Psi_1^s \Psi_1^{\vphantom{s}}
  +i\bar w_{-\pi\pi}\Psi_{\bar 1}^s\Psi_{\bar 1}^{\vphantom{s}}\;,\nonumber\\
  &&{\cal L}_D^2 = 
  2 \Psi_2^s (v\bar \partial^{(2)}+iw_{00}) \Psi_{\bar 2} + 2 \bar 
  \Psi_{\bar 2}^s (v\partial^{(2)}+i\bar w_{00})\bar \Psi_2 \nonumber\\
  &&\hspace{2.0cm} +iw_{-\pi-\pi} \Psi_2^s \Psi_2^{\vphantom{s}} +
  iw_{\pi\pi} \Psi_{\bar 2}^s \Psi_{\bar 2}^{\vphantom{s}}\nonumber\\
  &&\hspace{2.0cm} +i\bar w_{-\pi-\pi} \bar \Psi_2^s 
  \bar\Psi_2^{\vphantom{s}} +i\bar w_{\pi\pi}\bar\Psi_{\bar 2}^s
  \bar\Psi_{\bar 2}^{\vphantom{s}}\;,\nonumber\\ 
  &&{\cal L}_D^{12} = 2i\big(w_{0-\pi} \bar \Psi_1^s \Psi_2^{\vphantom{s}}
  + w_{\pi 0} \bar\Psi_1^s \Psi_{\bar 2}^{\vphantom{s}} \nonumber\\
  &&\hphantom{{\cal L}_{D}^{12} = }
  ~+ w_{-\pi 0} \bar \Psi_{\bar 1}^s \Psi_2^{\vphantom{s}} +
  w_{0\pi} \bar\Psi_{\bar 1}^s \Psi_{\bar2}^{\vphantom{s}}\nonumber\\
  &&\hphantom{{\cal L}_D^{12} = }
  ~+ \bar w_{0-\pi} \Psi_1^s \bar\Psi_2^{\vphantom{s}} 
  + \bar w_{\pi 0} \Psi_1^s \bar\Psi_{\bar 2}^{\vphantom{s}}\nonumber\\
  &&\hphantom{{\cal L}_D^{12} = }
  ~+ \bar w_{-\pi 0} \Psi_{\bar 1}^s \bar\Psi_2^{\vphantom{s}}
  + \bar w_{0\pi} \Psi_{\bar 1}^s \bar\Psi_{\bar2}^{\vphantom{s}}\big)\;,
\end{eqnarray}
where ${\cal L}_{D}^1, {\cal L}_{D}^2$, and ${\cal L}_{D}^{12}$ govern
the pairs of nodes $(1,\bar1)$, $(2,\bar2)$, and the coupling between
them, respectively.  The factors of two in these expressions arise
from combining terms: $\Psi'^s A \Psi \pm \Psi^s A \Psi' = 2\Psi'^s A
\Psi$ for a symmetric (or antisymmetric) operator $A$.

Equations (\ref{fun_int_basic}) and (\ref{L_basic}) cast the dirty
$d$-wave problem into the field-theoretical form that all subsequent
analysis will be based on.

\subsection{Symmetries of the Gaussian field theory}
\label{sec:symmetries_ft}

We next explore the symmetries of the Lagrangian (\ref{L_basic}).  The
outcome will pre-determine the structure of the low-energy effective
field theories that derive from it.

To prepare the stage, we recall that the word ``symmetries'' has two
aspects to it.  Firstly, there exists a symmetry group, which acts by
transformations that leave the Lagrangian invariant.  This is the
proper meaning of the word ``symmetry''.  Secondly, the degrees of
freedom of the theory, the fields, take values in a target manifold.
In current speak the latter, too, is sometimes referred to as the
``symmmetry'' of the theory, although it is of course different from
the symmetry group.  The remark we wish to make is the trivial and yet
important statement that these two meanings must be kept apart.  In
the present section we are going to discuss the first aspect, namely
the symmetry group of the action functional.  The structure of the
target manifold will be discussed in Section \ref{sec:proceed}, after
the introduction of the relevant low-energy degrees of freedom.

We begin by considering the case of highest symmetry where all of the
disorder except for $w_{00}$ is switched off.  Symmetry breaking
caused by hard scattering $w_{pq}$ and the perturbations that break
time-reversal invariance, will be discussed afterwards.

{\it Class $A{\rm III}$} (forward scattering, ${\rm T}$-invariance):
on setting $E = 0$ and $w_{pq} = 0$ (for $pq \not= 00$), the
Lagrangian reduces to four independent terms:
\begin{eqnarray*}
  {\cal L} &=&
  2\bar \Psi_1^s (v\partial^{(1)}+iw_{00})\bar \Psi_{\bar 1} +
  2\Psi_{\bar 1}^s (v \bar\partial^{(1)}+i\bar w_{00}) \Psi_1 \\
  &+& 2\Psi_2^s (v\bar\partial^{(2)}+iw_{00})  \Psi_{\bar 2} + 
  2\bar\Psi_{\bar 2}^s (v\partial^{(2)}+i\bar w_{00})\bar \Psi_2 \;,
\end{eqnarray*}
which are distinguished by the dichotomy of holomorphic $(\Psi)$
versus antiholomorphic $(\bar\Psi)$ fields, and by the grouping into
pairs of nodes: $(1,\bar 1)$ and $(2,\bar 2)$.  This Lagrangian
describes (anisotropic) Dirac fermions in a random Abelian vector
potential $A = w_{00}$, $\bar A = \bar w_{00}$.  Since the four terms
are identical in structure, but involve different partial derivatives
$\partial^{(1)} \not= \partial^{(2)}$ for non-vanishing anisotropy
$\gamma \not= 1$, the symmetry group will be a Cartesian product of
four copies of the same group, which is readily identified as the
supergroup ${\rm GL}(2|2)$.  Indeed, looking at, say, the first two
terms, and noting that $w_{00}({\bf r})$ and $\bar w_{00}({\bf r})$
are just complex numbers, we see that a local transformation
\begin{eqnarray}
  &&\Psi_{\bar 1}^s({\bf r}) \mapsto \Psi_{\bar 1}^s({\bf r})
  T(z_1)^{-1} \;, \quad \Psi_1({\bf r}) \mapsto T(z_1) \Psi_1({\bf r})
  \;, \nonumber \\ &&\bar\Psi_1^s({\bf r}) \mapsto \bar\Psi_1^s({\bf
    r}) \bar T(\bar z_1)^{-1} \;, \quad \bar\Psi_{\bar 1}({\bf r})
  \mapsto \bar T(\bar z_1) \bar\Psi_{\bar 1}({\bf r})\;,
  \label{GL_sym}
\end{eqnarray}
where $z_1 \equiv \gamma x_1 + i x_2$ and $\bar z_1 \equiv \gamma x_1
- i x_2$, leaves the action functional invariant.  Because $\Psi_1,
\Psi_{\bar 1}$ comprise two commuting and two anticommuting
components, the invertible supermatrix $T$ is of size $(2+2) \times
(2+2)$.  Hence the symmetry group of the present case is ${\rm
  GL}(2|2)$ or, rather, four independent copies thereof.  As we shall
see, the local nature of the symmetry group makes the present theory
exactly solvable.  Note that here, and throughout this section, the
word ``symmetry group'' means the {\it complexified} \cite{complex}
symmetry group, which ignores the relation that exists between the
bosonic field components and their complex conjugates.

We mention in passing that the symmetry group of the totally clean
limit ($w_{00} = \bar w_{00} = 0$) described by free fields, is four
copies of the orthosymplectic supergroup ${\rm OSp}(4|4)$.  This will
come to play a role when we turn to the method of non-Abelian
bosonization.

{\it Class $A$} (soft scattering, no {\rm T}-invariance): as we saw in
Section \ref{sec:d_wave_ham}, the Hamiltonian $H_{\rm e}$ ceases to be
off-diagonal in {\sc ph} space (while remaining diagonal in {\sc cc}
and node space) when time-reversal symmetry is broken.  Hence the
breaking of ${\rm T}$ introduces terms into the Lagrangian that mix
the two sectors of holomorphic and antiholomorphic fields.
Consequently, invariance of the Lagrangian requires the actions of
${\rm GL}(2|2)$ in these sectors to be related to each other.  This
``locking'' of group actions is the only effect ${\rm T}$-breaking has
on the symmetries, and therefore the net result is that the four
copies of ${\rm GL}(2|2)$ get reduced to two copies, corresponding to
the two pairs of nodes $(1,\bar1)$ and ($2,\bar 2)$.

Let us look at the locking of ${\rm GL}(2|2)$ group actions in more
detail.  As we recalled in Section \ref{sec:d_wave_ham}, the
supercurrent flow in the mixed state acts as a scalar potential $(v^A
+ v^B) 1 \otimes \tau_3$.  After transformation to the canonical
representation, $H_e = U H U^\dagger$, this reads $(v^A + v^B)
\sigma_3 \otimes \tau_3$.  Consider therefore the ${\rm T}$-breaking
perturbation
\begin{eqnarray*}
  \psi^T (\sigma_3 \otimes \tau_3) \phi = \tilde\beta^T \tau_3 \gamma
  - \beta^T \tau_3 \tilde\gamma = 2 \tilde\Psi^s (\sigma_3^{\sc cc}
  \otimes \tau_3) \Psi \;.
\end{eqnarray*}
In order for this perturbation to remain unchanged under the
transformation (\ref{GL_sym}), we must require $\bar T^s = \eta T^{-1}
\eta$ where $\eta = \sigma_3^{\sc cc} \otimes \tau_3$.  We will use
this relation to identify the target manifold for class $A$ in Section
\ref{sec:proceed}.

{\it Class $C{\rm I}$} (hard scattering, ${\rm T}$-invariance): we
next consider the situation where the hard scattering channels are
switched on.  The matrix elements $w_{pq} \not= 0$ then mix fields
pertaining to different nodes, whence the four copies of ${\rm GL}
(2|2)$ for class $A{\rm III}$ no longer act independently but become
locked to each other.  Moreover, in order for terms in the Lagrangian
such as $\Psi_1^s \Psi_1^{\vphantom{s}}$ to be invariant under
$\Phi_1({\bf r}) \mapsto T \Phi_1({\bf r})$, we require $T^s T = 1$,
where the supermatrix transpose $T^s$ is defined by $(T \Phi)^s =
\Phi^s T^s$.  From (\ref{pspinor}) we read off that $\Phi^s$ is
obtained from $\Phi$ by taking the ordinary transpose, and then
exchanging the two commuting (anticommuting) components by 
multiplication with $\sigma_1$ ($i\sigma_2$).  The equation $g^T 
\sigma_1 g = \sigma_1$ defines an orthogonal group ${\rm O}(2,
C)$, and $g^T \sigma_2 g = \sigma_2$ the symplectic group ${\rm 
Sp}(2,C)$.  Thus the condition $T^s T = 1$ determines an 
orthosymplectic subgroup ${\rm OSp} (2|2)$ of ${\rm GL}(2|2)$.  The 
full symmetry group consists of two copies of ${\rm  OSp}(2|2)$, 
since the holomorphic fields of the pair $(1,\bar 1)$ and 
antiholomorphic fields of $(2,\bar 2)$ can be transformed 
independently of the remaining fields.

The following remark may be helpful.  Given the physical distinction
between the {\sc bb} and {\sc ff} sectors, there are two versions of
${\rm OSp}$ to consider.  The first one acts by symplectic
transformations in the former sector and by orthogonal transformations
in the latter.  (We refer to this as ``symplectic'' bosons and
``orthogonal'' fermions for short.)  The ${\rm OSp}(4|4)$ symmetry
group of the free-field problem is of this kind.  In the second version
of ${\rm OSp}$, which is the relevant one here, the roles of the
orthogonal and symplectic groups are interchanged (``orthogonal''
bosons and ``symplectic'' fermions).

As another aside, we mention that en route from soft to hard
scattering one could imagine a scenario where the scattering couples
each node only with its conjugate \cite{ntw}.  Such a scenario would
lead to a different symmetry group and eventually to a different
target manifold \cite{fk}.  However, we see no physical room for this
theoretical possibility and will not pursue it here.

{\it Class $C$} (hard scattering, no {\rm T}-invariance): the symmetry
group for this final case can be approached from class $A$ by
including hard scattering, or from class $C$I by breaking ${\rm T}$.
In either way, one finds that the remaining symmetry is just a single
copy of the group ${\rm OSp}(2|2)$ with orthogonal bosons and
symplectic fermions.  This is the orthosymplectic supersymmetry
omnipresent in class $C$ \cite{rss,bcsz1,bcsz2,bele}.

For future reference, the symmetry groups for the four different cases
are summarized in the following table, where the notation $\times^n G$ 
means $n$ independent copies of the group $G$.
\begin{center}
  \begin{tabular}{{|c|c||l|}}\hline
    ${\rm T}$& hard scattering& symmetry group\\ \hline
    + & -- & $\vphantom{\big|}\times^4 {\rm GL}(2|2)$ \\
    + & + & $\vphantom{\big|}\times^2 {\rm OSp}(2|2)$ \\
    -- & -- & $\vphantom{\big|}\times^2 {\rm GL}(2|2)$\\
    -- & + & $\vphantom{\big|}{\rm OSp}(2|2)$\\
    \hline
  \end{tabular}
\end{center}

\subsection{Disorder average}
\label{sec:dis_av}

We now carry out the disorder average.  By integrating the Gaussian
generating functional over the distribution of the matrix elements
$w_{pq}$, which are subject to the correlation laws stated in Section
\ref{sec:d_wave_ham}, we obtain
\begin{eqnarray*}
  {\cal Z}_{\rm av}[0]\equiv \langle{\cal Z}[0]\rangle = \int {\cal
    D}\Psi {\cal D}\bar\Psi \,{\rm e}^{-\int ({\cal L}_{\rm D} + {\cal
      L}_E) d^2r} \;,
\end{eqnarray*}
where 
\begin{eqnarray}
  \label{L_average}
  &&{\cal L}_{\rm D} = 2v {\cal L}_0 + {\cal L}_{\rm dis} \;,\\ 
  &&{\cal L}_0 = \bar\Psi_1^s \partial^{(1)}\bar \Psi_{\bar 1} +
  \Psi_{\bar 1}^s \bar \partial^{(1)} \Psi_1 + \bar\Psi_{\bar 2}^s
  \partial^{(2)}\bar \Psi_2 + \Psi_2^s \bar \partial^{(2)} \Psi_{\bar
    2} \;,\nonumber\\ 
  &&{\cal L}_{\rm dis} = g {\cal O}_{00} + g' {\cal O}_{00}' + g_{\pi
    0} {\cal O}_{\pi 0} + g_{\pi \pi} {\cal O}_{\pi\pi} \;, \nonumber
\end{eqnarray}
and all perturbations ${\cal O}_{pq}$ are quartic in the fields:
\begin{eqnarray*}
  {\cal O}_{00} &=& 4 \bar \Psi_1^s \bar \Psi^{\vphantom{s}}_{\bar 1}
  \Psi_1^s \Psi^{\vphantom{s}}_{\bar 1} + 4 \Psi_2^s
  \Psi^{\vphantom{s}}_{\bar 2} \bar \Psi_2^s \bar
  \Psi^{\vphantom{s}}_{\bar 2}\;,\\ 
  {\cal O}_{00}'&=& 4 \bar\Psi_1^s \bar\Psi^{\vphantom{s}}_{\bar 1}
  \bar\Psi_2^s \bar\Psi^{\vphantom{s}}_{\bar 2} + 4 \Psi_1^s
  \Psi^{\vphantom{s}}_{\bar 1} \Psi_2^s \Psi^{\vphantom{s}}_{\bar
    2}\;,\\ 
  {\cal O}_{\pi 0} &=& 4 \bar\Psi_1^s \Psi^{\vphantom{s}}_2 \Psi_1^s
  \bar\Psi^{\vphantom{s}}_2 + 4 \bar\Psi_{\bar 1}^s
  \Psi^{\vphantom{s}}_{\bar 2} \Psi_{\bar 1}^s \bar
  \Psi^{\vphantom{s}}_{\bar 2}\;,\\ 
  &+& 4 \bar\Psi_{\bar 1}^s \Psi^{\vphantom{s}}_2 \Psi_{\bar 1}^s \bar
  \Psi^{\vphantom{s}}_2 + 4 \bar\Psi_1^s \Psi^{\vphantom{s}}_{\bar
    2}\Psi_1^s \bar\Psi^{\vphantom{s}}_{\bar 2}\;,\\ 
  {\cal O}_{\pi\pi} &=& \bar\Psi_{\bar 1}^s
  \bar\Psi^{\vphantom{s}}_{\bar 1} \Psi_{\bar 1}^s
  \Psi^{\vphantom{s}}_{\bar 1} + \bar\Psi_1^s
  \bar\Psi^{\vphantom{s}}_1\Psi_1^s \Psi^{\vphantom{s}}_1 \\ &+&
  \Psi_{\bar 2}^s \Psi^{\vphantom{s}}_{\bar 2} \bar \Psi_{\bar 2}^s
  \bar\Psi^{\vphantom{s}}_{\bar 2} + \Psi_2^s \Psi^{\vphantom{s}}_2
  \bar \Psi_2^s \bar \Psi^{\vphantom{s}}_2 \;.
\end{eqnarray*}
In the initial stage of the calculation, disorder averaging produces
expressions of the form
\begin{eqnarray*}
  \int d^2r \int d^2r' \, f(|{\bf r}-{\bf r}'|/\xi) 
  \, (\Psi^s\Psi)({\bf r}) \, (\Psi^s \Psi) ({\bf r}') \;,
\end{eqnarray*}
where $f$ is the correlation function of the disorder.  To arrive at
the above expressions for ${\cal O}_{pq}$, we omitted the finite
spread of this function due to a non-zero value of $\xi$.  This
approximation is justified by the fact that all corrections from
expansion around the local limit ${\bf r} - {\bf r}' = 0$ carry at
least two derivatives, which renders them irrelevant in the
renormalization group (RG) sense.  Note that all of the perturbations
${\cal O}_{pq}$ are marginal by power counting at the free-fermion
point.

Eq.~(\ref{L_average}) defines the disorder-averaged theory.  Before
going further, we will inquire into the nature of the RG flow caused
by the perturbations ${\cal O}_{pq}$.

\subsection{Renormalization group}
\label{sec:RG}

In this section we perform a one-loop renormalization group analysis
to explore the relevance of the five couplings $g_{pq}$.  There exists
a standard formula \cite{zamoRG,cardysbook} to use for that purpose,
which refers to the operator product expansion (OPE) of the
perturbating operators.  The general statement is that, if the
short-distance expansion for a set of marginal perturbations ${\cal
  O}^{(i)}$ has the form
\begin{eqnarray*}
  \int\limits_{\ell < |{\bf r} | < \ell + \delta\ell} {\cal
    O}^{(i)}({\bf r}) {\cal O}^{(j)}(0) d^2r = 2\pi {\delta\ell \over
    \ell} \sum_k c_k^{ij} {\cal O}^{(k)}(0) + \dots
\end{eqnarray*}
the corresponding couplings $g_k$ renormalize according to the
equation
\begin{eqnarray*}
  {d g_k \over d \ln \ell} = \beta_k ({\bf g}) 
  = -\pi \sum_{ij} c_k^{ij} g_i g_j \;,
\end{eqnarray*}
where $\ell$ is the short-distance cutoff, $\beta_k$ are the beta
functions, and $c_k^{ij}$ are called the structure constants of the
algebra of operators ${\cal O}^{(i)}$.

To apply this formula to the problem at hand, we need the OPE for the
fundamental fields.  Apart from an overall multiplicative constant,
which can be removed by a conformal rescaling of the fields, these are
\begin{eqnarray*}
  \bar \Psi_{\bar 1,A}(x_1,x_2) \bar \Psi_{1,B}(0,0) &=&
  {\delta_{AB} \over \alpha x_2 - i \alpha^{-1} x_1}\;,\\ 
  \Psi_{\bar 1,A}(x_1,x_2) \Psi_{1,B}(0,0) &=& {-\delta_{AB}
    \over \alpha x_2 +i \alpha^{-1} x_1}\;,\\ \Psi_{\bar 2,A}
  (x_1,x_2) \Psi_{2,B}(0,0) &=& {\delta_{AB} \over \alpha x_1 -i
    \alpha^{-1} x_2}\;,\\ \bar \Psi_{\bar 2,A} (x_1,x_2) \bar
  \Psi_{2,B}(0,0) &=& {-\delta_{AB} \over \alpha x_1 +
    i\alpha^{-1} x_2} \;.
\end{eqnarray*}
Here $\alpha = \sqrt{\gamma} = \sqrt{t/\Delta}$ is the (square root of
the) anisotropy parameter, and $A,B$ is a composite index built
from the {\sc bf} and {\sc cc} indices of the fields.  As usual in
this context, the above relations have to be understood as identities
that {\it hold under the functional integral sign}.

By using the free-field expansions in conjunction with Wick's theorem,
we can now work out the OPE for the set of composite operators ${\cal
  O}_{pq}$.  By a straightforward if tedious calculation, this yields the 
RG flow equations
\begin{eqnarray}
  \dot g'&=& 0 \nonumber \;,\\ \dot g &=& \textstyle{{1\over 2}}
  g_{\pi\pi}^2 + g_{\pi 0}^2 \nonumber \;,\\ \dot g_{\pi 0} &=& g \,
  g_{\pi 0} + \textstyle{{1\over 4}} g_{\pi \pi} g_{\pi 0} \nonumber
  \;,\\ \dot g_{\pi\pi} &=& 2(g\,g_{\pi\pi} + g_{\pi 0}^2 )\;,
  \label{RG_flow}
\end{eqnarray}
where $\dot g_i$ stands for $d g_i/d\ln\ell$ (times some unimportant
constant which we do not specify).  Notice the following features:

\begin{itemize}
\item The beta functions contain no terms of linear order in the 
  couplings, which expresses the fact that we are dealing with a set of 
  marginal perturbations.
\item In the physical regime of positive couplings, the beta functions
  are never negative, so none of the couplings decreases under the RG
  flow.
\item None of the couplings supports its own flow. In particular, for
  the model with only forward scattering ($g_{pq}=0$) the coupling
  constant $g$ is truly marginal.
\item For generic initial data, the nature of the flow is marginally
  relevant, i.e. the couplings (save for the constant $g'$)
  increase under the flow, although the {\it rate} of increase
  vanishes in the limit of weak disorder $g_i \searrow 0$.
\item The anisotropy parameter $\gamma$ does not enter the one-loop RG
  equations.
\end{itemize}

The fact that the anisotropy has disappeared from the RG equations can
be understood heuristically as follows.  The OPE scheme for the set
$\{ {\cal O}_{pq} \}$ encodes the one-loop renormalization of the
four-fermion vertices of the theory.  These vertices are related to
the fermion self energy through a Ward identity.  To explore the
effects of the anisotropy, one may therefore directly analyse the
self-energy diagrams.  The latter have been shown \cite{ntw} to
be unaffected by the anisotropy, to leading (or one-loop) order.  The
physical reason is that the first-order self energy diagram measures
the Born scattering rate between the four low-energy sectors.  This
rate is not affected by the ellipsoidal shape (expressing the
anisotropy) of the low-energy lobes in momentum space; it only depends
on the total phase volume available.  In higher orders of perturbation
theory, the situation changes and the self energy begins to be
affected by the anisotropy in momentum space.  In particular,
higher-order scattering between neighbouring nodes is suppressed
\cite{ntw}.  Accordingly, we expect that the beta functions at
higher loop orders do depend on the anisotropy of the model.

Let us stress two messages that emerge from the current section: (i)
owing to the quadratic dependence of the beta functions on their
arguments, the RG flow is marginal (or very slow) in the limit of
small couplings (or weak disorder), and (ii) for generic values of the
disorder-generated couplings, the renormalization group flow drives
the system away from the free-fermion theory.  In combination with the
phenomenological input that the inter-node couplings $g_{pq}$ are 
small as compared to the intra-node coupling $g$, the first message 
leads us to expect a prolonged crossover intervening between a 
ballistic regime at short scales, and the diffusive regime governed by 
a non-linear sigma model at large scales.

\section{How to proceed?}
\label{sec:proceed}

Anticipating that a direct perturbative analysis of the theory
(\ref{L_average}) will not capture all of the important physics, one
is tempted to employ the ``standard'' scheme for dealing with weakly
disordered non-interacting particles: Hubbard-Stratonovich
transformation followed by a saddle-point analysis and gradient
expansion. It turns out, however, that this approach is beset with a
number of problems, none of which appears in systems with a
non-relativistic kinetic energy.  These difficulties will eventually
force us to adopt a strategy where elements of the standard scheme are
supplemented, and even superseded, by an alternative scheme tailored
to relativistic fermions: non-Abelian bosonization.

For pedagogical reasons, we will here proceed in a conservative
fashion and continue somewhat further along the much-trodden path
(Hubbard-Stratonovich transformation and so on), introducing along the
way a few concepts of general validity.  In particular, we will
identify the degrees of freedom of the hierarchy of low-energy
effective theories, and the way they are associated with the global
symmetries discussed in Section \ref{sec:symmetries_ft}.  As a
by-product, we will be able to make the connection to previous 
self-consistent approaches.

To begin with, we set the couplings $g_{pq}$ to zero and focus on the
functional integral
\begin{eqnarray*}
  Z_{\rm fs}[0] \equiv Z_{\rm av}[0] \big|_{g_{pq} = 0} \;,
\end{eqnarray*}
containing only the largest perturbation, $g{\cal O}_{00} + g' {\cal
  O}_{00}'$.  The first step of the standard approach is to make a
Gaussian transformation, called the ``Hubbard-Stratonovich
transformation''.  Introducing two auxiliary fields $P$ and $Q$, and
exploiting the equality of the unrenormalized coupling constants $g =
g'$, it is straightforward to show that
\begin{eqnarray*}
  && Z_{\rm fs}[0] = \int {\cal D}\Psi {\cal D}\tilde\Psi \int {\cal
    D} Q{\cal D}P \; {\rm e}^{-{1\over g}\int {\;\rm STr\;}(Q^2 +
    P^2)}\\ &&\hspace{1.0cm} \times {\rm e}^{-\int ({\cal L}_E + 2v
    {\cal L}_0 + \tilde \Psi^s(iQ-P)\Psi + \Psi^s \tau_1 (iQ+P)
    \tau_1 \tilde \Psi)}\\ && \hspace{0.5cm} = \int {\cal D}Q{\cal
    D}P \; {\rm e}^{-\frac{1}{g}\int {\;\rm STr\;}(Q^2 + P^2)}\\ 
  &&\hspace{0.2cm} \times \exp \, - {\textstyle{1 \over 2}}\, {\rm
        STr} \ln \left[ {E \over i}+ \pmatrix{iQ-P & iZ_0 \cr 
            i\bar Z_0 & \tau_1(iQ+P)\tau_1} \right] \;,
\end{eqnarray*}
where the values of $P$ and $Q$ are supermatrices of linear size $4
\times 2 \times 2$ acting in nodal space and internal space (the
latter being the tensor product of charge conjugation and
boson-fermion space).  For simplicity we here ignore the issue of
convergence of integrals, deferring a more strict treatment until
Section \ref{sec:boso}.

We mention in passing that it can be understood at this stage why the
restriction to equally distributed disorder in the normal and order
parameter channels makes no essential difference: one can relax this
assumption by introducing two more Hubbard-Stratonovich fields, $R$
and $S$, coupling to the {\sc ph} matrices $\sigma_1$ and $\sigma_2$,
respectively.  On varying the action with respect to these fields, one
finds that they vanish on the saddle-point level.  Since a
non-vanishing saddle point is a necessary condition for the formation
of Goldstone modes, which control the infrared behaviour, we conclude
that these fields and hence the choice of (un)equally distributed
disorder, are of no importance.

With the aim of subjecting the functional integral to a
stationary-phase analysis, we now vary the action with respect to the
fields $Q$ and $P$ to generate a set of saddle-point equations.  We
first look for solutions $(P_0,Q_0)$ that are diagonal in both nodal
and internal space.  For vanishing energy $E$, the solutions are
\begin{eqnarray*}
  P_0 &=& 0,\\ Q_0 &=& -i\kappa \;,
\end{eqnarray*}
where the real parameter $\kappa$ is defined implicitly through
\begin{equation}
  \label{spe}
  {2\over g}=\int {d^2k\over(2\pi)^2}\,\frac{1}{\kappa^2+\epsilon(k)^2} \;.
\end{equation}
This equation, which can be regarded as the Dirac analog of the
SCBA equation for disordered normal-conducting systems (with $-Q_0$
representing the self energy), has been discussed in the literature
(see, e.g., Refs.~\cite{Lee,ntw}).  Its solution is 
\begin{eqnarray*}
  \kappa^2 = \mu^2 {\rm e}^{- 8\pi \Delta t/g} \;,
\end{eqnarray*}
where $\mu$ represents a UV cutoff that regularizes the integral over
momenta, which would otherwise diverge logarithmically.  The parameter
$\kappa$ can be regarded as a self-consistently determined self
energy.  This interpretation is strengthened by re-instating sources
$j,k$ into ${\cal Z}_{\rm fs}[0]$, and then using this generating
functional to compute matrix elements of the physical Green function.
Doing so, one finds that these are given by the inverse of the
operator
\begin{equation}
  \pmatrix{\kappa - iE & iZ_0 \cr i\bar Z_0 & \kappa -iE \cr} \;,
\end{equation}
and the information content of the above saddle-point approximation
coincides with that of the diagrammatic SCBA approach. 

However, as we know from experience with disordered normal-conducting
systems, there exists much physics that does not unfold at the
saddle-point level, but resides in the fluctuations of $Q$.  The same
is true here.  The saddle point $(Q_0,P_0) = (-i\kappa,0)$ breaks the
symmetry $\times^4 {\rm GL}(2|2)$ for class $A{\rm III}$ as recorded
in the table at the end of section \ref{sec:symmetries_ft}.  By the
Goldstone mechanism, this leads to the appearance of massless modes,
and one expects that it is the fluctuations in these ``soft'' modes,
rather than any single saddle point, that determines the behaviour of
the system at large scales or low energies.

Let us therefore ask what happens when the functional integral $Z_{\rm
  fs}[0]$ is subjected to a global symmetry transformation taken from
$\times^4 {\rm GL}(2|2)$.  (As we saw, the symmetry is actually {\it
  local}, but we should hesitate to draw any conclusions from that, as
the axial part of the symmetry is anomalous; see Appendix
\ref{chiral_anomaly}).  Since these transformations do not mix the
nodal sectors $(1,\bar 1)$ and $(2,\bar 2)$, we can concentrate on one
of them, say $(1,\bar 1)$.  Consider the action functional evaluated
at the saddle point,
\begin{eqnarray*}
  {1\over 2} {\rm STr} \ln \pmatrix{\kappa &-iv\tau_2
    \partial^{(1)}\cr iv\tau_2\bar{\partial}^{(1)} &\kappa\cr}\;,
\end{eqnarray*}
where $\tau_2$ denotes the second Pauli matrix acting in nodal space,
and the energy $E$ has temporarily been set to zero.  Then recall the
transformation (\ref{GL_sym}) with the matrices $(T,\bar T) \in {\rm
  GL}(2|2) \times {\rm GL} (2|2)$, which are now taken to be constant
in space.  By transferring this transformation to the argument of the
logarithm, we obtain the ``rotated'' action
\begin{eqnarray}
\label{S_rotated}
  {1 \over 2}{\rm STr} \ln \pmatrix{ \kappa M^s &0 &0
    &-v\partial^{(1)} \cr 0 &\kappa M^{-1} &v\partial^{(1)} &0\cr 0
    &v\bar\partial^{(1)} &\kappa M &0\cr -v\bar\partial^{(1)} &0 &0
    &\kappa (M^s)^{-1} \cr} \;,
\end{eqnarray}
where $M = T \bar T^s$.  This expression states that not only the
diagonal matrix $i\kappa$, but in fact any ${\rm GL}(2|2)$
configuration $i\kappa \times {\rm diag}(M^s,M^{-1},M,(M^s)^{-1})$ is
a solution of the saddle-point equation.  Put differently, the
saddle-point manifold for the nodal sector $(1,\bar 1)$ is isomorphic
to ${\rm GL}(2|2)$.  On including an identical factor for the other
sector, $(2,\bar 2)$, the total saddle-point manifold of the model
with only forward scattering becomes ${\rm GL}(2|2) \times {\rm
  GL}(2|2)$.  (Note that a complete description of the saddle-point
approximation would have to specify the real submanifold to be
integrated over.  The present discussion gives only the {\it complex}
saddle-point manifold for simplicity.)

We note in passing that what we are encountering here is reminiscent
of the phenomenon of chiral symmetry breaking in quantum
chromodynamics.  The appearance of a chiral symmetry group $G_L \times
G_R = {\rm GL} (2|2) \times {\rm GL}(2|2)$ (still maintaining the
focus on a single nodal sector), with independent left and right
factors, is directly connected with the {\it oddness} of the
Hamiltonian under conjugation by $\sigma_3$ in {\sc ph} space:
$\sigma_3 H \sigma_3 = - H$.  The latter symmetry is broken by the
saddle point $Q_0 = -i\kappa$, which is {\it even} under conjugation
by $\sigma_3$.  In four space-time dimensions, the role of $i\sigma_3
= \sigma_1 \sigma_2$ is played by $i\gamma_5 = \gamma_0 \gamma_1
\gamma_2 \gamma_3$, for massless fermions one has $\gamma_5 H \gamma_5
= - H$, and $M$ becomes the pion field.  In QCD, as in the present
case, the saddle-point manifold is given by a diagonal subgroup $G_L =
G_R$ (the ``axial'' symmetry transformations) of the chiral symmetry
group.  The main difference turns out to be that chiral symmetry is
not truly broken (although its axial part is anomalous) in the vacuum
of our 2d theory.

For the case of forward scattering with ${\rm T}$-invariance (class
$A{\rm III}$) the above discussion answers the question, raised at the
beginning of Section \ref{sec:symmetries_ft}, concerning the
relationship between the global symmetries of the field theory and its
degrees of freedom.  The former are represented by $T$ and $\bar T$,
and the latter by $M = T \bar T^s$.  The next step of the standard
scheme would be to allow $M$ to vary slowly and carry out a gradient
expansion of the action functional.  However, for reasons that are
spelled out below, we do not pursue this approach here but will switch
to an altogether different strategy in the next section.  Before doing
so, we extend the construction of saddle-point manifolds to the other
symmetry classes: $C{\rm I}$, $A$, and $C$.  The procedure is always
the same:  we let the global symmetry group act on the diagonal
saddle-point $Q_0 = -i\kappa$, and the saddle-point manifold is then
simply the result of this group action.

{\it Class} $C{\rm I}$ (hard scattering, ${\rm T}$-invariance): Recall
from Section \ref{sec:symmetries_ft} that the couplings $g_{pq}$,
which mix the nodes and lead from class $A{\rm III}$ to $C{\rm I}$,
reduce the symmetry group from four copies of ${\rm GL}(2|2)$ to two
copies of ${\rm OSp}(2|2)$.  Because there are no other changes, the
above form of the symmetry-transformed action functional continues to
hold, with $M$ still given by $M = T \bar T^s$.  The orthosymplectic
group property, $T^s = T^{-1}$ and $\bar T^s = \bar T^{-1}$, entails
$M^s = M^{-1}$.  As $T$ and $\bar T$ vary over ${\rm OSp}(2|2)$, so
does $M$, and hence the (complex) saddle-point manifold is isomorphic
to that group.  The symmetry group is still chiral, acting on the
field $M({\bf r})$ independently on the left and right by $M({\bf r})
\mapsto T M({\bf r}) \bar T^s$.

{\it Class} $A$ (soft scattering, no ${\rm T}$-invariance): Again, we
focus without loss on the pair $(1,\bar 1)$.  According to the
discussion in Section \ref{sec:symmetries_ft}, breaking of ${\rm
  T}$-invariance by a supercurrent flow locks the left and right
actions of ${\rm GL}(2|2)$ to each other by $\bar T^s = \eta T^{-1}
\eta$ with $\eta = \sigma_3^{\sc cc} \otimes \tau_3$.  Because $M$ and
$T$ are node-diagonal, this equation simplifies to $\bar T^s =
\sigma_3^{\sc cc} T^{-1} \sigma_3^{\sc cc}$. The expression for $M = T
\bar T^s$ thus becomes $M = T \sigma_3^{\sc cc} T^{-1} \sigma_3^{\sc
  cc}$, which is invariant under translations $T \mapsto Th$ by
elements $h$ that leave $\sigma_3^{\sc cc}$ fixed: $h \sigma_3^{\sc
  cc} h^{-1} = \sigma_3^{\sc cc}$.  These are easily seen to form a
subgroup ${\rm GL}(1|1) \times {\rm GL}(1|1)$ of ${\rm GL}(2|2)$.
Hence the complex saddle-point manifold is isomorphic to
\begin{displaymath}
  {\rm GL}(2|2)/({\rm GL}(1|1) \times {\rm GL}(1|1)) \;.
\end{displaymath}
This coset space is very familiar \cite{efetov} as the target manifold
of the non-linear sigma model for systems in the Wigner-Dyson class
with unitary symmetry (alias class $A$).  The standard symbol for the
field in this case is $Q = M \sigma_3^{\sc cc}$.  Identical statements
apply to the pair $(2,\bar 2)$.

{\it Class} $C$ (hard scattering, no ${\rm T}$-invariance): As before, 
the constraint $\bar T^s= \sigma_3^{\sc cc} T^{-1} \sigma_3^{\sc cc}$ 
from ${\rm T}$-breaking leads to $Q = T \sigma_3^{\sc cc} T^{-1}$.
What is different now is that the field is completely locked in node 
space, and $T \in {\rm OSp}(2|2)$ obeys the condition $T^s = T^{-1}$.
Again, the field $Q$ does not change when $T$ is multiplied on the right
by an element $h$ that fixes $\sigma_3^{\sc cc}$ under conjugation.
Such elements turn out to form a subgroup ${\rm GL}(1|1)$, and $Q$ 
therefore parametrizes the coset space ${\rm OSp}(2|2)/{\rm GL}(1|1)$.

This concludes our discussion of saddle-point manifolds.  A summary of
the essential results was tabulated in the introductory section.  Let
us mention that major elements of that table have appeared in the
literature, although in the framework of the fermion-replica trick:
\begin{itemize}
\item Prior to the celebrated discovery that the order parameter of
  high-temperature superconductors has $d_{x^2-y^2}$ symmetry,
  Oppermann \cite{Oppermann} had studied a system he called the
  ``superconducting Ising glass'', and derived for it a non-linear
  sigma model over the compact group ${\rm Sp}(2r)$, with $r = 0$
  owing to the use of the replica trick.  Oppermann's superconducting
  Ising glass is a time-reversal invariant system with local
  superconducting order and conserved spin.  Thus it has all the
  prerogatives of class $C{\rm I}$, and the target manifold ${\rm
    Sp}(2r)$ is just what one expects on the basis of the general
  classification \cite{rss}.  For the important case of $d$-wave
  superconductivity, the ${\rm Sp}(2r)$ non-linear sigma model was
  recently rediscovered by Senthil et al.~\cite{sfbn}.
\item According to Nersesyan, Tsvelik, and Wenger \cite{ntw}, the
  effect of impurity scattering on an isolated node of a $d$-wave
  superconductor is described by a non-linear sigma model (more
  precisely, by a WZW model) on ${\rm U}(r)$.  We will discuss a
  supersymmetric variant of that model in some detail later.
\item The non-linear sigma model with target manifold ${\rm Sp}(2r) /
  {\rm U}(r)$ was identified as the low-energy effective theory for
  dirty $d$-wave superconductors in class $C$ by Senthil et
  al.~\cite{sfbn}.  The supersymmetric extension of that theory had
  appeared in \cite{ast_sc} (to describe the physics of low-energy
  quasi-particle excitations in disordered superconductor/normal metal
  junctions) and \cite{bcsz1} (as a description of quasi-particles in
  the core of a disordered vortex.)
\item The appearance of the Wigner-Dyson class of unitary symmetry
  (class $A$) for $d$-wave superconductors in the mixed state is
  implicit to several recent papers \cite{volovik,ye,ft}.
\end{itemize}
The merit of the present work is that it assembles the various
symmetry classes and field theories into a single coherent scheme.

After this extensive tour of symmetries and their field-theoretical
realization, we are now ready to attack the concrete goal of giving a
bona fide construction of the low-energy effective theories.  Equipped
with a strong background in the supersymmetric theory of disordered
metals, one would deem the strategy to follow quite obvious: one
should decompose the Hubbard-Stratonovich fields $P,Q$ into Goldstone
(or massless) modes $M$ and a complementary set of massive modes.  One
would then integrate over the latter in Gaussian approximation and,
finally, expand to lowest order in the gradients $\nabla M$, measuring
the energy cost due to spatial variations in the massless fields.  It
came as a surprise to us that this program is corrupted by two sources
of severe difficulty:

\begin{itemize}
\item In the course of carrying out the gradient expansion we are
  confronted with a two-dimensional version of the chiral anomaly (see
  Appendix \ref{chiral_anomaly}).
\item For the physically interesting limit of systems with ${\rm
    T}$-invariance and forward scattering only (class $A{\rm III}$),
  we are unable to isolate and eliminate the massive fields in a
  controlled manner.
\end{itemize}

The presence of an anomaly leads to disastrous results when the
gradient expansion is carried out in the most straightforward way.
This, however, is a surmountable problem; once the nature of the
anomaly has been understood, a less naive and properly regularized
expansion scheme can be set up to yield the correct answer.  The
second problem is more serious.  Its origin is that Dirac fermions are
perturbed by a random Abelian vector potential (which is the concrete
form taken by the disorder in the present realization of class $A{\rm
  III}$) in a truly marginal way.  What this means is that the system
retains the conformal invariance of the free-fermion theory, with {\it
  no mass scale} being generated.  As a result, any attempt to
integrate out the ``massive'' modes perturbatively is doomed to fail.
In principle, one could try to go beyond the Gaussian saddle-point
approximation for the massive modes but, in practice, this seems
unfeasible.

In this situation it comes as a relief that an alternative approach to
the problem exists: the method of non-Abelian bosonization due to
Witten \cite{Witten84}.

\section{Non-Abelian bosonization}
\label{sec:boso}

The utility of non-Abelian bosonization as a tool to construct the
low-energy effective action for the quasi-particles of a $d$-wave
superconductor was first recognized by Nersesyan, Tsvelik, and Wenger
\cite{ntw}.  These authors used the fermion-replica trick to deal with
Dirac fermions perturbed by various types of disorder. By bosonizing
the disorder-averaged fermion-replica theory, they arrived at a
Wess-Zumino-Novikov-Witten (WZW) model.  This model had the attractive
feature of exact solvability, which enabled NTW to predict some exact
values for the scaling exponent of the low-energy density of states.
As discussed in the introduction, however, the original approach of
NTW has a number of weak spots, one of which, namely the neglect of
the fact that the multi-valued term in the WZW action changes sign
when the orientation of position space is reversed, bears drastic
consequences.  Following NTW's original strategy, it is the goal of
this section to introduce a bosonization scheme in which (i) all
symmetries of the system are fully included, (ii) the dependence of
the WZW action on orientation is accounted for, and (iii) the physics
of the crossover regimes in between the limiting cases of hard and
soft scattering is included. We will begin by reviewing a few elements
of the general bosonization approach to supersymmetric theories, and
then discuss the application of these concepts to the case at hand.

Originally, non-Abelian bosonization was introduced to construct
bosonic representations of relativistic two-dimensional fermion models
with continuous internal symmetries \cite{Witten84}.  More recently,
the approach has been extended \cite{bsz} to the supersymmetric case
of relativistic fermions supplemented by a bosonic ghost system, which
is the setup needed to treat problems with disorder.  Below we will
review some elements of this approach, in a form tailored to the
$d$-wave application.  Readers who are familiar with bosonization will
find that there are no structural differences to the original,
fermionic version.  Readers who are not may wish to consult first an
introduction to standard non-Abelian bosonization, e.g.~Witten's
original and highly pedagogical article \cite{Witten84}.  A detailed
account of the principal supersymmetric extension of the approach can
be found in Ref.~\cite{bsz}.

Before turning to the problem of the $d$-wave superconductor, let us
begin with some preliminary considerations of a generic model:
consider the functional integral
\begin{eqnarray*}
  {\cal Z} \equiv \int {\cal D}\psi {\cal D}\bar\psi \, 
  {\rm e}^{-S_0[\psi,\bar\psi]} (\dots)_{\psi,\bar\psi} \;,
\end{eqnarray*}
where $\psi$ and $\bar\psi$ are fields with $2m$ bosonic and $2m$
fermionic components and the ellipses $(\dots)_{\psi,\bar\psi}$ stand
for certain operators constructed from these fields. The action
functional $S_0$ denotes the free supersymmetric action
\begin{equation}
  \label{nb_Spsi}
  S_0[\psi,\bar \psi] = \int d^2 r \, 
  (\bar\psi^t \partial \bar\psi + \psi^t \bar \partial\psi) \;,
\end{equation}
where $\psi \mapsto \psi^t$ is an ``orthosymplectic'' transpose which
{\it differs} from the $s$-operation introduced earlier in that it
defines a {\it skew} symmetric scalar product:
\begin{eqnarray*}
  \psi^t \psi' = - \psi'^t \psi \;.
\end{eqnarray*}
(Although skew symmetry is the only property that matters for our
purposes, it may be helpful to recall the concrete realization given
in \cite{bsz} for $n = 1$: $\psi^t = (\beta,\gamma,b,-c)$ and $\psi =
(\gamma,\beta,c,b)^{\rm T}$, where Greek and Roman letters denote
fermionic and bosonic components, respectively.)  It is implicitly
assumed that the existence of the functional integral is ensured by
the presence of some convergence generating term in the action. (For
an explicit example of such a term, see below.)

Furthermore, let $M$ be a field taking values in the (anomalous)
symmetry group of $S_0$, which is the complex supergroup ${\rm OSp}
(2m|2m)$ of matrices fulfilling $M^t M = 1$, with the matrix operation
$M \mapsto M^t$ defined by $(M\psi)^t = \psi^t M^t$.  The basic
statement of non-Abelian bosonization then is that the functional
integral ${\cal Z}$ defined above is equivalent to
\begin{eqnarray*}
  {\cal Z}'\equiv \int_{C|D} {\cal D}M\,{\rm e}^{-W[M]}(\dots)_M \;,
\end{eqnarray*}
where 
\begin{eqnarray*}
  W[M]\equiv -{1\over 16\pi} \int d^2r {\,\rm STr\,}(\partial_\mu
  M^{-1}\partial_\mu M) + {i\over 24\pi} \Gamma[M] \;,
\end{eqnarray*}
$\Gamma[M]$ is the WZW functional, and the ellipses stand for a
representation of $(\dots)_{\psi,\bar\psi}$ in terms of the matrix
field $M$ to be specified momentarily. The phrase ``equivalent'' here
means that, with the transcription $(\dots)_{\psi, \bar\psi}
\leftrightarrow (\dots)_M$ understood, the two functional integrals
${\cal Z}$ and ${\cal Z}'$ produce identical results.

Following Witten, an explicit expression for $\Gamma$ can be written
down by choosing some extension of the field $M$ to a one-parameter
family of fields $\tilde M(t)$ with the property $\tilde M(0) = 1$ and
$\tilde M(1) = M$. Then $\Gamma[M] = $
\begin{eqnarray*}
  \int d^2 r \int_0^1 dt \, \epsilon^{\lambda\mu\nu} {\,\rm STr\,}
  (\tilde M^{-1} \partial_\lambda \tilde M \tilde M^{-1} \partial_\mu
  \tilde M \tilde M^{-1} \partial_\nu \tilde M) \;,
\end{eqnarray*}
where $(\lambda,\mu,\nu)$ represents the coordinate triple $(t, x_1,
x_2)$, and $\epsilon^{\lambda\mu\nu}$ is the fully antisymmetric
tensor.

The bosonization dictionary needed for the transcription $(\dots)_
{\psi,\bar\psi} \leftrightarrow (\dots)_M$ is given as follows:
\begin{eqnarray}
  \psi \psi^t &\leftrightarrow& J \equiv (2\pi)^{-1} M \partial M^{-1}
  \;, \nonumber \\ \bar \psi \bar \psi^t &\leftrightarrow& \bar J
  \equiv (2\pi)^{-1} M^{-1} \bar \partial M\;, \nonumber \\ \psi \bar
  \psi^t &\leftrightarrow& \ell^{-1} M\;, \nonumber\\ \bar \psi \psi^t
  &\leftrightarrow& \ell^{-1} M^{-1}\;,
\label{dictionary}
\end{eqnarray}
where $\ell$ is some length scale serving to UV-regularize the Dirac
theory.

Finally, the subscript ``$C|D$'' in the definition of ${\cal Z}'$
indicates that the functional integration over $M$ does not really
extend over ${\rm OSp}(2m|2m)$. The reason is that the invariant
metric ${\rm STr} (M^{-1}{\rm d}M)^2$ on this group, as on any complex
group, is not Riemannian.  This in turn implies that a functional
integral controlled by the metric term ${\rm STr\,} (\partial_\mu
M^{-1}\partial_\mu M)$ cannot be defined by an unrestricted
integration over ${\rm OSp}(2m|2m)$.  (This term has indefinite sign,
which spoils the convergence of the integral.)  For ordinary groups
the obvious remedy is to restrict the complex group to a compact real
subgroup; one passes, for example, from ${\rm O}(2m,C)$ to ${\rm
  O}(2m)$.  No such remedy exists for supergroups such as ${\rm
  OSp}(2m|2m)$, as these do not possess subgroups on which the
invariant metric becomes of definite sign.  To rescue the functional
integral, one has to abandon the group structure and restrict ${\rm
  OSp}(2m|2m)$ to what is called a Riemannian symmetric superspace of
type $C|D$ in the terminology of Ref.~\cite{rss}.  Referring for a
detailed discussion to Ref.~\cite{bsz}, we here merely mention that
the {\sc bb} sector of this manifold is given by the non-compact coset
space ${\rm Sp}(2m,C)/{\rm Sp}(2m)$, while the {\sc ff} sector is the
compact group ${\rm O}(2m)$.

We next apply this general apparatus to the $d$-wave system. Firstly,
it is necessary to cast the free action $\int d^2 r {\cal L}_0$ (we
here set $v = 1$) of the system into the form of the action $S_0$
above.  To this end we define
\begin{eqnarray}
  \label{bo_psi_def}
  \bar \psi_1^t \equiv (\bar \Psi_1^s,-\bar \Psi_{\bar
    1}^s),\qquad&& \bar \psi_1 \equiv \left(\matrix{\hphantom{-}\bar
      \Psi_{\bar 1}\cr\hphantom{-} \bar \Psi_1}\right)\;,\nonumber\\ 
  \psi_1^t \equiv (\Psi_{\bar
    1}^s,\hphantom{-}\Psi_1^s),\qquad&& \psi_1 \equiv \left(\matrix{
      \hphantom{-}\Psi_1\cr -\Psi_{\bar 1}}\right)\;, \nonumber\\ 
  \bar\psi_2^t \equiv (\bar\Psi_{\bar 2}^s,-\bar
  \Psi_2^s),\qquad&& \bar \psi_2 \equiv \left(\matrix{\hphantom{-} \bar
      \Psi_2\cr \hphantom{-} \bar \Psi_{\bar 2}}\right)\;,\nonumber\\ 
  \psi_2^t \equiv (\Psi_2^s,\hphantom{-}\Psi_{\bar
    2}^s),\qquad&& \psi_2 \equiv \left(\matrix{ \hphantom{-}
      \Psi_{\bar 2}\cr - \Psi_2} \right) \;.
\end{eqnarray}
Note $\bar\psi_1^t \bar\psi_2 = - \bar\psi_2^t \bar\psi_1$ and
$\psi_1^t \psi_2 = - \psi_2^t \psi_1$ (skew symmetry), but $\bar
\psi_1^t \psi_2 = + \psi_2^t \bar\psi_1$.  One then verifies that the
free part of the Lagrangian can be represented as
\begin{eqnarray*}
  && {\cal L}_0 = \sum_{n=1,2}\left(\bar \psi_n^t \partial^{(n)}
    \bar\psi_n^{\vphantom{t}} + \psi_n^t \bar \partial^{(n)}
    \psi_n^{\vphantom{t}} \right),\\ &&{\cal L}_E = -2iE
  \sum_{n=1,2} \bar\psi_n^t \psi_n^{\vphantom{t}} \;.
\end{eqnarray*}
To make this Lagrangian amenable to a direct application of the
bosonization rules, we temporarily remove the anisotropy inherent to
the operators $\partial^{(n)}$.  This can be done by rescaling the
coordinates for the sector of nodes $(1,\bar 1)$ according to $x_1 \to
\gamma^{-1} x_1, x_2 \to x_2$ and the coordinates for the
complementary sector $(2,\bar 2)$ according to $x_1 \to x_1, x_2 \to
\gamma^{-1} x_2$.  The derivative operators then assume the isotropic
form, $\partial^{(1)} \to \partial_1 - i\partial_2, \bar\partial^{(2)}
\to \partial_2 - i\partial_1$. (It must be kept in mind, however, that
in the two sectors different scaling operations were performed.  This
type of scaling is only meaningful for those elements of the theory
which do not involve a coupling between neighbouring nodes.  We will
therefore undo the scaling immediately after the individual sectors
have been bosonized.)

The bosonization of the free Lagrangian ${\cal L}_0$ then leads to the
sum $W[M_1] + W[M_2]$ of two WZW actions, where $W[M_n]$ represents
the sector $(n,\bar n)$ and $M_n$ ($n=1,2$) are two independent fields
taking values in the Riemannian symmetric superspace of type $C|D$
associated with the supergroup ${\rm OSp}(4|4)$.  From the symmetry of
the orthosymplectic currents, $J_n = \psi_n \psi_n^t = - \tau_2 J_n^s
\tau_2$ and $\bar J_n = \bar\psi_n \bar\psi_n^t = - \tau_2 \bar J_n^s
\tau_2$, the field $M_n$ inherits the property
\begin{eqnarray*}
  M_n^{-1} = \tau_2 M_n^s \tau_2 \;.
\end{eqnarray*}
As before, $\tau_k$ ($k = 1,2,3$) denotes the Pauli matrices acting in
node space (presently, the two-component space underlying the
definition (\ref{bo_psi_def})).  To prepare the bosonization of the
remaining operators, including those that couple the nodes, we now
undo the above scaling operation.  This is achieved by making the
replacements
\begin{eqnarray*}
  W[M_1] \to W[M_1,\gamma] \;, \quad W[M_2] \to W[M_2,\gamma^{-1}],
\end{eqnarray*}
where
\begin{eqnarray*}
  &&W[M,\gamma] \equiv {i\over 24 \pi} \Gamma[M]-{1\over 16\pi} \int
  dx_1 dx_2 \times\\ &&\hspace{1.0cm}\times {\,\rm STr\,} \left(
    \gamma^{-1} \partial_1 M^{-1}\partial_1 M + \gamma \partial_2
    M^{-1} \partial_2 M \right)
\end{eqnarray*}
is the anisotropic analog of the WZW action above.  Notice that, owing
to its topological character, the WZW term $\Gamma[M]$ is not affected
by the scaling operation.

We now bosonize the remaining content of the theory, by making use of 
the dictionary.  We obtain
\begin{eqnarray*}
  {\cal O}_{00} &=& \sum_{n=1,2}(\bar \psi_n^t \tau_3 \bar \psi_n)
  (\psi_n^t \tau_3 \psi_n)\leftrightarrow \\ &\leftrightarrow&
  \sum_{n=1,2} {\rm STr}\,(\tau_3 \bar J_n) \,{\rm STr}\,(\tau_3
  J_n)\;,\\ 
  {\cal O}'_{00} &=& (\bar \psi_1^t \tau_3 \bar \psi_1) (\bar\psi_2^t
  \tau_3 \bar\psi_2)+(\psi_1^t \tau_3 \psi_1) (\psi_2^t \tau_3 \psi_2)
  \leftrightarrow \\ 
  &\leftrightarrow& {\rm STr\,}(\tau_3 \bar J_1)\,{\rm STr\,}(\tau_3
  \bar J_2) + {\rm STr\,}(\tau_3 J_1)\,{\rm STr\,}(\tau_3 J_2)\;,\\ 
  {\cal O}_{\pi\pi} &=& {1 \over 2} \sum_{n=1,2} \, \sum_{k=1,2}
  (\bar\psi_n^t \tau_k \bar\psi_n) (\psi_n^t \tau_k \psi_n)
  \leftrightarrow\\ &\leftrightarrow& {1 \over 2} \sum_{n=1,2}\,
  \sum_{k=1,2} {\rm STr\,}(\tau_k \bar J_n)\,{\rm STr\,} (\tau_k J_n)
  \;,\\ 
  {\textstyle{1\over 2}} {\cal O}_{\pi 0} &=& \bar\psi_1^t \psi_2
  \bar\psi_2^t \psi_1 + \sum_{k = 1,2,3} \bar\psi_1^t \tau_k \psi_2
  \bar\psi_2^t \tau_k \psi_1 \leftrightarrow \\ 
  &\leftrightarrow& \ell^{-2} {\rm STr\,} M_1 M_2 + \ell^{-2}
  \sum_{k=1,2,3} {\rm STr}\, M_1 \tau_k M_2 \tau_k \;,\\
  {\cal L}_E &=& -2iE \sum_{n=1,2} \bar\psi_n^t \psi_n^{\vphantom{t}}
  \leftrightarrow -2i {E\over \ell} \sum_{n=1,2} {\rm STr\,} M_n \;.
\end{eqnarray*}
The sum
\begin{eqnarray}
  &&S \equiv W[M_1,\gamma] + W[M_2,\gamma^{-1}] + \int d^2r \, {\cal
    L}_{\rm pert} \;,\\ &&{\cal L}_{\rm pert} = g{\cal O}_{00} +
  g'{\cal O}_{00}'+ g_{\pi\pi}{\cal O}_{\pi\pi} + g_{\pi 0}{\cal
    O}_{\pi0} + {\cal L}_E
  \label{action_boso}
\end{eqnarray}
represents our final result for the low-energy effective action of the
disordered $d$-wave superconductor: two WZW actions, which are coupled
by a number of marginally relevant operators due to inter-node and
intra-node scattering of the quasi-particles.  Note that the rotational
symmetry of Euclidean space (or Lorentz invariance of Minkowski space)
is broken by the term ${\cal O}_{00}'$, which is a relict of the order
parameter symmetry $d_{x^2-y^2}$ of the superconductor. In the
following subsections we are going to discuss some physical
consequences of the field theory (\ref{action_boso}).

\subsection{Hard scattering, T-invariance (class $C{\rm I}$)}
\label{sec:hard,T}

We first discuss the case where inter-node scattering is present, and
is strong.  By construction of the bosonized theory, the RG flow
equations for its couplings coincide with those for the perturbed
Dirac theory.  From (\ref{RG_flow}) we then know that all couplings,
including $g_{\pi 0}$, increase under renormalization.  While ${\cal
  O}_{00}$, ${\cal O}_{00}'$, and ${\cal O}_{\pi\pi}$ are
current-current perturbations, the operator ${\cal O}_{\pi 0}$ is seen
to act as a ``potential''.  We expect such a perturbation to make some
of the fields massive and remove them from the low-energy theory.  To
elucidate this effect, we parametrize the fields $M_n$ as follows:
\begin{eqnarray*}
  M_n = \exp \pmatrix{A_n^+ + A_n^- &B_n\cr C_n &-A_n^+ + A_n^- \cr}
  \quad (n = 1, 2) \;.
\end{eqnarray*}
The requirement $M_n^{-1} = \tau_2 M_n^s \tau_2 \in {\rm OSp} (4|4)$
is satisfied by imposing the conditions $(A_n^\pm)^s = \pm A_n^\pm$,
$B_n^s = B_n$, and $C_n^s = C_n$ for $n = 1, 2$.  In this
parametrization, the expansion of ${\cal O}_{\pi 0}$ around unity
reads
\begin{eqnarray*}
  \ell^2 {\cal O}_{\pi 0} &=& {\rm STr} \Big( (A_1^+)^2 + (A_2^+)^2 \\ 
  &+& (A_1^- + A_2^-)^2 + B_1 C_1 + B_2 C_2 \Big) + ... \;.
\end{eqnarray*}
It is now extremely important that, by construction \cite{bsz} of the
Riemannian symmetric superspace of type $C|D$, this potential energy
{\it is bounded from below by zero.} The low-energy configurations of
the fields $M_1 , M_2$ are those that minimize ${\cal O}_{\pi 0}$,
which implies
\begin{eqnarray*}
  A_1^+ = A_2^+ = A_1^- + A_2^- = B_n = C_n = 0 \quad (n = 1, 2)\;.
\end{eqnarray*}
By inserting the solution of this equation into the parametrization
for $M_n$, we see that the low-energy fields are scalar in node space:
\begin{eqnarray*}
  M_1 = \pmatrix{M &0\cr 0 &M\cr} = M_2^{-1} \;,
\end{eqnarray*}
where $M$, a supermatrix of size $(2+2)^2$, is subject to the
condition $M^{-1} = M^s$.  The last equation defines the
orthosymplectic supergroup ${\rm OSp}(2|2)$ with ``orthogonal'' bosons
and ``symplectic'' fermions.  (Technically speaking, $M$ takes values
in a Riemannian symmetric superspace of type $D|C$ inside ${\rm OSp}
(2|2)$.)  Inserting the constrained form of $M_1$ and $M_2$ into the
operators ${\cal O}_{00}$, ${\cal O}_{00}'$, and ${\cal O}_{\pi\pi}$,
we see that all of these vanish.  The effective action thus reduces to
\begin{eqnarray}
  S_{\rm eff} &=& 2 W[M;\gamma] + 2 W[M^{-1};\gamma^{-1}] \nonumber \\ 
  &=& - {\gamma + \gamma^{-1} \over 8\pi} \int d^2r \, {\rm STr} \,
  \partial_\mu M^{-1} \partial_\mu M \;.
  \label{nlsm_DC}
\end{eqnarray}
This is isotropic and independent of the disorder strength, and its
${\sc ff}$ part coincides with the replica field theory written down
by Senthil et al.~\cite{sfbn}.  The WZW terms have canceled because
$\Gamma[M_1] + \Gamma[M_2] = 0$ for $M_1 = M_2^{-1}$.

In view of recent statements to the contrary \cite{Fukui}, we
emphasize that the cancellation of WZW terms is a robust feature that
does not depend on any specific model assumptions made (as long as the
disorder is generic, placing the model in class $C{\rm I}$).  The
basic mechanism behind the cancellation is the dependence of the WZW
term on parity or, equivalently, the choice of orientation of
two-dimensional space: its sign gets reversed by the transformation
$x_1 \leftrightarrow x_2$.  Although this dependence by itself does
not forbid the presence of a WZW term for parity-invariant systems (as
the sign change can be absorbed by a target space isometry $M \mapsto
M^{-1}$), it does so for a $d$-wave superconductor in zero magnetic
field.  The crucial fact here is that the pure Dirac Hamiltonians for
the two pairs of nodes $(1,\bar 1)$ correspond to {\it opposite}
orientations (they map onto each other by the transformation $x_1
\leftrightarrow x_2$).  Therefore, in the bosonized theory they are
represented by WZW actions with topological coupling constants that
{\it still carry opposite signs}, provided that uniform conventions
for assigning Dirac bilinears to WZW fields are in force. (In the
above treatment, we arranged for the signs to be identical by choosing
our conventions to be different for the two nodal sectors.  This was
done for notational convenience.)  When any kind of scattering
(consistent with the generic symmetries of the system) between
neighbouring nodes is turned on, the two WZW fields $M_1$ and $M_2$
become locked in the low-energy theory, now by $M_1 = M_2$, and the
topological couplings with differing signs inevitably add up to zero.

The model (\ref{nlsm_DC}), with the WZW term being absent, is called
the principal chiral non-linear sigma model on ${\rm OSp}(2|2)$ (more
precisely, on a Riemannian symmetric superspace of type $D|C$, which
is a supermanifold based on the direct product of the non-compact space
$R^+ = {\rm SO}(2,C)/{\rm SO}(2)$ with the compact group ${\rm
  Sp}(2)$).  Referring for a more detailed discussion of its replica
analog to Ref.~\cite{sfbn}, we here review only one salient feature of
this theory.  According to Friedan \cite{friedan}, the one-loop beta
function of any 2d non-linear model is determined by the Ricci
curvature of the target manifold.  For symmetric superspaces, just
like for ordinary symmetric spaces, the Ricci curvature is
proportional to the metric tensor, and in the present case is easily
shown to be positive \cite{helgason}.  This means that the beta
function is positive, and the coupling $t = 8\pi/(\gamma +
\gamma^{-1})$ therefore increases under renormalization.  By plausible
extrapolation to strong coupling $(t \to \infty)$, one then expects
the theory to be in an insulating phase with vanishing (spin)
conductance $\sigma_s = 2/(\pi t) \to 0$.  This phase was named the
``spin insulator'' in Ref.~\cite{sfbn}.

The local density of states $\nu$ of the spin insulator is predicted
to vanish linearly, $\nu(E) \sim |E|$, at ultra-low energies, by a
scaling argument due to Senthil and Fisher \cite{sf}.  On the basis of
the supersymmetric field theory (\ref{nlsm_DC}), we can phrase the
argument as follows.  The growth of the coupling $t$ under
renormalization means that, while there is asymptotic freedom
(accompanied only by small field fluctuations) at short scales, the
fluctuations of $M$ grow strong for large wave lengths.  The scale for
crossover from weak to strong coupling is set by the localization
length, $\xi \sim {\rm e}^{\gamma+\gamma^{-1}}$ \cite{sfbn}.  Let us
therefore partition the system into blocks of linear size $\xi$.  To
mimic the crossover between short-range order and long-range disorder,
we take the fields on different blocks to be independent, and on each
individual block to be spatially constant.  By doing the field
integral in this simple approximation we find
\begin{eqnarray*}
  \nu(E) = \nu_0 f(E\nu_0 \xi^2) \;,
\end{eqnarray*}
where $\nu_0$ is the SCBA density of states, which depends only weakly
on energy, and the scaling function $f(x)$ (with asymptotic limit
$f(\infty) = 1$) has the small-$x$ expansion $f(x) = \pi x / 4 + {\cal
  O}(x^2)$.  Thus the local density of states goes to zero linearly at
$E = 0$, with the characteristic energy scale being given by $(\nu_0
\xi^2)^{-1}$, the level spacing for one localization volume.

\subsection{Soft scattering, T-invariance (class $A{\rm III}$)}
\label{sec:soft,T}

We now set the inter-node couplings $g_{\pi\pi}, g_{\pi 0}$ to zero.
What then remains is the free-fermion theory, as represented by the
sum of WZW actions $W[M_1] + W[M_2]$, marginally perturbed by the
operator $g {\cal O}_{00} + g^\prime {\cal O}_{00}^\prime$.  Two
observations simplify the analysis of this theory.  Firstly, we may
omit the term ${\cal O}_{00}^\prime$ when calculating the density of
states.  The physical reason is that in the absence of inter-node
coupling we can project on one pair of nodes, say $(1,\bar 1)$, and
take the other one into account by multiplying the final answer with a
factor of 2.  Thus, we may set $\psi_2 = \bar\psi_2 = 0$ and drop the
field $M_2$.  Alternatively, the reduction can be seen by reasoning
within the field-theoretical formulation, where it comes about because 
the functional integral giving the contribution to the density of states 
from the pair $(1,\bar 1)$ has an invariance under BRST transformations 
acting in the sector $(2,\bar 2)$.  Secondly, we may rescale the coordinates 
to make the field theory isotropic.  We are then facing an isotropic WZW 
model with a current-current perturbation:
\begin{equation}
  W[M] + g \int d^2r\,{\rm STr}(\tau_3 J)\, {\rm STr}(\tau_3\bar J)\;.
   \label{AIII_OSp}
\end{equation}
{}From our earlier computation of the RG beta functions we know that
the coupling constant $g$ does not flow.  This can be verified
directly from the WZW model by observing that the OPE of the Abelian
current ${\rm STr}\, \tau_3 J$ with itself does not generate any UV
singularities.  Thus the above action is a fixed point for the
renormalization group, and the theory is conformally invariant.

What are its properties?  A quick approach to the answer is to recall
that the physical problem we are dealing with here is the much studied
\cite{ntw,lfsg,mcw} problem of Dirac fermions in a random Abelian
vector potential.  This problem can doubtless be analysed via the
above action based on ${\rm OSp}(4|4)$.  Unfortunately, that
formulation does not seem to be the most efficient one to use and, in
any case, it is not the one used in related previous work.  What
forced us to introduce the ${\rm OSp}(4|4)$ target space was the fact
that we do not know of any other way of bosonizing the perturbations 
that couple the neighbouring nodes.  In the present problem, however,
where the coupling between the nodes is absent, the symmetry group
consists of two copies of ${\rm GL}(2|2)$, the subgroup of elements 
$h$ of ${\rm OSp}(4|4)$ with the property $h \tau_3 h^{-1} = \tau_3$.  
The excess of field degrees of freedom in ${\rm OSp}(4|4)$, as compared 
to ${\rm GL}(2|2)$, means that there is some redundancy.  An efficient 
method of solution will try to avoid this redundancy, and can avoid it as 
follows.

In standard non-Abelian bosonization (without boso\-nic ghosts), two
schemes are distinguished: bosonization takes the free-fermion theory
into a level-one WZW model either over the orthogonal group ${\rm
  O}(2n)$, or over the unitary group ${\rm U}(n)$.  While the two
bosonization schemes are equivalent in the free limit, the ${\rm
  O}(2n)$ version has the advantage of exposing the {\it full} set of
symmetries of the free theory of $n$ Dirac fermions (or $2n$ Majorana
fermions), whereas the ${\rm U}(n)$ version is better adapted to the
treatment of certain perturbations.  The latter scheme was developed
particularly by Affleck \cite{affleck} in his treatment of the 1d
Hubbard and Heisenberg models.

By straightforward transcription of Ref.~\cite{bsz}, we can generalize
the ${\rm U}(n)$ bosonization scheme to include a bosonic ghost
system.  The resulting theory is again a supersymmetric WZW model,
with an action functional of the same form as before, but the field
now takes values in the supergroup ${\rm GL}(2|2)$ (or, rather, in a
Riemannian symmetric superspace of type $A|A$ inside ${\rm GL}
(2|2)$).  Deferring the details of this generalization to Appendix
\ref{sec:CFT}, we here take a short cut.  To arrive at the bosonized
theory in the desired form, we start from the action functional
(\ref{AIII_OSp}) and simply insert for $M_1$ the reduced expression
\begin{eqnarray*}
  M_1 = \pmatrix{M &0\cr 0 &(M^s)^{-1}\cr} \;,
\end{eqnarray*}
where $M \in {\rm GL}(2|2)$ is parametrized, for example, by
\begin{eqnarray*}
  M = \exp \pmatrix{X_{\sc bb} &X_{\sc bf}\cr X_{\sc fb} &X_{\sc
      ff}\cr} , \quad X_{\sc bb}^\dagger = X_{\sc bb} \;, \quad
  X_{\sc ff}^\dagger = - X_{\sc ff} \;.
\end{eqnarray*}
The action functional for $M$ then becomes
\begin{eqnarray}
  S[M] &=& -{1 \over 8\pi} \int d^2r\,{\rm STr}\,\partial_\mu M^{-1}
  \partial_\mu M + {i \Gamma[M] \over 12\pi} \nonumber \\ &+& {g\over
    \pi^2} \int d^2r\,{\rm STr} (M^{-1} \partial M)\, {\rm STr} (M
  \bar\partial M^{-1}) \;.
  \label{marg_line}
\end{eqnarray}
The same result is obtained by direct application of the ${\rm GL}
(2|2)$ bosonization rules to the supersymmetric Dirac theory with the
marginal perturbation $g\int {\cal O}_{00} d^2r$.  The numerical (or
bosonic) part of this action has the important property of being
bounded from below (for $g$ not in excess of a certain critical
value), and hence leads to a well-defined functional integral.
(Positivity of the action is ensured by the construction of the field
manifold as a Riemannian symmetric superspace.)

Past solutions of the physical problem of Dirac fermions in a random
Abelian vector potential have used the equivalent representation by a
replica sine-Gordon model \cite{ntw}, or diagonalization of the
quantum Hamiltonian by a Bogoliubov transformation \cite{lfsg}, or the
study of Kac-Moody current algebras with ${\rm U}(1|1) \times {\rm
  U}(1|1)$ symmetry \cite{mcw}. A closely related model has been
investigated via sigma model techniques in \cite{gll}.  Here we are
going to take a different approach:  we will solve the problem by
direct manipulation of the functional integral.  Following Knizhnik
and Zamolodchikov \cite{kz}, this is possible by exploiting the local
(infinite-dimensional) symmetry $G_L \times G_R$ (with $G = {\rm
  GL}(2|2)$).  For better readability, the technical details of this
computation have been relegated to Appendix \ref{sec:CFT}, and we now
just give a summary of the most important points.

The basic idea is to study the response of the functional integral to
field variations of the form 
\begin{eqnarray*}
  \delta_X M({\bf r}) \equiv -X({\bf r}) M({\bf r}) \;,
\end{eqnarray*}
where $X({\bf r})$ takes values in the Lie algebra ${\rm gl}(2|2)$.  
As usual, variation of the action yields the Noether current:
\begin{eqnarray*}
  \delta_X S &=& - {1 \over \pi} \int d^2r \, J_{\bar\partial X} \;, \\ 
  J_X &=& - {\rm STr}(XM\partial M^{-1}) + {2g\over \pi} {\rm STr}(X)
  {\rm STr}(M\partial M^{-1}) \;.
\end{eqnarray*}
{}From the stationarity condition $\delta_X S = 0$, one infers that
$J$ is a holomorphic conserved current, i.e.~$\bar\partial J_A = 0$
for spatially constant $A$ and on solutions of the equations of
motion.  In addition, the theory has a conserved current $\bar J$
which is antiholomorphic: $\partial \bar J_A = 0$.  The expression for
it is determined by the change of $S$ in response to variations
$\delta_Y M({\bf r}) = M({\bf r}) Y({\bf r})$.  In the following we
concentrate on the holomorphic sector.

By exploiting the invariance of the functional expectation value
$\langle ... J_B (0) \rangle$ (where the ellipses indicate additional
operator insertions) w.r.t. the variable change $M \to {\rm e}^{-X} 
M = M + \delta_X M + ...$ with $\delta_X M = -XM$, one deduces 
the operator product expansion (OPE) for the currents:
\begin{eqnarray}
  J_A(z) J_B(0) &=& {f(A,B) \over z^2} + {J_{[A,B]}(0)\over z} +
  ...\;, \label{OPE_current} \\ f(A,B) &=& -{\rm STr}(AB) +
  {2g\over\pi} {\rm STr}(A) {\rm STr}(B) \nonumber \;.
\end{eqnarray}
Like any operator product expansion, it is to be understood as an 
identity that holds under the functional integral sign.  The dots 
indicate terms which remain finite in the limit $z \to 0$.

The behaviour of correlation functions under scale changes or, more
generally, under conformal transformations $z \mapsto \varepsilon(z)$,
is determined by the OPE of the fields with the holomorphic component
of the stress-energy-momentum tensor, $T(z)$.  To obtain the latter,
one demands that the OPE between $T(z)$ and the currents starts as
$T(z) J_A(0) = J_A(0)/z^2 + ...$, which expresses the fact that
$J_A(z)$ has conformal dimension $(1,0)$.  Assuming $T(z)$ to be of
the Sugawara form und using Eq.~(\ref{OPE_current}), one then finds
\begin{eqnarray}
  T(z) = {\kappa^{ij} \over 2} J_i(z) J_j(z) + {1 - 2g/\pi \over
    2} J_e(z) J_e(z) \;.
\label{sugawara}
\end{eqnarray}
The notation here means the following: choosing some basis $\{e_i\}$ of the Lie
superalgebra ${\rm gl}(2,2)$, we set $J_i = J_{e_i}$, and $\kappa_{ij}
= - {\rm STr}\,e_i e_j$, and we raise the indices of the metric tensor by 
$\kappa^{ij} \kappa_{jk} = \delta_k^i$.  The current $J_e$, whose
square gives the second term in the formula for $T(z)$,  represents
the generator $e = {\rm id}$ (unit matrix).

The observable we are interested in is the local density of
quasi-particle states at low energy.  Recall from (\ref{fun_int_basic})
that a finite energy $E$ is accounted for in the field theory by a
perturbation $-2iE\int d^2r\,(\bar\Psi_1^s \Psi_1^{\vphantom{s}} +
\bar \Psi_{\bar 1}^s \Psi_{\bar 1}^{\vphantom{s}})$.  This bosonizes
to $-2i(E/\ell) \int d^2r \, {\rm STr} (M + M^{-1})$.  Moreover, the
local density of states at ${\bf r} = 0$ is given by the expectation
value of the bosonic (or fermionic) part of $\bar\Psi_1^s(0)
\Psi_1^{\vphantom{s}} (0) + \bar \Psi_{\bar 1}^s(0) \Psi_{\bar
  1}^{\vphantom{s}}(0)$, which bosonizes to $(2\ell)^{-1}{\rm Tr}\,
\left( M(0) + M(0)^{-1} \right)$.

Thus, the density of states is determined by the expectation value of
the fundamental field $M$ and its inverse $M^{-1}$.  To work out its 
scaling behaviour, we need the conformal dimension of $M$, which is 
the number $\Delta_M$ multiplying the leading singularity in the OPE 
of $T(z)$ with $M(0)$:
\begin{eqnarray*}
  T(z) M(0) = \Delta_M M(0) / z^2 + ... \;.
\end{eqnarray*}
Given the Sugawara form (\ref{sugawara}) of the stress-energy-momentum
tensor, the value of $\Delta_M$ follows from the OPE
\begin{eqnarray*}
  J_A(z) M(0) = - A M(0)/ z + ... \;,
\end{eqnarray*}
which reflects the fact that, by construction, $J_A$ represents the
generator of left translations $M \mapsto {\rm e}^{-A} M$.  Using that
the quadratic Casimir in the fundamental representation of ${\rm
  GL}(2|2)$ vanishes ($\kappa^{ij} e_i e_j = 0$), one finds $\Delta_M
= {1 \over 2} - g/\pi$.  The total scaling dimension of $M$ is the sum
(from holomorphic and antiholomorphic sectors) 
\begin{eqnarray*}
  \Delta = \Delta_M + \bar\Delta_M = 1-2g/\pi \;.
\end{eqnarray*}
To calculate the local density of states, $\nu(E)$, we add
\begin{eqnarray*}
  -2iE \int d^2r \, {\rm STr}(M+M^{-1}) +
  \lambda {\rm Tr} \left( M(0) + M(0)^{-1} \right)
\end{eqnarray*}
to the action functional and differentiate with respect to the source
$\lambda$ at $\lambda = 0$.  If the short-distance cutoff is raised by
a renormalization group transformation ${\bf r} \mapsto b {\bf r}$,
the energy scales as $E \mapsto b^{2-\Delta} E$ and
the local source as $\lambda \mapsto b^{-\Delta}\lambda$.
Consequently, $\nu(E)$ obeys the homogeneity relation
  $$
  \nu(E) = b^{-\Delta} \nu(b^{2-\Delta}E) \;.
  $$
The solution is a power law
\begin{equation}
  \nu(E) \propto |E|^\alpha \;, \quad \alpha = {\Delta \over 2-\Delta}
  = {1 - 2g/\pi \over 1 + 2g/\pi}\;.
  \label{ntw_dos}
\end{equation}
Thus the density of states is expected to vary algebraically with a
non-universal exponent that depends on the disorder strength $g$.
This is in agreement with a result first found by Nersesyan, Tsvelik,
and Wenger \cite{ntw}.

Let us mention in passing that the critical line for Dirac fermions in
a random Abelian vector potential has been argued \cite{mcw} to be
unstable with respect to an infinite number of relevant perturbations.
In our opinion, this instability is not a true effect but disappears
when the theory is formulated on a target manifold with Riemannian
structure (see Appendix \ref{sec:WZW_AA}). 

\subsection{Realistic scenario?}

Having reviewed the two extreme cases of strongly coupled and
completely decoupled nodes, it is pertinent to address what kind of
scenario will prevail under realistic conditions. By the very nature
of the question, no clear-cut answer exists. Yet assuming that the
basic modeling by the microscopic Hamiltonian (\ref{hatH}) has some
finite overlap with experimental reality, we believe that two basic
scenarios are conceivable:

{\it Systems with a significant amount of large-momentum transfer
  scattering on the microscopic level.} Here the adjective
``significant'' refers to a regime in which the strength of the
inter-node coupling is in excess of the other characteristic energy
scales of the problem (such as temperature, experimental resolution,
etc., which couple to the action via the energy parameter $E$).  In
that case, relative fluctuations of the two fields $M_1$ and $M_2$ are
negligible, and it is only the isotropic content of the theory,
described by the action (\ref{nlsm_DC}), that matters.

A more complicated situation arises in {\it systems whose microscopic
  Hamiltonian contains only soft scattering}: in systems where the
inter-node scattering is a small perturbation, we run into a crossover
scenario.  The bare theory is given by two WZW actions, weakly
perturbed by inter-node scattering.  As we saw, renormalization makes 
the inter-node couplings grow.  Therefore, the theory flows to strong 
coupling on asymptotically large scales and the fields for the 
different nodes will again be locked.  However, that fixed-point regime 
may be preceded by a large intermediate region in which the critical 
properties of the WZW theory prevail.  In that case we expect for the
DoS a power law which is both non-trivial and non-universal, as 
predicted by (\ref{ntw_dos}).

Summarizing, the $d$-wave system renormalizes to a universal limit
described by the rotationally and parity invariant action
(\ref{nlsm_DC}).  There may, however, be an extended crossover region
in which the system of decoupled single-node actions dominates. The
question of which type of scenario prevails in real and numerical
experiments is system specific and not for the present approach to
decide.

\subsection{Soft scattering, broken T-invariance (class $A$)}
\label{sec:soft-scatt-brok}

We have argued in Section \ref{sec:symmetries} that perturbations
breaking time-reversal invariance reduce the symmetry of the $d$-wave
superconductor without inter-node scattering from class $A$III to $A$.
In the mixed state, as we recall from Section \ref{sec:d_wave_ham},
T-breaking is due primarily to the supercurrent flow, which gives rise
to a scalar potential in the quasi-particle Hamiltonian projected on a
single node.  This issue is resumed in the present subsection, where
we address the perturbation of the class $A$III Dirac Lagrangian (for
the pair of nodes $1,\bar 1$) by
\begin{eqnarray*}
  {imv \over 2} \, v_s \left( \bar\Psi_1^s \sigma_3^{\sc cc}
    \Psi_1^{\vphantom{s}} + \Psi_{\bar 1}^s \sigma_3^{\sc cc}
    \bar\Psi_{\bar 1}^{\vphantom{1}} \right) \;,
\end{eqnarray*}
where $v_s = v_2^A + v_2^B$ is the second component of the
supercurrent velocity.  According to the bosonization rules derived in
Appendix \ref{sec:CFT}, this perturbation transforms into
\begin{eqnarray*}
  \bar\Psi_1^s\sigma_3^{\sc cc}\Psi_1^{\vphantom{s}} + \Psi_{\bar 1}^s
  \sigma_3^{\sc cc} \bar\Psi_{\bar 1}^{\vphantom{s}} \longrightarrow
  \ell^{-1} {\rm STr}\, \sigma_3^{\sc cc} (M + M^{-1}) \;.
\end{eqnarray*}
As before, $M$ takes values in (the Riemannian symmetric superspace of
type $A|A$ inside) ${\rm GL}(2|2)$.  Thus the complete action
functional is
\begin{eqnarray*}
  S[M] + {imv \over 2\ell} \int d^2r \, v_s {\rm STr} 
  \, \sigma_3^{\sc cc} (M + M^{-1}) \;,
\end{eqnarray*}
where $S[M]$ was defined in Eq.~(\ref{marg_line}).  The perturbation
has positive scaling dimension $2 - \Delta = 1 + 2g/\pi$ and is
therefore relevant.

We expect such a perturbation to cause a reduction of the field
manifold.  Unlike the symmetry-breaking terms we considered earlier,
the present one is {\it imaginary}-valued, giving rise to oscillations
in the functional integral.  The argument for field reduction
therefore cannot be the rigorous kind of potential-energy argument
that applied to the crossover from class $A$III to $C$I.  What we must
rely on now is stationary phase, i.e.~the fact that contributions from
field configurations with $M$-dependent phases tend to cancel out.
Starting from the identity element $M = 1$ and conjugating by a
symmetry transformation for class $A$, $M \mapsto T M \sigma_3^{\sc
  cc} T^{-1} \sigma_3^{\sc cc}$ with $T \in {\rm GL}(2|2)$, we get a
stationary-phase manifold parametrized by
\begin{eqnarray*}
  M \sigma_3^{\sc cc} = \sigma_3^{\sc cc} M^{-1} = T \sigma_3^{\sc cc}
  T^{-1} \equiv Q \;.
\end{eqnarray*}
By construction, the above ${\rm T}$-breaking perturbation vanishes
on this manifold,
\begin{eqnarray*}
  {\rm STr\,} \sigma_3^{\sc cc}(M+M^{-1}) \to 0 \;.
\end{eqnarray*}
As for the other terms in the Lagrangian, it is straightforward to
show that they reduce to
\begin{eqnarray*}
  &&{\rm STr\,} \partial_\mu M^{-1} \partial_\mu M \to {\rm STr\,}
  \partial_\mu Q \partial_\mu Q \;,\\ &&{\rm STr}(M \partial M^{-1})
  \, {\rm STr}(M^{-1} \bar\partial M) \to 0 \;.
\end{eqnarray*}
The fate of the WZW term $\Gamma[M]$ under the field reduction is less
trivial.  Doing the same steps as in Section 7 of Ref.~\cite{bsz}, we
find that $\Gamma[M]$ reduces to a topological theta term on the
manifold of $Q$-matrices:
\begin{eqnarray*}
  {i\Gamma[M] \over 12\pi} \to S_{\rm top}[Q] &\equiv& i\theta \int
  d^2r \, {\cal L}_{\rm top} \;, \\ {\cal L}_{\rm top} &=&
  {\epsilon^{\mu\nu} \over 16\pi i} {\rm STr\,} Q \, \partial_\mu Q \,
  \partial_\nu Q \;,
\end{eqnarray*}
with $\theta = \pm\pi$.  The ambiguity in the sign of $\theta$ results
from the multivaluedness of the WZW term: to calculate $\Gamma[M]$ one
must extend the field $M$ to a 3-space bounded by the two-dimensional
position space, and there is no unique way of doing that.  However, if
the position space is closed (as is implied by its being a boundary)
the ambiguity in the sign of $\theta$ has no consequence.  In that
case it can be shown that the topological term is a winding number or,
in other words, the integral of the topological density ${\cal L}_{\rm
  top}$ over any closed surface takes quantized values in $2\pi i Z$,
so that a shift of $\theta/2\pi$ by any integer $n$ is unobservable in
the functional integral.  For the pair of nodes $(2,\bar 2)$ the
situation is similar.

To summarize, we find that the low-energy effective action for the
pair of nodes $(n,\bar n)$ of a $d$-wave superconductor in symmetry
class $A$ is
\begin{eqnarray*}
  &&S^{(n,\bar n)}[Q] = \pm S_{\rm top}[Q] \\ &-& \frac{1}{8\pi}\int
  d^2r\, {\rm STr}\left( \gamma^{-s_n} \partial_1 Q \partial_1 Q +
    \gamma^{s_n} \partial_2 Q\partial_2 Q \right) \;,
\end{eqnarray*}
where $s_1 = 1$ and $s_2 = -1$.  In the isotropic limit $\gamma = 1$,
this action functional is familiar from the integer quantum Hall
effect, where it is known as Pruisken's action \cite{pruisken} with
longitudinal electrical conductivity $\sigma_{xx} = 1/\pi$ and Hall
electrical conductivity $\sigma_{xy} = \theta / 2\pi = \pm 1/2$.  The
physics associated with this action is (i) a finite and smooth density
of states at $E = 0$, and (ii) critical behaviour in the transport
coefficients.

At first sight, one might be suspicious of our nonzero result for the
coupling $\sigma_{xy} = \theta / 2\pi$.  What is puzzling about it is
this.  Since the Hall conductivity transforms as a pseudoscalar,
i.e.~has its sign reversed by the exchange of Cartesian coordinates $x
\leftrightarrow y$, a nonvanishing Hall response breaks parity.  On
the other hand, the massless Dirac theory is invariant under parity
(if the transformation $x \leftrightarrow y$ is accompanied by the
exchange of the left-moving and right-moving fields) and so is the WZW
model (if $x \leftrightarrow y$ is combined with $M \leftrightarrow
M^{-1}$).  It goes without saying that a parity-invariant theory
cannot generate pseudoscalar observables.  On these grounds, recent
papers have stated \cite{sl,ye,vmz} that the thermal Hall conductivity
of a $d$-wave superconductor in the mixed phase vanishes in the linear
(or Dirac) approximation.

Does it follow from this symmetry argument that the reduction to the
nonlinear sigma model and/or our nonzero result for the coupling
$\sigma_{xy}$ must be false?  It does not.  Indeed, recall the
invariance of $\exp i\theta \int d^2r {\cal L}_{\rm top}$ under shifts
$\theta \to \theta \pm 2\pi$, when the position space has no boundary.
This invariance means that $\sigma_{xy} = 1/2$ and $\sigma_{xy} =
-1/2$ give equivalent field theories for the case of a closed system.  Thus,
although $\sigma_{xy}$ is a pseudoscalar, the nonlinear sigma model at
$\sigma_{xy} = \pm 1/2$ enjoys the property of being parity-invariant,
and there does not exist any conflict with the parity invariance of
the massless Dirac theory (or the WZW model).

At the same time, it would be a preposterous proposition to claim that
Pruisken's nonlinear sigma model for the integer quantum Hall effect
(IQHE) at $\sigma_{xy} = \pm 1/2$ describes a parity-invariant
physical system with vanishing Hall response!  How do we then
reconcile parity invariance at $\sigma_{xy} = \pm 1/2$ with IQHE
phenomenology?  The discrepancy is resolved by observing that the Hall
conductivity is not a Fermi-edge quantity, which is to say that it is
not determined by the Green's functions at a single energy $E = E_F$
only, but is given by a sum over energies.  This feature of the Hall
conductivity, which distinguishes it from the longitudinal
conductivity $\sigma_{xx}$, poses a difficulty to the nonlinear sigma
model formulation: since the field theory is derived for a fixed
energy, all the information needed to reconstruct the Hall response in
the bulk cannot be contained in it.  In other words, the topological
coupling $\sigma_{xy}$ of the nonlinear sigma model does not coincide
with the total Hall conductivity of the bulk, and hence the parity
invariance of the field theory at $\sigma_{xy} = \pm 1/2$ does not
imply that the Hall response of the quantum Hall system vanishes.

Nonetheless, as was pointed out by Pruisken \cite{pruisken}, it is
possible to monitor the Hall response field-theoretically by studying
systems that have a {\it boundary}.  There exist chiral boundary
currents at the edges of quantum Hall systems, and the topological
term of the nonlinear sigma model (for a fixed energy $E = E_F$) does
capture the physics of these boundary currents.  In fact, when a
boundary is present, the invariance under shifts $\sigma_{xy} \to
\sigma_{xy} \pm 1$ no longer holds, parity symmetry is broken for all
nonzero values of $\sigma_{xy}$ including $\sigma_{xy} = \pm 1/2$, and
the topological term gives rise to long-range current-current
correlations at the edges of the system.  The message from this is
that, in order to get an unambiguous field-theoretic view of thermal
and spin Hall transport in a $d$-wave superconductor, we are well
advised to consider geometries with a boundary.

Having understood that, we run into another difficulty: A consistent
extension of a {\it single} WZW model to a system with boundaries is
not possible!  Instead, a geometric construction detailed in Appendix
\ref{sec:wzw-theory-systems} will show the following:
\begin{itemize}
\item The modeling of a system $\Sigma$ with boundary $\partial
  \Sigma$ requires, at least, {\it two} WZW fields $M_1$ and $M_2$.
\item These fields are coupled at the boundary through a condition
  $(M_1)^{-1}\big|_{\partial \Sigma}=M_2\big|_{\partial \Sigma}$ (plus
  additional conditions on the derivatives of the fields to ensure
  current conservation.) This mechanism reflects the fact that
  specular reflection at the boundaries of a finite systems couples
  quasi-particle species of different parity. (Specular reflection
  inverts {\it one} of the momentum components while leaving the other
  invariant: a parity operation.)
\item As far as the Hall conductivity of the coupled system is
  concerned, everything depends on the {\it masses} $m_{1,2}$ of the
  two field species: (i) if both fields are massless, they enter the
  theory with equal weight and the Hall conductivity vanishes by
  symmetry.  (ii) In highly anisotropic cases, e.g. $m_1 \gg m_2$, the
  field $M_1$ decouples from the low energy sector of the theory.
  Under these conditions, $\sigma_{xy} = \pm 1/2$ depending on which
  field has become heavy, and the sign of $m_{1,2}$.
  Finally, (iii) if the masses $|m_1| \simeq |m_2|$ are approximately
  equal, it is the signs of $m_{1,2}$ that determine the Hall
  conductivity. If the signs are the same, we expect to get
  $\sigma_{xy} = \pm 1$; if they differ, we expect $\sigma_{xy}=0$.
\end{itemize}

What can we infer from these observations about the problem at hand?
Recall that the two pairs of conjugate nodes $(1, \bar 1)$ and
$(2,\bar 2)$ are represented by the WZW fields $M_1$ and $M_2$
respectively, with our conventions being such that the total WZW term
is $\Gamma[M_1] + \Gamma[M_2]$.  Given these low-energy effective
fields, we impose the boundary condition $(M_1)^{-1} \big|_{\partial
  \Sigma} = M_2 \big|_{\partial\Sigma}$.  (Alternatively, the boundary
conditions could mix in high-energy fields corresponding to excited
quasi-particle bands of the superconductor.)  From what we said
before, this boundary condition is forced by the requirement that the
WZW functional be mathematically well-defined.  If the parity
invariance of the microscopic Hamiltonian is broken by the presence of
a magnetic field in the form of vortices and by the supercurrent flow,
there exists no fundamental symmetry principle that would protect the
fields from acquiring (small) masses $m_{1,2}$ in the low-energy
effective theory.

More detailed analysis would then be required to decide which scenario
(i-iii) applies in the case of a realistic $d$-wave superconductor
(for the most complete investigation to date, see \cite{vmz}).
However, as far as theory is concerned, let us re-iterate that the WZW
model and the original Dirac fermion model represent equivalent
descriptions of the system. In view of the consistency problems
outlined above, and their resolution by an extended theory,
one expects that a faithful description of Hall transport in the
Dirac fermion language, too, has to involve {\it all} four nodal
fermion species simultaneously. Therefore, symmetry-based
statements that the Hall conductivity of a {\it single} Dirac fermion
system vanishes\cite{ye} should be taken with a grain of salt.

To avoid confusion, we elaborate on one other subtle point that may
seem paradoxical.  Recall that what we set out to calculate was the
single-particle Green function, giving the density of states.  Since
the field theory we have derived for it contains massless (Goldstone)
fields $Q$, one might be misled into thinking that our theory predicts
singular corrections to the density of states.  In the present
instance, this is not so.  The reason why we obtained massless fields
is that, in the course of the derivation, we chose to carry out a {\it
  doubling} of field variables (introducing charge conjugation space,
so as to accommodate the symplectic symmetry of the Gorkov
Hamiltonian, which becomes important as soon as the coupling between
nodes is turned on).  This means that we are effectively working in
the two-particle sector.  Put differently, the present theory involves
{\it all} of the fields for the nodes $(1,\bar 1)$, although the basic
quasi-particle Hamiltonian connects only half of them.  As a result,
when we compute the density of states the source fields couple only to
a subset of the Grassmann fields.  Thus there is a global fermionic
symmetry, which causes the field integral to collapse by the usual
BRST mechanism and yield a vanishing correction to the density of
states.  The field theory stays alive only for the two-particle Green
function, viz.~transport.

\subsection{Hard scattering, broken T-invariance (class $C$)}

The final case to consider is the simultaneous presence of inter-node
scattering and breaking of time-reversal invariance.  Having switched
to the ${\rm GL}(2|2)$ bosonization scheme for convenience of
presentation and solution in the previous two subsections, we must now
return to the more general ${\rm OSp}(4|4)$ scheme in order to
accommodate the inter-node couplings.  The T-breaking perturbation by
a supercurrent flow is rewritten in terms of $\psi_n, \bar\psi_n$ as
\begin{eqnarray*}
  i \tilde\Psi^s ( \sigma_3^{\sc cc} \otimes \tau_3^{\vphantom{\sc
      cc}} ) \Psi = i {\rm STr} \, (\sigma_3 ^{\sc cc} \otimes
  \tau_3^{\vphantom{\sc cc}}) \left( \psi_1^{\vphantom{t}}
    \bar\psi_1^t + \bar\psi_2^{\vphantom{t}} \psi_2^t \right) \;.
\end{eqnarray*}
By the ${\rm OSp}(4|4)$ bosonization rules, this bosonizes to
\begin{equation}
  i \ell^{-1} {\rm STr} \, (\sigma_3^{\sc cc} \otimes
  \tau_3^{\vphantom{\sc cc}}) (M_1^{\vphantom{-1}} + M_2^{-1}) \;.
  \label{vs_pert}
\end{equation}
Now recall that the effect of inter-node scattering is to lock the
fields ($M_1 = M_2^{-1}$) and to reduce $M_1$ to a scalar in node
space,
\begin{eqnarray*}
  M_1 = \pmatrix{M &0\cr 0 &M\cr} \;,
\end{eqnarray*}
with $M^s = M^{-1}$.  Inserting this expression for $M_1$ and using
${\rm Tr}\,\tau_3 = 0$, we find that the perturbation disappears.
Thus, in the linear approximation used here, the breaking of
time-reversal symmetry by the supercurrent flow ceases to be effective
in the node-locked limit.

While this may seem surprising at first, it is not hard to comprehend
in physical terms.  The perturbation we are considering represents the
Doppler shift due to the motion of quasi-particles relative to the
supercurrent flow ${\bf v}_s$.  Because the node $1 (2)$ and its
conjugate $\bar 1 (\bar 2)$ are located at opposite momenta, the
corresponding quasi-particle states experience opposite Doppler
shifts.  Therefore, scattering between opposite notes causes the
Doppler effect to alternate randomly in time.  It is clear that the
node-locked limit will be attained over scales large enough in order
for a quasi-particle to undergo many scattering events between the
nodes.  On such scales, and in modes of long-range quantum
interference that are stable with respect to disorder averaging, the
Doppler shift averages to zero.

However, this does {\it not} mean that the presence of the
supercurrent flow is of no consequence.  At the free-fermion point,
the field $M$ has scaling dimension one, which makes the perturbation
(\ref{vs_pert}) more relevant in the RG sense than the marginal
perturbations that represent inter-node scattering.  (Precisely
speaking, this is true only for wave lengths smaller than the spacing
between vortices.  On large scales, the spatial oscillations of the
supercurrent velocity cause the perturbation to average to zero.)  We
should therefore evaluate the consequences of the perturbation
(\ref{vs_pert}) {\it before} locking the nodes.  Doing so, and
appealing to the same argument as in the previous subsection, we
expect the perturbation (\ref{vs_pert}) to reduce the field $M_n$ ($n
= 1, 2$) to stationary-phase manifolds which now are given by
\begin{eqnarray*}
  M_1 &=& T_1^{\vphantom{-1}} (\sigma_3^{\sc cc} \otimes
  \tau_3^{\vphantom{\sc cc}}) T_1^{-1} (\sigma_3^{\sc cc} \otimes
  \tau_3^{\vphantom{\sc cc}}) \;, \\ M_2 &=& (\sigma_3^{\sc cc}
  \otimes \tau_3^{\vphantom{\sc cc}}) T_2^{\vphantom{-1}}
  (\sigma_3^{\sc cc} \otimes \tau_3^{\vphantom{\sc cc}}) T_2^{-1} \;,
\end{eqnarray*}
with $T_n \in {\rm OSp}(4|4)$, the diagonal subgroup of the chiral
symmetry group of the $C|D$ WZW model.

Next, we take the effects of inter-node scattering into account (with
the plausible assumption that reduction of the field manifold does not
change the relevant nature of the perturbations from inter-node
scattering.)  By the mechanism described in Section \ref{sec:hard,T},
the fields will be locked:
\begin{eqnarray*}
  T_1 = T_2 = \pmatrix{T &0\cr 0 &T\cr} \;,
\end{eqnarray*}
where $T$ now belongs to the other variant of ${\rm OSp}(2|2)$ 
(orthogonal bosons, symplectic fermions).  By inserting the locked
fields into the WZW action, we arrive at a non-linear sigma model with
action
\begin{equation}
  S[Q] = - {1 \over t} \int d^2r\, {\rm STr}\, \partial_\mu Q
  \partial_\mu Q \;, \quad t = {8\pi\over\gamma + \gamma^{-1}}\;,
\label{nlsm_D3C1}
\end{equation}
where $Q = T \sigma_3^{\sc cc} T^{-1}$ parametrizes a Riemannian
symmetric superspace of type $D|C$.  Note that in the linear order of
approximation used, there is no topological theta term in this
non-linear sigma model.  Precisely the same field theory (without a
definite value for the coupling $t$) was obtained in \cite{bcsz2} from
a more phenomenological approach.  The known physics of this system is
a Zeeman-field dependent weak-localization correction to the thermal
conductivity \cite{bcsz2}, and a density of states that vanishes as
$E^2$ at ultra-low energies \cite{sf}.

Rather than elaborating on these properties, we turn to another way of
breaking time-reversal symmetry, which is to make the order parameter
of the superconductor {\it complex} by including an imaginary
component.  Motivated by recent developments that were reviewed in the
introduction, we consider an $id_{xy}$ component with momentum
dependence $i \lambda \sin (k_x a) \sin(k_y a)$.  In the Gorkov
Hamiltonian, such a term enters as the Pauli matrix $\sigma_2$ acting
in particle-hole space.  If we again rotate by ${\rm e}^{i \pi
  \sigma_1/4}$ and linearize around the nodes, the term becomes
proportional to $\sigma_3$ and acts as a Dirac mass term, with the
mass parameter being positive for one pair of nodes, say $(1,\bar 1)$,
and negative for the other pair.  In the field theory, upon
bosonization, the perturbation is represented by the operator
\begin{equation}
  {i \lambda\over\ell}\, {\rm STr}\,\sigma_3^{\sc cc}
  (M_1^{\vphantom{-1}} - M_2^{-1}) \;,
\end{equation}
which is relevant at the free-fermion point. The observation we made
earlier for the case of a supercurrent flow applies here as well: if
we assume the limit of locked fields $(M_1 = M_2^{-1})$, the
perturbation vanishes.  This is again plausible, since the Dirac
masses resulting from $\sin(k_x a) \sin(k_y a)$ carry opposite signs
for neighbouring nodes, which, as we recall, transform into each other
by $k_x \mapsto -k_x$ or $k_y \mapsto -k_y$.  Hence the Dirac mass of
a quasi-particle subject to inter-node scattering is zero on average.
As before, however, the correct procedure is first to deal with the
relevant mass perturbation, and only then to impose field locking.
Proceeding in this way, we would again be led to the action
(\ref{nlsm_D3C1}) for the low-energy effective theory.

We should, however, be suspicious of the last result, as it does not
conform with what is expected on symmetry grounds.  The point is that
the action of the non-linear sigma model (\ref{nlsm_D3C1}) is invariant
under a parity transformation $x_1 \mapsto x_1$ and $x_2 \mapsto -x_2$,
whereas the perturbed WZW theory we are starting from is not.  Indeed,
if we combine the parity transformation with field inversion $M_i
\mapsto M_i^{-1}$, so as to preserve the WZW action, the sign of the
perturbation gets reversed:
\begin{eqnarray*}
  {\rm STr}\, \sigma_3^{\sc cc} M_1^{-1} = {\rm STr}\, \sigma_3^{\sc
    cc} M_1^t = - {\rm STr}\, M_1 \sigma_3^{\sc cc} \;,
\end{eqnarray*}
where the last equality follows from ${\rm STr}(AB) = {\rm STr} (AB)^t
= {\rm STr}(B^t A^t)$ and the fact that $\sigma_3^{\sc cc}$ is odd
under the orthosymplectic transpose defined by $(\sigma_3^{\sc cc}
\psi)^t = \psi^t (\sigma_3^{\sc cc})^t$.  Thus, parity is broken in
the initial theory, and it is reasonable to expect that it will still
be broken in the final theory.

As was first pointed out by Pruisken et al.~\cite{llp,pruisken} in the
related context of the integer quantum Hall effect, the way to break
parity invariance in the non-linear sigma model is to add a theta term
with topological angle $\theta \notin \pi Z$ to the action.  From what
has been said, we expect such a term in the low-energy effective
action when a secondary $i d_{xy}$ component of the order parameter is
present.  To compute the angle $\theta$, we abandon non-Abelian
bosonization and revert to the standard method: Hubbard-Stratonovich
transformation and saddle-point approximation, followed by a gradient
expansion.  (Note that, although non-Abelian bosonization is perfectly
suited to describe the flow of the field theory away from the
free-fermion point, it is a less reliable tool for understanding how
the theory arrives at and flows into the non-linear sigma model.)  The
last step is readily performed using the method of heat kernel
regularization reviewed in Appendix \ref{sec:heatkernel}.  As a result
we obtain the action presented in Eq.~(\ref{nlsm_DIIICI}), with
couplings
\begin{eqnarray*}
  \sigma_{11}^0 = {\gamma + \gamma^{-1} \over \pi} \;, \qquad
  \sigma_{12}^0 = {2\lambda \over \pi \kappa} \;.
\end{eqnarray*}
A summary of the physics of this model, and references to 
the relevant literature, were given in the introduction.

\section{Comparison with other approaches}
\label{sec:comparison}

In this section, we comment on the second controversy mentioned in the
introduction: the incompatibility of field theory and self-consistent
approaches.  Several arguments have been put forward to explain the
existence of conflicting results for observables as basic as the mean
DoS (finite at zero energy in the self-consistent approaches,
vanishing in field theory).  It has, for example, been argued that the
vanishing of the DoS in field theory might be due to an ill-defined
continuum limit \cite{zhh_reply}. On the other hand, it has been
pointed out \cite{nt_comment} that the notorious appearance of
logarithmic divergences in the diagrammatic results hints at an
insufficient UV-regularization.  
However, in hindsight it looks like these, UV related aspects of the problem
carry minor if any significance. The origin of the discrepancy between
field theory and early self diagrammatic aproaches to the problem lies
somewhere else:

First of all, it is important to realize that the vanishing of the DoS
obtained in the field-theoretical description is due to a mechanism
which can simply not be included into diagrammatic approaches.
Indeed, we have seen in Section \ref{sec:proceed} that the breaking of
the chiral symmetry of the Hamiltonian on the saddle-point level leads
to the appearance of Goldstone modes.  In early perturbative
approaches to the problem~\cite{Gorkov,Lee,hwe}, these modes had not
been included at all.  In the field theory, however, they are
responsible for much of the physical behaviour of the system.  In
particular, it is the strong fluctuation of these modes in
low-dimensional systems that leads to a vanishing of the DoS at $E =
0$.

It is tempting to explain this vanishing by the Mermin-Wagner-Coleman
theorem as was done by NTW.  Indeed, a non-vanishing DoS would have
meant that a continuous symmetry of the field theory was spontaneously
broken.  (As with normal disordered systems, the DoS plays the role of
an order parameter in the field theory.)  The Mermin-Wagner-Coleman
theorem essentially states that, as a consequence of unbounded
Goldstone mode fluctuations, spontaneous breaking of a continuous
symmetry does not occur in dimensions $d\le 2$.  There is, however, a
caveat with that argument, which is that the Mermin-Wagner-Coleman
theorem explicitly refers to {\it compact field manifolds}.  In the
non-perturbative supersymmetry approach at hand, we are {\it not}
dealing with such types of field manifold.  This is not an academic
point, as is shown by counterexamples: there exist cases, namely
two-dimensional systems in class $D$ or $D$III, where the naive
application of the Mermin-Wagner-Coleman theorem leads to the
erroneous prediction of a vanishing DoS. (The non-linear sigma model
for weakly disordered 2d systems in these classes predicts a
zero-energy DoS which diverges in the thermodynamic limit
\cite{sfD,bsz}.)  However, for the systems in the classes $A$III,
$C$I, and $C$ studied in the present paper, the inclusion of the
Goldstone modes indeed leads to a vanishing DoS.

It is very instructive to explore the phenomenon in the
zero-dimensional case (i.e.~the case of ergodic systems, where spatial
fluctuations of the Goldstone modes are frozen out.)  The reason why
$d = 0$ is a good case to study is that the zero-dimensional
non-linear sigma model integral can be performed explicitly (c.f.,
e.g., Ref.~\cite{ast}).  It turns out that the results, including the
vanishing of the DoS at zero energy, agree with the phenomenological
random-matrix theory approach to systems of class $C$, $C$I, and
$A$III \cite{Altland,v1,v2}.  Moreover, the simplicity of the
$0d$-case makes it particularly straightforward to deduce what is
missing in the diagrammatic approach.  After all, it cannot be that
perturbative diagrammatic approaches are completely oblivious to the
existence of the Goldstone modes mentioned above.  In fact, it has
been shown in Ref.~\cite{Altland} that relevant classes of diagrams
are missed within the SCBA approach to computing the Green function.
``Relevant'' here means that the diagrams in question diverge as the
energy approaches zero.  This behaviour is indicative of the fact that
these diagram classes represent the perturbative implementation of the
Goldstone modes discussed above (very much like the standard diffuson
mode is the first-order perturbative contribution to the Goldstone
modes induced by the spontaneous breaking of the symmetry between
retarded and advanced Green functions for disordered metals).  More
recent diagrammatic approaches\cite{Gornyi} have included these modes
into the perturbative theory of the disordered $d$-wave
superconductor. In intermediate energy regimes, where the IR
singularity of the Goldstone modes has not yet become virulent, the
extended diagrammatic formulation represents an alternative to the
field-theoretical approach.

Summarizing, we find that (i) the inclusion of Goldstone modes is
crucial for a correct description of the quasi-particles of a $d$-wave
superconductor (in particular its DoS), (ii) that within diagrammatics
these modes are represented by singular diagram classes which (iii)
can be brought under control in intermediate, but not in the lowest
energy regimes.

Finally we comment on work where a non-vanishing zero-energy DoS has
been obtained in a non-perturbative manner.  In particular, Lee
\cite{Lee} considered the DoS on the background of an ensemble of
impurities at the unitarity limit.  The strength of these impurities
makes a comparison with our analysis difficult.  (The construction of
the field theory crucially relies on the existence of a parametric
separation between the energetic extension of the nodal region and the
much smaller width of the disorder distribution.)

Another type of non-perturbative approach has been put forward by
Ziegler, Hettler and Hirschfeld \cite{zhh}.  In their work, a
non-vanishing DoS was obtained on the basis of rigorous estimates
applied to the single-particle Green function.  Upon close inspection,
however, their line of reasoning can be dismissed by its lack of
relevance for superconductors.  Indeed, the Hamiltonian they consider
is extremely special and non-generic in class $C{\rm I}$.  It is
formulated on a lattice and, in the notation of Ref.~\cite{zhh}, reads
\begin{eqnarray*}
  H = - (\nabla^2 + \mu)\,\sigma_3 + \hat\Delta^d \sigma_1 \;,
\end{eqnarray*}
where the kinetic energy $-\nabla^2$ and the non-local $d$-wave order
parameter $\hat\Delta^d$ are taken to act by
\begin{eqnarray*}
  (\nabla^2 \psi)({\bf r}) &=& \psi({\bf r} + 2e_1) + \psi({\bf r} -
  2e_1) \\ &+& \psi({\bf r} + 2e_2) + \psi({\bf r} - 2e_2) \;, \\ 
  (\hat\Delta^d \psi) ({\bf r}) &=& \Delta\left( \psi({\bf r} + e_1)+
    \psi({\bf r} - e_1) \right. \\ &-& \left. \psi({\bf r} + e_2) -
    \psi({\bf r} - e_2) \right) \;,
\end{eqnarray*}
and $\mu$ is a random chemical potential. The choice for $\nabla^2$
has the artificial feature that it allows hopping only between {\it
  third-nearest} neighbours, which leads to a conservation law alien
to superconductors: $H$ commutes with the operator $D\sigma_3$, where
$D_{{\bf r},{\bf r}^\prime} = (-1)^{x_1 + x_2} \delta_{{\bf r} , {\bf
    r}^\prime}$ multiplies by plus one on the sites of an $A$
sublattice ($x_1 + x_2$ even) and by minus one on the sites of the $B$
sublattice ($x_1 + x_2$ odd).  Since $D\sigma_3$ has two eigenvalues
$\pm 1$, the Hilbert space decomposes into two sectors not coupled by
$H$.  The first one consists of all particle states on the $A$
sublattice and hole states on the $B$ sublattice, while the second
sector contains all the complementary states.  Without loss of
generality we may restrict the Hamiltonian to one sector.  The
particle-hole degree of freedom then becomes redundant --- we can make
a particle-hole transformation on the $B$ sites, so that the first
sector becomes all particles and the second sector all holes --- and
the Hamiltonian reduces to
\begin{eqnarray*}
  \tilde H = - (\nabla^2 + \mu)\Big|_A + (\nabla^2 + \mu)\Big|_B
  + \hat\Delta^d \;,
\end{eqnarray*}
where $-(\nabla^2 + \mu)\big|_A$ denotes the restriction of $-(
\nabla^2 + \mu)$ to the $A$ sublattice.  $\tilde H$ is a generalized
discrete Laplacian augmented by a random potential.  It belongs to the
symmetry class $A{\rm I}$, the Wigner-Dyson class with $\beta = 1$,
which is to say that the Hamiltonian matrix is real symmetric, and the
large-scale behaviour of its two-particle Green's function is
controlled by the cooperon and diffuson modes well known from the
theory of disordered metals.  There exist {\it no} quantum
interference modes affecting the single-particle Green's function.
Thus the quasi-particle Hamiltonian of Ziegler, Hettler and Hirschfeld
models a time-reversal invariant {\it normal metal} rather than a
superconductor!  As one would expect, it can be proved \cite{zhh} that
such a Hamiltonian has a non-zero density of states at $E = 0$ for a
wide class of distributions of the random potential $\mu$.  However,
this result tells us nothing about a disordered $d$-wave
superconductor.  The built-in $D\sigma_3$ conservation law eliminates
all the modes of quantum interference that are characteristic of the
superconductor and act to suppress the density of states at zero
energy.  As a corollary, we conclude that the ZHH model provides no
test of the accuracy of the self-consistent $T$-matrix approximation
widely used for superconductors.

To preclude any misunderstanding, let us emphasize that the ZHH model
is built on a superfluid condensate, which may of course exhibit the
Meissner effect and other phenomena associated with superconductivity.
However, the issue at hand is not the nature of charge transport by
the superconducting condensate.  As was emphasized in the
introduction, we are not studying the condensate, but rather its {\it
  quasi-particle} excitations, their spectral statistics and their
transport properties, which can be probed experimentally via spin or
thermal transport.  From this perspective the ZHH Hamiltonian, albeit
built on a superconducting ground state, models a normal metal.  (More
precisely, it gives the behaviour a thermal insulator, since quantum
interference effects ultimately drive the model to strong localization
in two dimensions.)

\section{Numerical Analyses of the Quasi-Particle Spectrum}
\label{sec:numer-analys-quasi}

Much of the early work on quasi-particles in disordered $d$-wave
superconductors was analytical.  More recently, a number of numerical
investigations exploring the effects of disorder scattering appeared
(see, e.g., Refs.~\cite{ahm,zst,huckalt}).  Taking the soft and hard
scattering regimes in turn, the present section reviews elements of
these works, and relates them to the results disussed in previous
sections.

\subsection{Hard Scattering}
\label{sec:hard-scattering}

A comprehensive analysis of quasi-particle spectra in time-reversal
invariant $d$-wave superconductors with point-like scatterers
(symmetry class $C$I) appeared in Refs.~\cite{ahm}.  Going beyond the
mere diagonalization of the lattice Hamiltonian (\ref{hatH}), these
papers determined the order parameter self-consistently.  Moreover,
the role of a nesting symmetry of particle-hole type, which is present
in the case of a half-filled band and refers to momentum transfers $q
= (\pm \pi/a,\pm \pi/a)$, was explored.  Without going into
quantitative detail, the main results of these papers can be
summarized as follows:
\begin{itemize}
\item The self-consistent $T$-matrix approximation fails to correctly
  describe the DoS below a certain energy scale, the value of which
  increases with disorder.
\item At zero energy the DoS vanishes in all cases but the extreme one
  of scatterers at the unitarity limit.  In that particular case,
  spectral weight accumulates at the band center, reflecting the
  creation of impurity bound states.  For the special limit of zero
  chemical potential, corresponding to fully realized particle-hole
  nesting symmetry, the low-energy DoS diverges logarithmically in
  accord with the analysis of Lee and P\'epin \cite{Pepin}.
\item Away from zero energy (and for generic scatterers), a regime of
  linearly increasing DoS, tentatively identified as the DoS profile
  predicted by Senthil and Fisher \cite{sf}, is observed.  We must
  caution, however, that this identification does not convince us for
  weakly disordered systems: as discussed earlier, the linear
  suppression appears in an insulating phase of separated localization
  volumes.  Given the large size of the localization length in weakly
  disordered two-dimensional systems, it is not clear whether the
  separation of characteristic length scales can be realized on
  lattice sizes accessible to numerical computation.
\item Self consistency leads to a further suppression of the DoS,
  in particular in systems with binary-alloy type scatterers close to
  the unitarity limit.
\item As was explained in Ref.~\cite{Gornyi}, the nesting symmetry is
  of little relevance except for the case of unitary scatterers.
\end{itemize}

\subsection{Soft Scattering}
\label{sec:soft-scattering}

\begin{figure}
  \begin{center}
    \epsfxsize=7.6cm \leavevmode \epsffile{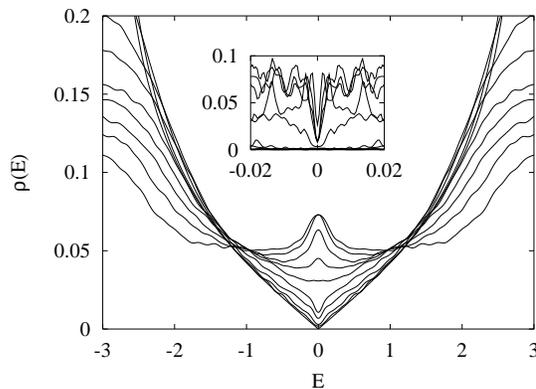}
    \caption{Density of states for $\Delta = 1$, correlation length
      $\xi = 2$, and disorder strengths $W = 0, 1, \dots, 8, 10$
      (bottom to top at $E = 0$).  All energies are measured in units
      of the hopping matrix element.  The system size is $L = 33$ and
      the level broadening (introduced so as to suppress oscillations
      on the scale of the mean level spacing) $\Gamma = 0.05$.  The
      inset shows the same data on a smaller scale with $\Gamma =
      0.0005$.  The finite DoS at $E = 0$ is due to the finite level
      broadening $\Gamma$.}
    \label{fig:1}
  \end{center}
\end{figure}

The quasi-particle spectrum of time-reversal invariant $d$-wave
superconductors with soft scatterers (class $A$III) has been
investigated numerically in Ref.~\cite{huckalt}.  We will now review
the results of that work in some detail.  The starting point was the
usual lattice Hamiltonian defined in Eq.~(\ref{hamil}), but with the
assumption of long-range correlated disorder so as to stay within the
soft-scattering regime.  Specifically, the on-site disorder potential
was defined by
\begin{eqnarray*}
  \epsilon_i = {W\over \sqrt{\Sigma}}\sum_j f_j \exp\left[ -{|{\bf
        r}_i-{\bf r}_j|^2 \over \xi^2}\right]
\end{eqnarray*}
where $\Sigma = \sum_j\exp(-2|{\bf r}_j^2|/\xi^2)$, and $f_j$ were
independent random variables drawn from a uniform distribution on
$[-1/2,1/2]$.  The results are shown in Fig.~\ref{fig:1}.

The suppression of inter-node scattering by smoothing the potential is
very strong: for a potential correlated over just two lattice spacings
($\xi = 2$), the inter-node scattering matrix elements are reduced by
a factor of about $10^{-8}$ as compared to the intra-node matrix
elements.  Although such matrix elements will become relevant at very
large scales, it is expected that potentials with $\xi \ge 2$ place
small systems of size up to $100 \times 100$ firmly inside the pure
symmetry class $A{\rm III}$.  In practice, this means that each of the
low-energy sectors associated with the four nodes in the dispersion
relation is described by Dirac fermions subject to pure gauge
disorder.

\begin{figure}
  \begin{center}
    \epsfxsize=7.6cm \leavevmode \epsffile{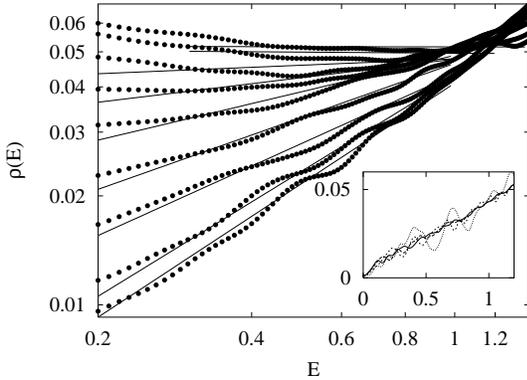}
    \caption{Double logarithmic plot of the density of states of
      Fig.~\ref{fig:1}. Disorder ranges from $W=1$ to 10. Dots
      ($\bullet$) represent data and lines power law fits in the
      respective intervals. Inset: Density of states for $W=2$ and
      $L=15$ (dotted), 25 (short-dashed), 35 (long-dashed), and 45
      (solid). Note that the numerical uncertainties are considerably
      smaller than the amplitude of the fluctuations.}
    \label{fig:2}
  \end{center}
\end{figure}

\begin{figure}[tbp]
  \begin{center}
    \epsfxsize=7.6cm \leavevmode \epsffile{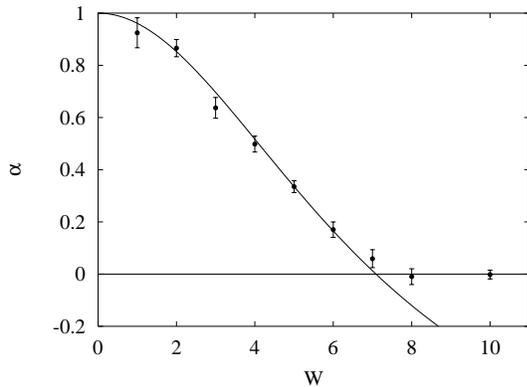}
    \caption{Exponents $\alpha$ extracted from the fitted curves in
      Fig.~\ref{fig:2} as a function of the disorder strength $W$ for
      $\Delta=1$. The solid curve is the result of NTW,
      Eq.~(\ref{ntw_dos}).}
    \label{fig:4}
  \end{center}
\end{figure}

\begin{figure}[tbp]
  \begin{center}
    \epsfxsize=7.6cm
    \leavevmode
    \epsffile{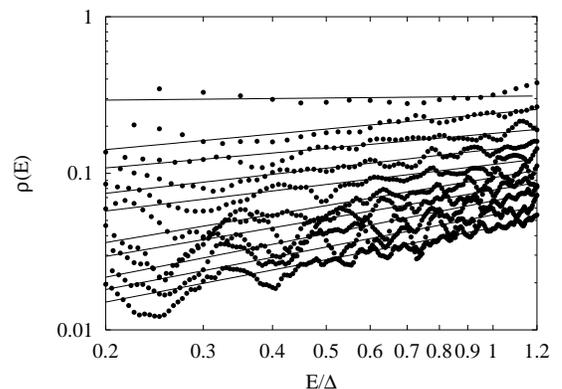}
    \caption{Density of states for values of the order parameter $\Delta 
      = 0.1$ to 1.0 (top to bottom) and disorder strength $W = 3$.
      Each curve is shifted by a factor of 1.2 for clarity.}
    \label{fig:7}
  \end{center}
\end{figure}

Figure \ref{fig:1} shows the DoS of the system for various disorder
strengths.  The quantitative analysis of the spectral data shows that
three different regimes can be distinguished: (i) for low energies,
$\epsilon < E_{\rm min}$, the structure of the spectrum is dominated
by finite size effects.  (For the lattice analysed in
Ref.~\cite{huckalt} $E_{\rm min} \simeq E_0/10$ where $E_0$ denotes
the total width of spectrum.)  The most apparent of these effects is a
disorder and system size dependent bump in the DoS.  At very low
energies, the DoS vanishes, as is seen in the high resolution inset of
Fig.~\ref{fig:1}.  (ii) For high energies, $\epsilon > E_{\rm max}
\approx E_0/2$, the structure of the spectrum depends on non-universal
lattice effects.  (iii) Most interesting is the intermediate regime,
$E_{\rm min} < \epsilon < E_{\rm max}$.  In this region, the
energy-dependent DoS exhibits power law behaviour
(c.f.~Fig.~\ref{fig:2}.)  Of course, an energy window of width $E_{\rm
  max} / E_{\rm min} \sim 5$ provides a rather poor statistical basis
for establishing power law behaviour.  Nevertheless, the procedure
seems justified as it is not just one power law with a single exponent
but rather a two-parameter family of exponents $\alpha(W,\gamma)$ that
is analysed.  Here $W$ measures the strength of the disorder while
$\gamma = t/\Delta$ is the anistropy parameter.  In Fig.~\ref{fig:4},
the exponents obtained by fitting the DoS determined numerically are
compared with the NTW prediction (\ref{ntw_dos}) for the isotropic
case $\gamma = 1$.  Specifically, for the present system,
\begin{displaymath}
  \rho(E) \sim |E|^\alpha, \qquad \alpha = \frac{1-2g/\pi}{1+2g/\pi},
  \qquad g = \frac{W^2}{32\Delta t} \;.
\end{displaymath}
Similarly, Fig.~\ref{fig:7} displays exponents obtained for fixed
disorder strength but different anisotropy parameters.  Notice that
the comparison between the numerical data and the family of analytical
exponents does not involve undetermined fit parameters.  The
applicability of the NTW scaling law is limited to small disorder
strengths, $g < 1$.  In Ref.~\cite{gurarie} it has been argued that
for larger values of $g$, the DoS becomes energy-independent.  This
prediction is supported by the numerical data.

To summarize, numerical analysis of the quasi-particle spectra in
$d$-wave superconductors reveals the need to distinguish between three
different types of disorder: scatterers at the unitarity limit,
non-unitary point-like impurities, and soft scattering potentials.
Although the comparison with analytical predictions is impeded by
finite size effects, there exists reasonable agreement for each of
these types.

\begin{figure}[tbp]
  \begin{center}
    \epsfxsize=7.6cm
    \leavevmode
    \epsffile{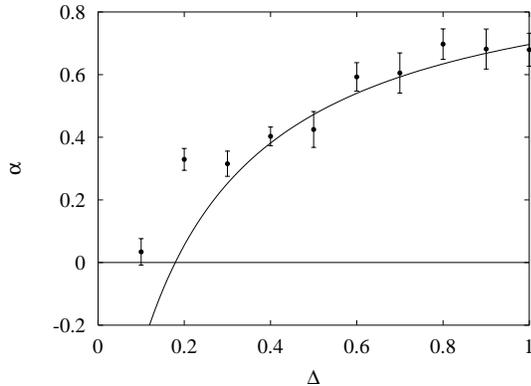}
    \caption{Exponents $\alpha$ for $W = 3$ as a function of the order
      parameter $\Delta$.  The solid curve is the result of NTW, 
      Eq.~(\ref{ntw_dos}).}
    \label{fig:8}
  \end{center}
\end{figure}

\section{Discussion}
\label{sec:discussion}

This concludes our survey of the influence of disorder on the
quasi-particle properties of disordered $d$-wave superconductors.
Since we included the majority of the discussion in the introduction,
we will limit our remarks to some key points: broadly speaking, the
analysis above emphasized that the low-energy transport properties of
the model $d$-wave system depend sensitively on the nature of the
impurity potential.  In two dimensions, a potential which is
short-ranged in space places the system in the spin insulator phase,
where all quasi-particle states are localized.  On the other hand, a
potential which contains only forward-scattering components leads to a
marginally perturbed WZW theory in which the zero-energy
quasi-particle states are critical, and the density of states vanishes
as a power law.  The low-energy theory in this case belongs to a
one-parameter family of fixed points, each identified through a
different value of the disorder-coupling strength.  This is reflected
in a low-energy density of states which varies with energy as a power
law with an exponent that depends on the strength of disorder.

Experimentally, the relevant scattering phenomenology is likely to be
somewhere in between the cases considered above: at intermediate
energy scales, signatures of the critical theory may well be visible
in thermal transport measurements, although the behaviour at very low
energies must, ultimately, be that of the spin insulator.

The collapse of the critical theory in the presence of strong disorder
was attributed to a cancellation of the multi-valued WZW terms arising
from the different nodal sectors of the theory.  In fact, such
cancellations are guaranteed by symmetry to occur in lattice models
which exhibit a Dirac-node structure at the Fermi level (such as the
random $\pi$-flux model).  In such lattice models, the nodes arise in
pairs related by parity.  Under the same parity transformation, the
WZW term is mapped onto a partner with opposite sign.  Therefore, when
the fields belonging to each nodal sector are locked by strong
disorder, the different WZW terms add and cancel pairwise.

Finally, the formalism developed and investigated above is not
entirely specific to $d$-wave superconductors.  The global structures
of the theory relied only on the existence of a Dirac-like spectrum of
the clean system.  We believe that the general scheme outlined above
could be applied in the investigation of other model systems with
a gapless linear density of states such as gapless semiconductors, and
superfluids.

{\sc Acknowledgments}: We would like to acknowledge useful discussions
with Patrick Lee, Catherine P\'epin, and Alexei Tsvelik. Furthermore, we 
are particularly grateful to Bodo Huckestein for providing access to the 
numerical data presented in section~\ref{sec:numer-analys-quasi}.

\appendix

\section{Gradient Expansion and the Chiral Anomaly}
\label{sec:gradient}

To complement our derivation of the effective action for the 
soft-scattering limit from non-Abelian bosonization, we include here 
a derivation of the same action from the gradient expansion.  The 
motivation is largely of pedagogical nature:  to facilitate comparison 
with existing works in the literature on weakly disordered fermion 
systems, it is useful to present more than a single route to the 
construction of the critical theory. We also wish to point at some 
unexpected difficulties that are encountered in the standard 
approach to deriving the low-energy effective action of the 
$d$-wave superconductor (or for that matter any disordered 
{\it relativistic} fermion system).
 
To be specific we will formulate the gradient expansion for a
soft-scattering system that is time-reversal invariant (class $A$III).
The inclusion of perturbations driving the model to any of the other
three classes is straightforward. However, since this appendix mainly
serves a pedagogical purpose, we will limit ourselves to the
discussion of just one class.  As discussed in Section
\ref{sec:basics}, the independence of the nodes entails a decoupling
of the low-energy theory into two pairs of nodal sectors $(1,\bar{1})$
and $(2,\bar{2})$.  We discuss one specific sector, say $(1,\bar{1})$,
anticipating that the full theory can later be obtained by
straightforward combination of both sub-sectors.

Our starting point is the soft-mode action for the $(1,\bar{1})$
sector derived in Section \ref{sec:proceed} and given by
Eq.~(\ref{S_rotated}).  Rearranging matrix blocks, the action can be
brought into the simpler form
\begin{eqnarray}
  S[M] = {\bf STr} \ln \left(\matrix{\kappa M^{-1} & \partial \cr
      \bar{\partial} &\kappa M}\right),
\label{node1_action2}
\end{eqnarray}
where the block decomposition is in ${\sc ph}$-space and we have
omitted the superscript $(1)$ on the derivative operator for
notational simplicity.  To further simplify the notation, we have set
the two characteristic velocities $v_i$ ($i=1,2$) {\em temporarily} to
unity (i.e.~$v=1$, and $\gamma=1$).  (In fact, some authors attempt to
get rid of these scales altogether by means of a coordinate rescaling
$x_i \to v_i x_i$.  However, as we are going to discuss below, this
seemingly innocuous manipulation may lead to inconsistencies once the
nodes are coupled.  Moreover, the influence of such a rescaling on the
unspecified source components of the action must be treated with
caution.  We will therefore re-instate the scales $v_i$ towards the
end of this section.)  The bold-face notation ${\bf STr}$ means that
we are taking the (super)trace over both superspace and Hilbert space.

To compute a low-energy action from the above expression, we can
follow one of at least three different routes:

\begin{itemize}

\item The most direct approach would be to introduce coordinates on
  the field manifold, say by $M = {\rm e}^X$, to expand around unity:
  $M= \openone + X +\dots$, and then to derive a low-energy action for
  the $X$'s via a straightforward gradient expansion.  Owing to the
  overall ${\rm GL} (2|2)$ invariance of the model, such an approach
  determines the low-energy action not just in the vicinity of unity
  but rather on the entire manifold.

\item Alternatively, as in the main body of the text, one may resort
  to an entirely symmetry-oriented approach and obtain the structure
  of the low-energy action by means of current algebra and non-Abelian
  bosonization.  This was the route taken by NTW \cite{ntw}.

\item Finally, a third option is not to introduce coordinates on the
  field manifold but to attempt a gradient expansion directly based on 
  the original degrees of freedom $M$.  While such schemes are standard 
  in applications of non-linear sigma models to disordered metallic
  systems, we here run into difficulties, caused by the appearance of
  ill-defined momentum integrals.  The way to overcome this problem is 
  first to subject (\ref{node1_action2}) to a UV-regularization scheme 
  and only then to expand in the spatial fluctuations of the 
  fields.  This will be our method of choice in this section.  Its main 
  advantages are that it is computationally efficient and better exposes
  the global structures of the theory than a coordinate-based approach 
  does.
\end{itemize}

\subsection{Chiral Anomaly}
\label{chiral_anomaly}

Before subjecting the action functional to a gradient expansion, let
us first make some pedagogical remarks.  For the time being, let $M$
be a field taking values in some matrix group $G$, and let us consider
the functional determinant
\begin{eqnarray*}
  {\cal Z}[M] &=& {\bf Det} D_M = {\rm e}^{{\bf Tr} \ln D_M} = {\rm
    e}^{-S[M]} \;, \\ D_M &=& \pmatrix{\kappa M^{-1} &\partial\cr
    \bar\partial &\kappa M \cr}\;.
\end{eqnarray*}
(When $G$ is a group of supermatrices, ${\bf Det}$ has to be replaced
by ${\bf SDet}^{-1}$.)   Our goal is to expand $\ln(1/{\cal Z})$ in 
gradients to produce a low-energy effective action for $M$.

Now, if $G$ is a group of unitary matrices $M^{-1} = M^\dagger$, the 
determinant is not real:
\begin{eqnarray*}
  \bar{\cal Z} &=& {\bf Det} \pmatrix{\kappa {M^{-1}}^\dagger
    &-\partial\cr -\bar\partial &\kappa M^\dagger \cr} \\ &=& {\bf
    Det} \pmatrix{\kappa M^{-1} &\bar\partial\cr \partial &\kappa M
    \cr} \not= {\cal Z} \;,
\end{eqnarray*}
so $\ln {\cal Z}$ has an imaginary part.  On general field-theoretic
grounds, we expect this imaginary part to be a multi-valued functional
of WZW type.  How can we compute ${\rm Im}\, \ln {\cal Z}$?

A natural idea is to ``take the square root'':  $M \equiv T_2^{
  \vphantom{-1}} T_1^{-1}$, and manipulate the determinant as follows:
\begin{eqnarray*}
  &&{\bf Det} \pmatrix{\kappa T_1^{\vphantom{-1}}T_2^{-1} &\partial\cr
    \bar\partial &\kappa T_2^{\vphantom{-1}}T_1^{-1} \cr} = \\ &&{\bf
    Det} \, \pmatrix{T_1 &0 \cr 0 &T_2\cr} \pmatrix{\kappa
      &T_1^{-1} \partial \, T_1^{\vphantom{-1}}\cr T_2^{-1}\bar\partial
      \, T_2^{\vphantom{-1}} &\kappa\cr} \pmatrix{T_2^{-1} &0\cr 0
      &T_1^{-1}\cr} .
\end{eqnarray*}
One might now be tempted to assume the multiplicativity of ${\bf
  Det}$, which would lead to ${\cal Z}$ being equal to
\begin{eqnarray*}
  {\cal Z}^\prime = {\bf Det} \pmatrix{\kappa &T_1^{-1}\partial \,
    T_1^{\vphantom{-1}}\cr T_2^{-1}\bar\partial \, T_2{\vphantom{-1}}
    &\kappa \cr} \;.
\end{eqnarray*}
Dropping the diagonal factors on the left and right seems especially
innocuous when $T_1$ and $T_2$ are unitary.

The motivation for trying to pass from ${\cal Z}$ to ${\cal Z}^\prime$
is that the latter can be computed exactly by a standard procedure
(see, e.g., Ref.~\cite{Alvarez} and the next subsection) in the limit
of small $\kappa$.  The result,
\begin{eqnarray*}
  \lim_{\kappa \to 0} \ln {\cal Z}^\prime
  [T_1,T_2] = +2W[T_1^{\vphantom{-1}} T_2^{-1}] \;,
\end{eqnarray*}
is expressed by the celebrated WZW functional:
\begin{eqnarray*}
  W[M] &=& {1 \over 16\pi} \int d^2r \, {\rm Tr} \, \partial_\mu
  M^{-1} \partial_\mu M + {i\Gamma[M] \over 24\pi} \;, \\
  \Gamma[M] &=& \int d^3r \, \epsilon_{\mu\nu\lambda} {\rm Tr} \,
  M^{-1} \partial_\mu M\, M^{-1}\partial_\nu M \,
  M^{-1}\partial_\lambda M \;.  \nonumber
\end{eqnarray*}
Note, however, the inequality ${\rm Re} \, W[M] \ge 0$ for unitary
$M$.  Thus, if ${\cal Z}$ were equal to ${\cal Z}^\prime$, the
constant fields $M({\bf r}) = M_0$ would {\it minimize} rather than
maximize the Boltzmann weight ${\cal Z}[M]$.  We would then be forced
to conclude that the field theory with action $S[M] = - \ln {\cal Z}
[M]$ is unstable with respect to spatial fluctuations and does not
exist.  By extending the argument to the supersymmetric setting, we
would find the theory with action (\ref{node1_action2}) to be sick.
On the other hand, we know (e.g. from non-Abelian bosonization) that
this is not the case, so there must be something wrong with the
present argument.  Where is the error?

The answer is that the manipulation taking $\cal Z$ into ${\cal Z}^
\prime$ disregards the existence of the notorious {\it chiral anomaly}
and is correct only for $T_1 = T_2$, the case of a pure gauge
transformation.  In other words, for a gauge transformation with an
axial component ($T_1 \not= T_2$) the passage from ${\cal Z}$ to
${\cal Z}^\prime$ is accompanied by a Jacobian different from unity.
Indeed, for $G = {\rm U}(1)$, $M = {\rm e}^{i\varphi}$,
straightforward application of the method of Abelian bosonization
\cite{zinnjustin} gives
\begin{eqnarray*}
  - \ln {\cal Z}[{\rm e}^{i\varphi}] = {1 \over 8\pi} \int d^2r \,
  (\partial_\mu \varphi)^2 = \ln {\cal Z}^\prime [{\rm e}^{i\varphi}]
  \;.
\end{eqnarray*}
By analogy, we expect that also in the non-Abelian case, correct
evaluation of ${\cal Z}[M]$ yields a stable theory with the proper
sign of the coupling.  A safe way of computing the gradient expansion
is to first UV regularize the Dirac operator $D_M$ after which axial
gauge transformations can readily be performed.

Returning to our original problem, we notice that a technically
convenient way of regularizing in the ultraviolet is to add to the
action (\ref{node1_action2}) a term
\begin{eqnarray*}
  - {\bf STr}\, \ln \pmatrix{\varepsilon M^{-1} & \partial \cr
    \bar{\partial} & \varepsilon M \cr} \;,
\end{eqnarray*}
which vanishes by supersymmetry when $\varepsilon$ is taken to be a
positive infinitesimal ($\varepsilon \to 0+$).  The resulting
expression,
\begin{eqnarray*}
  S[M] = {\bf STr}\, \ln \left[ \pmatrix{ \kappa M^{-1}&\partial \cr
      \bar\partial &\kappa M \cr} \pmatrix{\varepsilon M^{-1}
      &\partial\cr \bar\partial &\varepsilon M \cr}^{-1} \right] \;,
\end{eqnarray*}
is indeed manifestly well-behaved in the ultraviolet.  (The difference
between $\varepsilon M$ and $\kappa M$ becomes negligible for large
eigenvalues of the Dirac operator, in which case the two matrix
factors cancel each other and the action approaches zero.)  Setting $M
= T_2^{\vphantom{-1}} T_1^{-1}$ and using the cyclic invariance of the
trace, we rewrite the action functional as
\begin{eqnarray*}
  S &=& {\bf STr}\, \ln \pmatrix{ \kappa &T_1^{-1} \partial
    T_1^{\vphantom{-1}} \cr T_2^{-1} \bar\partial T_2^{\vphantom{-1}}
    &\kappa \cr} \\ &-& {\bf STr}\, \ln \pmatrix{ \varepsilon
    &T_1^{-1} \partial T_1^{\vphantom{-1}} \cr T_2^{-1} \bar\partial
    T_2^{\vphantom{-1}} &\varepsilon \cr} \;.
\end{eqnarray*}
Because $\kappa$ now acts as a mass, the low-energy limit of the
theory is captured entirely by the second term.  The first
contribution becomes appreciable only for momenta larger than
$\kappa$, where it cancels the second term.  Thus the role of the
first contribution has been relegated to that of a UV regulator.  We
are, of course, at liberty to replace it by some other UV
regularization scheme.  Doing so and expanding the second term in
gradients, we safely arrive at the WZW action, now with the correct
overall sign.  This will be demonstrated in more detail in the next
subsection.

\subsection{Heat kernel regularization}
\label{sec:heatkernel}

For our purposes, it is convenient to use Schwinger's proper-time
regularization (see, e.g., Ref.~\cite{Alvarez} where the same
procedure was applied to the non-Abelian Schwinger model,
i.e.~massless $1+1$ dimensional Dirac fermions coupled to an ${\rm
  SU}(N_c)$ gauge field.)  Without loss, we equate $T \equiv T_1 =
T_2^{-1}$ and rewrite the action in the form
\begin{eqnarray*}
  S &=& -{\bf STr}\, \ln \pmatrix{\varepsilon &T^{-1}\partial T \cr T
    \bar\partial T^{-1}&\varepsilon \cr} \;, \\ &=& -{\bf STr}\, \ln
  \left(\varepsilon^2-T\bar\partial T^{-2}\partial T\right) \;,
\end{eqnarray*}
where a UV cutoff at the momentum scale $\kappa$ is implied.

The proper-time regularization scheme for an elliptic operator $H$ is
implemented by
\begin{eqnarray*}
  - {\bf Tr} \ln H = \int_{1/\mu}^\infty \frac{ds}{s} \, {\bf Tr} \,
  {\rm e}^{-s H} \;.
\end{eqnarray*}
Notice that the lower integration bound $1/\mu$ cuts off the
contributions to $\ln {\bf Det} H$ from eigenvalues of $H$ greater
than $\mu$ and thus regularizes in the ultraviolet.  Applying this
scheme to our action (with cutoff $\mu = \kappa^2$), we obtain
\begin{eqnarray*}
  S = \int_{1/\kappa^2}^\infty \frac{ds}{s} \, {\bf STr} \, {\rm e}^{
    -s (\varepsilon^2 - T\bar\partial T^{-2} \partial T) } \;.
\end{eqnarray*}
This expression is both UV and IR finite and could in principle serve
as the starting point for a gradient expansion.  Much easier than the
direct evaluation of $S$, however, is the evaluation of its variation
$\delta S$.  We will therefore proceed by varying $S$ with respect to
some parameter, say $t$; then we will compute $\delta S \equiv \dot
S$, and finally we will reconstruct $S$ by integrating $\dot S$ with
respect to $t$.

Thus we consider some one-parameter family of fields $T({\bf r},t)$
with $T({\bf r},0) = \openone$ and $T({\bf r},1) = T({\bf r})$, and we
differentiate with respect to $t$.  This results in
\begin{eqnarray*}
  \dot S &=& \int_{1/\kappa^2}^\infty ds \, {\bf STr} \Big( \dot T
  \bar\partial T^{-2} \partial T + T \bar\partial T^{-2}\partial\dot T
  \\ &-& T \bar \partial (T^{-1} \dot T T^{-2} + T^{-2} \dot T T^{-1})
  \partial T \Big) {\rm e}^{-s(\varepsilon^2 - T \bar\partial T^{-2}
    \partial T)} .
\end{eqnarray*}
We next use the cyclic invariance of the supertrace to convert the
integrand into a total derivative with respect to the integration
variable $s$:
\begin{eqnarray*}
  \dot S &=& \int_{1/\kappa^2}^\infty ds \, {\rm e}^{-\varepsilon^2 s}
  \, {\bf STr} \, (T^{-1}\dot T + \dot T T^{-1}) \\ &&\times
  \frac{d}{ds} \left( {\rm e}^{s T \bar\partial T^{-2} \partial T} -
    {\rm e}^{s T^{-1} \partial T^2 \bar\partial T^{-1}}\right) \;.
\end{eqnarray*}
Performing the integral over $s$, setting the infinitesimal
$\varepsilon$ to zero, and making the integration over real space
$(\int d^2r)$ explicit, we obtain the expression
\begin{eqnarray*}
  \dot S = \int d^2r\, &&{\rm STr} \, (T^{-1}\dot T + \dot T T^{-1})
  ({\bf r}) \\ &&\times \left\langle {\bf r} \Big| {\rm e}^{-
      \kappa^{-2} H_1} - {\rm e}^{- \kappa^{-2} H_2} \Big| {\bf r}
  \right\rangle \;,
\end{eqnarray*}
where
\begin{eqnarray*}
  H_1 &=& - T^{-1} \partial \circ T^2 \bar\partial \circ T^{-1} \;, \\ 
  H_2 &=& - T \bar\partial \circ T^{-2} \partial \circ T \;,
\end{eqnarray*}
and the symbol $\circ$ means composition of operators.

Our next task is to compute the diagonal parts of the heat kernels
$\langle {\bf r} | {\rm e}^{-s H_i} | {\bf r}^\prime \rangle$
$(i=1,2)$, for small values of the dimensionful parameter $s =
1/\kappa^2$.  This is a standard exercise in semiclassical analysis,
and its solution can be found in textbooks \cite{bgv}.  Re-expressing
$H_1$ in the form
\begin{eqnarray*}
   H_1 = - {\textstyle{1\over 4}} (\partial_\mu -  i A_\mu)^2 + B \;, 
\end{eqnarray*}
where the non-Abelian gauge potential $A_\mu$ and field strength $B$
are functions of $T$ and its derivatives (which for brevity we do not
specify here), we have the standard short-time expansion
\begin{eqnarray*}
  \left\langle{\bf r}\Big| {\rm e}^{- sH_1}\Big| {\bf r} \right\rangle
  = {1 \over \pi s} - {B({\bf r}) \over \pi} + {\cal O}(s) \;.
\end{eqnarray*}
The same can be done for $H_2$ instead of $H_1$.  By taking the
difference of the two expansions, we obtain
\begin{eqnarray*}
  &&\left\langle {\bf r} \Big| {\rm e}^{-\kappa^{-2} H_1} - {\rm
      e}^{-\kappa^{-2} H_2} \Big| {\bf r} \right\rangle = -
  \frac{1}{\pi} \Big( [T^{-1}\partial T, \bar\partial T T^{-1}] \\ &&+
  \partial (\bar\partial T T^{-1}) + \bar\partial (T^{-1}\partial T)
  \Big)({\bf r}) + {\cal O}(1/\kappa^2) \;.  \nonumber
\end{eqnarray*}
On dimensional grounds, the term ${\cal O}(1/\kappa^2)$ must involve
four derivatives and therefore becomes negligible for wave lengths
much larger than the short-distance cutoff $1/\kappa$.

The above expansion is now substituted into the expression for $\dot
S$.  We then arrive at
\begin{eqnarray*}
  \dot S &=& - \frac{1}{\pi} \int d^2r\, {\rm STr}\, (T^{-1}\dot T +
  \dot T T^{-1}) \Big( \bar\partial (T^{-1} \partial T) \\ &+&
  \partial (\bar\partial T T^{-1}) + T^{-1}\partial T \bar\partial T
  T^{-1} - \bar\partial T T^{-2} \partial T \Big) \;.
\end{eqnarray*}
It is not hard to verify that, on making the identification $M =
T^{-2}$, this expression coincides with
\begin{eqnarray*}
  && \dot S = -\frac{1}{8\pi} \frac{d}{dt} \int d^2r\, {\rm STr}
  \left( \partial_\mu M \partial_\mu M^{-1} \right) \\ && +
  \frac{i}{4\pi} \int d^2r\, \epsilon^{\mu\nu} {\rm STr} \left(
    \partial_t M M^{-1} \partial_\mu M M^{-1} \partial_\nu M M^{-1}
  \right) \;.
\end{eqnarray*}
Integrating over time and noticing that $S = \int_0^1 \dot S dt$, we
obtain the WZW functional given in (\ref{marg_line}), with $g = 0$.
Finally, we undo the rescaling made at the beginning of the
calculation, and arrive at the anisotropic effective action
\begin{eqnarray*}
  S[M,\gamma] &=& {i\over 12\pi}\Gamma[M] - \frac{1}{8\pi} \int d^2r
  \times \\ &\times& {\rm STr}\left(\gamma^{-1} \partial_1 M^{-1}
    \partial_1 M + \gamma^{-1} \partial_2 M^{-1} \partial_2 M \right)
  \;.
\end{eqnarray*}
The corresponding result for the other pair of nodes $(2,\bar 2)$ is
obtained by exchanging coordinates $x_1 \leftrightarrow x_2$.

Notice that this result has the peculiar feature of being independent
of the disorder.  In the isotropic case $\gamma = 1$, the model
becomes completely universal in the sense that its coupling constants
assume fixed values.  By the arguments reviewed in Appendix
\ref{sec:AIII_solution}, this WZW model is equivalent to free
relativistic fermions, i.e.~our original model in the absence of
disorder.  But this raises the question where the information on the
presence of impurities was lost.  After all it is hard to conceive
that the disorder strength $g$ should be reflected only in the value
of the UV momentum cutoff $\kappa \sim {\rm e}^{- 4\pi t \Delta /g}$.

Our analysis using non-Abelian bosonization described in the main text
reveals the fact that the action above should be supplemented by a
current-current interaction with coupling given by $g$.  This begs the
question how the existence of this term got lost in the standard
scheme of saddle-point approximation plus gradient expansion.  The key
to the answer of this question lies in the presence of massive modes
in the model, a fact we have ignored thus far.  Indeed, we had
immediately reduced the Hubbard-Stratonovich field $Q$ to its
Goldstone-mode content, $Q \to i\kappa M$.  In fact, however, the
field $Q$ also contains massive modes, i.e.~modes $P$ which are not
compatible with the chiral symmetry of the Hamiltonian and, therefore,
fluctuate at a finite energy cost.  One way of handling the situation
would be to put $Q = i\kappa PM$, where the $P$'s are not just set to
unity but integrated out.  This procedure results in the appearance of
an additional current-current interaction
\begin{eqnarray*}
  {\rm STr}(M^{-1} \bar\partial M) \, {\rm STr} (M \partial M^{-1})
  \;.
\end{eqnarray*}
Unfortunately, it turns out to be impossible to reliably determine the
value of the coupling constant of this perturbation within the
standard scheme: to obtain the additional term, one has to integrate
out massive fluctuations, and yet the mass gap characterizing the
modes $P$ does not suffice to justify a Gaussian approximation to this
integral.  The reason for the last fact is that the disorder-generated
perturbation of the field theory is strictly RG marginal, which means
that there is no dynamically generated mass scale in the problem.
Thus, no intrinsic mechanism stabilizing the Gaussian approximation
exists, fluctuations are important, and to determine the coupling
constant, the $P$-integral must be performed exactly.  It goes without
saying that this is difficult to do in practice.  Ultimately it is
this deficiency of the standard scheme which forces us to build our
theory on the less standard approach of non-Abelian bosonization.

\section{Dirac fermions in a random vector potential}
\label{sec:CFT}

As was reviewed in the main text, the low-energy quasi-particles of a
dirty $d$-wave superconductor behave, in the single-node
approximation, as Dirac fermions in a random vector potential.  We
have argued that non-Abelian bosonization takes that theory into a
supersymmetric WZW model, from which we rederived the critical
exponent for the density of states.  In the present appendix we
elaborate on this issue and supply more of the technical details.

\subsection{WZW model of type $A|A$}
\label{sec:WZW_AA}

Let us begin by establishing the general context with a few remarks.
The characteristic feature of a WZW model is the multi-valued term
$\Gamma[M]$.  Multi-valued functionals of this type were studied in
the context of Hamiltonian mechanics by Novikov \cite{novikov}, and
became firmly established in field theory through Witten's celebrated
paper \cite{Witten84} on non-Abelian bosonization. Ever since, WZW
models defined on compact groups have been an object of intense study.
There exists a vast amount of literature on them, and they have been
solved in great detail.  To a much lesser extent, field theorists have
also studied WZW models of the {\it non-compact} type.  The target
spaces of these models are {\it not} groups but are non-compact
symmetric spaces, the simplest example being ${\rm SL}(2,C) / {\rm
  SU}(2)$.  One of the rare occurrences of such a model is found in
\cite{ctt}.

The target space of the WZW model we will be concerned with transcends
the classical setting in that it is a {\it superspace}.  It turns out
that the proper mathematical construction of a WZW model with
superspace target is a delicate matter.  Indeed, to define a
functional integral on Euclidean space-time, one needs a target space
with a {\it Riemannian} structure, providing for an action functional
that is bounded from below.  Sadly, the invariant geometry on
supergroups, such as ${\rm GL}_C(n|n)$, ${\rm U}(n|n)$, ${\rm
  GL}_R(n|n)$, ${\rm OSp}_R(2n|2n)$ etc., or even on symmetric
quotients such as ${\rm GL}_C(n|n) / {\rm U}(n|n)$, is never
Riemannian, but always of indefinite signature.  Therefore, WZW
models, and non-linear sigma models in general, {\it do not exist on
  supergroups}, at least not in the literal sense (i.e.~without some
procedure of analytic continuation of the fields).

One can easily appreciate this point by looking, for example, at the
complex Lie supergroup ${\rm GL}_C(1|1)$, with the standard
(bi-)invariant metric given by $\kappa = - {\rm STr}\,{\rm d}g^{-1}
{\rm d}g$, $g \in {\rm GL}_C(1|1)$.  To understand the properties of
this metric tensor, it suffices to examine it on the tangent space at
the group unit, which is the complex Lie algebra ${\rm gl}_C(1|1)$.
Elements of this Lie algebra are written as
\begin{eqnarray*}
  X = \pmatrix{a &\alpha\cr \beta &b\cr} \;,
\end{eqnarray*}
with commuting $a,b$ and anticommuting $\alpha,\beta$.  Via the
exponential mapping $g = {\rm e}^X$, the invariant metric $\kappa$
restricts to
\begin{eqnarray*}
  \kappa\big|_{X = 0} = {\rm STr} \, {\rm d}X {\rm d}X = {\rm d}a^2 -
  {\rm d}b^2 + 2{\rm d}\alpha {\rm d}\beta \;.
\end{eqnarray*}
Clearly, the numerical part ${\rm d}a^2 - {\rm d}b^2$ is of indefinite
sign, and this does not change on passing to ${\rm gl}_R(1|1)$ by
restricting the complex number field to the reals.  The same remains
true if we choose a compact real form ${\rm u}(1|1)$ by setting $a =
i\varphi_1$ and $b = i\varphi_2$, with real $\varphi_{1,2}$.  Thus no
matter which real form of ${\rm gl}_C(1|1)$ we choose, the numerical
part ${\rm d}a^2 - {\rm d}b^2$ of the invariant metric will always be
of indefinite sign.  Note also that it is pointless to replace the
compact supergroup ${\rm U}(1|1)$ by its non-compact analog ${\rm
  GL}_C(1|1)/{\rm U}(1|1)$.  This just reverses the overall sign of
the metric, leaving it as indefinite as before.

One might object that if the target space were taken to be the compact
supergroup ${\rm U}(1|1)$, the range of all field variables would be
finite, and one could then say that compact integrals always exist, no
matter what is the sign of the exponent of the Boltzmann weight.  This
reasoning would be valid if we were dealing with a finite number of
compact integrals.  However, the system at hand is a {\it functional}
integral over infinitely many degrees of freedom, and to ensure the
existence of this functional integral, we do need a target space with
Riemannian structure.  Indeed, if the metric of the Euclidean field
theory were Lorentzian rather than Riemannian, the energy of a field
configuration could be diminished below any bound by shortening the
wave length of some field.  For the case of ${\rm U}(1|1)$ with
Boltzmann weight $\exp - \int {\rm STr} (g^{-1} \partial_\mu g)^2$
this would be the field $a = i\varphi_1$.  If one attempted to define
the field theory by its Feynman diagrams, one would encounter the
problem that the free-field approximation to the Boltzmann weight for
a fluctuation $\varphi_1(k)$ with wave number $k$ is $\exp( |k|^2
|\varphi_1(k)|^2 )$, which {\it increases} with the strength of the
fluctuation!  Thus the field theory would be unstable
w.r.t.~short-wave length fluctuations in $\varphi_1$.  Of course one
could rescue the stability by putting the theory on a lattice with,
say, nearest-neighbour coupling for the ${\rm U}(1|1)$ field.
However, the state of lowest energy of such a lattice theory would be
ferromagnetic in $\varphi_2$, but antiferromagnetic in the
$\varphi_1$.  As a result the ground state would not be invariant
under global supersymmetry transformations.  Even worse, the partition
function of the theory (in the absence of sources) would be
identically zero and hence not normalizable to unity.  This follows
essentially from the fact \cite{berezin} that ${\rm U}(1|1)$ has
vanishing volume w.r.t.~its invariant Berezin-Haar measure.  All these
considerations make it rather unlikely that one could ever make sense,
for our purposes, of a Euclidean space-time nonlinear sigma model or
WZW model with target space ${\rm U}(1|1)$ or more generally ${\rm
  U}(n|n)$.

The construction that solves the difficulty was described in a general
mathematical setting in Ref.~\cite{rss}.  Its field-theoretic
implementation was discussed in some detail in Section 6 of
Ref.~\cite{cftiqhe} and Section 7.1 of Ref.~\cite{bsz}.  The basic
idea is easily understood.  To make the numerical part of the metric
tensor ${\rm d}a^2 - {\rm d}b^2$ positive definite, we should take $a
\equiv x$ from the real numbers, and $b \equiv iy$ from the imaginary
numbers.  By returning to group level via exponentiation, we then get
manifolds ${\cal M}_{\sc f} = {\rm U}(1)$ and ${\cal M}_{\sc b} = {\rm
  GL}(1,C)/{\rm U}(1)$.  In the general case of $n$ Green functions,
we obtain ${\cal M}_{\sc f} = {\rm U}(n)$ and ${\cal M}_{\sc b} = {\rm
  GL}(n,C)/{\rm U}(n)$.  These are Riemannian symmetric spaces of
compact and non-compact type, respectively.  On incorporating the
fermions into the exponential mapping, we are led to some sort of
superspace.  An equivalent procedure is to start from the complex
supergroup ${\rm GL}_C(n|n)$ and restrict the bosonic degrees of
freedom to ${\cal M}_{\sc b} \times {\cal M}_{\sc f}$.  (This
construction does not let us impose any reality constraints on the
fermions \cite{ast}.  Fortunately, there exists no fundamental
principle that would force us to do so.)  Let ${\bf X}_n$ denote the
resulting space.  ${\bf X}_n$ is not a group, but belongs to the
category of Riemannian symmetric superspaces \cite{rss}.  It is called
type $A|A$, which tells us that both the {\sc bb} and the {\sc ff}
sector are symmetric spaces belonging to the $A$ (or ``unitary'')
series.  Its main property is that the metric tensor $\kappa = - {\rm
  STr}\,{\rm d}g^{-1} {\rm d}g$ restricts to a Riemannian structure on
${\cal M}_{\sc b} \times {\cal M}_{\sc f}$.

Thus, the field theory to be studied has the target space ${\bf X}_n$.
We denote the field by $M$, and set $n = 2$ (although nothing
essential depends on that choice).  By slight generalization of the
action functional (\ref{marg_line}), we consider
\begin{eqnarray*}
  S[M] &=& k W[M] \\ &+& {k\lambda \over \pi^2} \int d^2r \, {\rm
    STr}(M\partial M^{-1}) \, {\rm STr}(M^{-1}\bar\partial M)
\end{eqnarray*}
with $W[M]$ being the WZW functional for ${\rm GL}(2|2)$:
\begin{eqnarray*}
  W[M] = - {1 \over 8\pi} \int d^2r \, {\rm STr} \, \partial_\mu
  M^{-1} \partial_\mu M + {i\Gamma[M] \over 12\pi} \;.
\end{eqnarray*}
This is the theory we shall now analyse.  As a by-product we will get
(by setting $\lambda = 0$) a justification of the bosonization rules
that lead to (\ref{marg_line}).

\subsection{Functional integral solution}
\label{sec:AIII_solution}

The main tool for dealing with the WZW functional is a relation due to
Polyakov and Wiegmann \cite{pwrel}:
\begin{eqnarray*}
  W[gh] = W[g] + W[h] + {1 \over \pi} \int d^2r \, {\rm STr} \,
  g^{-1}\bar\partial g \, \partial h \, h^{-1} \;.
\end{eqnarray*}
Using it, one sees that $W[M]$ is invariant under {\it local}
transformations
\begin{eqnarray*}
  M(z,\bar z) \mapsto g(z) M(z,\bar z) h(\bar z)^{-1} \;,
\end{eqnarray*}
where $g(z)$ and $h(\bar z)$ take values in ${\rm GL}_C (2|2)$.  We
will see that a deformation of this symmetry survives in the presence
of a non-vanishing coupling $\lambda$.

Consider now the expectation value $\langle {\cal F}[M] \rangle$ where
${\cal F}[M]$ is some functional of the field $M$ concentrated at a
set of points ${\bf r}_1, {\bf r}_2, ..., {\bf r}_N$.  Exploiting the
invariance of the functional integration measure under left
translations, we make a change of integration variables $M({\bf r})
\mapsto {\rm e}^{-X({\bf r})} M({\bf r})$, and denote the first
variation of the field by $\delta_X M({\bf r}) = - X({\bf r}) M({\bf
  r})$.  The resulting variation of the action defines the current
$J$:
\begin{eqnarray*}
  &&\delta_X S = - {1 \over \pi} \int d^2r \, J_{\bar\partial X} \;,
  \\ &&J_X = - k {\rm STr}(XM\partial M^{-1}) + {2k\lambda \over \pi}
  {\rm STr}(X) {\rm STr}(M\partial M^{-1}) \;.
\end{eqnarray*}
To derive this expression for the current, we made use of the
Polyakov-Wiegmann formula, and we used the identity
\begin{eqnarray*}
  \partial \, {\rm STr}(M^{-1}\bar\partial M) = \bar\partial \, {\rm
    STr}(M^{-1}\partial M)
\end{eqnarray*}
in conjunction with two partial integrations to cancel some terms.
(The boundary terms produced by these partial integrations at the
points ${\bf r}_1,\ldots, {\bf r}_N$ cancel each other.)

The invariance of the expectation value $\langle {\cal F}[M] \rangle$
w.r.t. the variation $\delta_X M$ implies
\begin{eqnarray*}
  \langle \delta_X {\cal F}[M] + {\cal F}[M] {1 \over \pi} \int d^2r
  \, J_{\bar\partial X} \rangle = 0 \;.
\end{eqnarray*}
{}From this relation it immediately follows that $\bar\partial J_A =
0$, called the equation of motion, holds under the functional integral
sign (away from the points of support of ${\cal F}[M]$) for any
spatially constant $A \in {\rm gl}_C(2|2)$.  As an additional
consequence one has
\begin{eqnarray*}
  {1 \over 2\pi i} \oint_\gamma dz \, \langle J_X(z) \cdot {\cal F}[M]
  \rangle = \langle \delta_X {\cal F}[M] \rangle \;,
\end{eqnarray*}
where $\gamma$ is any closed contour that circles once around the
points of support of ${\cal F}[M]$.  By specializing to ${\cal F}[M] =
M(0) {\cal F}_1[M]$, where ${\cal F}_1$ is some other functional, one
gets the operator product expansion between the current and the field:
\begin{eqnarray*}
  J_A(z) M(0) = - {A M(0) \over z} + ... \;,
\end{eqnarray*}
and by taking ${\cal F}[M] = J_B(0) {\cal F}_1[M]$ one gets the OPE
for the currents themselves:
\begin{eqnarray*}
  J_A(z) J_B(0) &=& {k f(A,B) \over z^2} + {J_{[A,B]}(0) \over z} +
  ...\;, \\ f(A,B) &=& -{\rm STr}(AB) + {2\lambda\over \pi} {\rm
    STr}(A){\rm STr}(B) \;.
\end{eqnarray*}
The dots indicate terms that remain finite in the limit $z \to 0$.

Identical considerations can be made starting from field variations
$\delta_Y M({\bf r}) = M({\bf r}) Y({\bf r})$, which generate right
translations $M({\bf r}) \mapsto M({\bf r}) {\rm e}^{Y({\bf r})}$.
They lead to a conserved current
\begin{eqnarray*}
  \bar J_Y &=& - k {\rm STr}(YM^{-1}\bar\partial M) + {2k\lambda \over
    \pi}{\rm STr}(Y) {\rm STr}(M^{-1}\bar\partial M) \;,
\end{eqnarray*}
which is antiholomorphic, i.e.~satisfies $\partial \bar J_A = 0$ for
spatially constant $A$.

The behaviour of correlation functions under conformal transformations
is determined by the stress-energy-momentum tensor, written in
components as
\begin{eqnarray*}
  T_{\mu\nu} dx^\mu dx^\nu = T_{zz} dz^2 + T_{\bar z \bar z} d\bar z^2
  \;.
\end{eqnarray*}
Conformal invariance implies $T_{z\bar z} = 0 = T_{\bar z z}$, and
$\bar\partial T_{zz} = 0 = \partial T_{\bar z \bar z}$ (on solutions
of the equations of motion).  We focus on the holomorphic sector and
set $T(z) = 2\pi T_{zz}(z)$.  Classical considerations based on the
general formula $T_{\mu\nu} = (\partial_\mu \varphi^i) \partial {\cal
  L}/\partial(\partial_\nu \varphi^i) - \delta_{\mu\nu} {\cal L}$
suggest
\begin{eqnarray*}
  T_{\rm cl} = {k \over 2} {\rm STr}\,(M\partial M^{-1})^2
  - {k\lambda \over \pi} {\rm STr}^2 (M\partial M^{-1}) \;,
\end{eqnarray*}
but this, it turns out, is modified by quantum fluctuations.  The
correct expression for $T(z)$ is obtained by demanding that the
leading singularity in the operator product $T(z) J_A(0)$ be $z^{-2}
J_A(0)$, as results from $J_A$ being a holomorphic conserved current.

The explicit form of the stress-energy-momentum tensor is obtained by
the Sugawara construction, which represents $T(z)$ as a quadratic form
in the currents.  Let $\{e_i\}$ be some basis of the Lie superalgebra
${\rm gl}(2,2)$ with metric $\kappa_{ij} = -{\rm STr}\, e_i e_j$.
Indices are raised by $\kappa^{ij} \kappa_{jk} = \delta_k^i$.  We
write $J_i \equiv J_{e_i}$ for short, and $J_e = J_{\rm id}$ for the
current corresponding to the unit matrix.  Setting
\begin{eqnarray*}
  \tilde T(z) = {1 \over 2k} \kappa^{ij} : J_i(z) J_j(z) :
\end{eqnarray*}
where the colons mean normal ordering (i.e. subtraction of the
short-distance singularities of the operator product at coinciding
points), and using the OPE for the currents and the associativity of
the operator product algebra, we get
\begin{eqnarray*}
  \lim_{z\to 0} z^2 \tilde T(z) J_A(0) &=& J_A(0) - {2\lambda \over
    \pi} {\rm STr}(A) J_e(0) \\ &+& {\kappa^{ij} \over 2k}
  J_{[e_i,[e_j,A]]}(0) \;.
\end{eqnarray*}
The last term is a quantum correction, which vanishes in the limit $k
\to \infty$.  For ${\rm gl}(2,2)$ one verifies the relation
$\kappa^{ij} [e_i, [e_j, A]] = 2e {\rm STr}A$.  This allows to rewrite
the right-hand side of the preceding relation as
\begin{eqnarray*}
  J_A(0) + \left({1\over k} - {2\lambda\over\pi} \right) {\rm STr}(A)
  J_e (0) \;.
\end{eqnarray*}
To obtain the required form of the operator product $T(z) J_A(0)
= J_A(0)/z^2 + ...$, we must cancel the second term, which is 
achieved by defining
\begin{eqnarray*}
  T(z) = \tilde T(z) + {1 - 2\lambda k/\pi \over 2k^2} : J_e(z) J_e(z)
  : \;.
\end{eqnarray*}
On setting $k = 1$ and renaming $\lambda$ to $g$, we arrive at the
expression for $T(z)$ claimed in Eq.~(\ref{sugawara}).  Note that the
symbol for normal ordering can be dropped, as the expression is
already finite as it stands \cite{mcw}.

Our final step is the computation of the OPE of $T(z)$ with the
fundamental field $M(0)$.  This is readily done from the formula
for $T(z)$ and the OPE between the currents and $M(0)$.  Using 
the fact that the quadratic Casimir invariant evaluated in the
fundamental representation of ${\rm gl}(2,2)$ vanishes $(\kappa^{ij}
e_i e_j = 0)$, we find
\begin{eqnarray*}
  T(z) M(0) = {\Delta_M M(0) \over z^2} + ... \;,
\end{eqnarray*}
where $\Delta_M = (1 - 2\lambda k/\pi) / (2k^2)$ is the (holomorphic)
scaling dimension of the field $M$.  The total dimension is 
\begin{eqnarray*}
  \Delta_M + \bar\Delta_M = {1 - 2\lambda k/\pi \over k^2} \;.
\end{eqnarray*}
Note that the dimension becomes negative for disorder strengths
$\lambda$ greater than $\pi/2k$.  If we substitute $M = \exp
\pmatrix{a &\alpha\cr \beta &b\cr}$ into the action $S[M]$ and isolate
the terms involving the boson-boson field $a = X_{\sc bb}$ only, we
get
\begin{displaymath}
  S[M] = {k \over 8\pi} \left( 1 - {2\lambda \over \pi} \right)
  \int d^2r \, (\partial_\mu a)^2 + \ldots \;.
\end{displaymath}
We see that the coupling constant turns negative, so the functional
integral becomes unstable, at $\lambda = \pi/2$.  For $k = 1$ the
onset of instability is located right at the value of $\lambda$ where
the dimension of $M$ becomes negative.  A plausible scenario for what
happens beyond that point (for $k = 1$) has been suggested by Gurarie
\cite{gurarie}.  For higher level $k > 1$, there exists an
intermediate regime $\pi/2k < \lambda < \pi/2$, where $\Delta_M +
\bar\Delta_M$ is negative while the functional integral is still
stable.  In this regime we expect the density of states (the
expectation value of the field $M$) to diverge at zero energy.

For $k = 1$ and $\lambda = 0$, the conformal dimensions of $M$ are
$(\Delta_M,\bar\Delta_M) = (1/2,1/2)$.  These coincide with the
conformal dimension of the bilinears $\Psi_1 \bar\Psi_{\bar 1}$ in the
supersymmetric Dirac theory with Lagrangian
\begin{eqnarray*}
  {\cal L} = \bar\Psi_1^s \partial \Psi_{\bar 1} + \Psi_{\bar 1}^s
  \bar\partial \Psi_1 \;.
\end{eqnarray*}
Since the OPEs involving the currents match, too, we expect that the
two theories are equivalent.  The bosonization rules are the same as
in (\ref{dictionary}) but for factors of two:
\begin{eqnarray*}
  \Psi_1^s \Psi_{\bar 1}^s &\leftrightarrow& \pi^{-1} M \partial M^{-1}
  \;, \nonumber \\ \bar\Psi_{\bar 1} \bar\Psi_1^s &\leftrightarrow&
  \pi^{-1} M^{-1} \bar \partial M\;, \nonumber \\ \Psi_1 \bar\Psi_1^s
  &\leftrightarrow& \ell^{-1} M\;, \nonumber\\ \bar\Psi_{\bar 1}
  \Psi_{\bar 1}^s &\leftrightarrow& \ell^{-1} M^{-1}\;.
\end{eqnarray*}

\section{WZW theory for systems with boundary}
\label{sec:wzw-theory-systems}

In the main text we had said that a single WZW theory represents too
narrow a framework to consistently describe systems with
boundaries: the WZW functional $\exp
i \Gamma[M]/12\pi$ for maps $M$ from the position space $\Sigma$ to
the target space ${\cal G}$ (the Riemannian symmetric superspace of
type $A|A$ in ${\rm GL}(2|2)$) makes sense only if $\Sigma$ is closed
(i.e. has no boundaries.)
The reason is that, to define $\Gamma[M]$ as an integral of the 3-form
${\rm STr}(M^{-1}{\rm d}M)^{\wedge 3}$, one needs to extend $M$ to a
map $\tilde M : B \to {\cal G}$ where $B$ is some 3-space.  But in
order for the WZW functional to be independent of the choice of
extension $\tilde M$, the boundary of $B$ must coincide with $\Sigma$
$(\partial B = \Sigma)$, which is possible only if $\Sigma$ is closed
($\partial \Sigma = \partial \partial B = 0$).  Thus the WZW
functional is ill-defined in the presence of a boundary.  To the
extent that the WZW model is a faithful representation of the Dirac
theory, we expect the implementation of a boundary to be problematic
in the latter, too.

How can we cure this disease and arrive at a meaningful functional
integral for systems with a boundary?  There exists only one solution
to the problem \cite{gawedzki}, and it requires having a minimum of
two WZW theories.  Let's assume there exists a pair of WZW fields $M$
and $M^\prime$, with the sum of WZW terms being $\Gamma[M] +
\Gamma[M^\prime]$.  The boundary condition we impose is $M^{-1}
\big|_{\partial\Sigma} = M^\prime \big|_{\partial \Sigma}$.  (To
ensure current conservation, we also need to impose a condition on the
normal derivatives of the fields at the boundary.)  This boundary
condition guarantees that the two maps $M : \Sigma \to {\cal G}$ and
$M^\prime : \Sigma \to {\cal G}$ combine to a single map $\Sigma \cup
(-\Sigma) \to {\cal G}$, where $\Sigma \cup (-\Sigma)$ consists of two
copies of the position space, glued together at the boundary $\partial
\Sigma$.  Because the double covering $\Sigma \cup (-\Sigma)$ is a
closed manifold, we can regard it as the boundary of a 3-space $B$, so
the independence of the choice of extension is restored and the total
WZW functional is again well-defined.

Having understood this, the question arises: where do we take the
second WZW theory from? Depending on the physics of the problem at
hand, a number of options are conceivable: If the original problem
mapped onto an {\it even number} of WZW species anyway -- this is the
case for our $d$-wave system -- it is natural to glue these fields
pairwise together at the boundaries.  This procedure is discussed in
the text.  However, one may also contemplate situations where the
low-energy sector of the original problem contains only a {\it single}
WZW theory. (In the language of fermions, this amounts to considering
a single Dirac fermion species.)  In this case, invocing a second WZW
field or, equivalently, a second Dirac species, changes the low-energy
content of the theory.  To repair this, we must give a {\it mass} to
the second species, so as to prevent it from contributing to the
long-range behaviour of the correlation functions.  Note the important
fact that the mass of the second species breaks parity, so we can no
longer conclude on symmetry grounds that pseudoscalar observables such
as the Hall conductivity vanish!  If we again carry out the low-energy
reduction to the nonlinear sigma model, we will get the same answer as
in section \ref{sec:soft-scatt-brok}, as the auxiliary second species
drops out in the low-energy limit.  Moreover, we expect the sign
ambiguity $\sigma_{xy} = \pm 1/2$ to be resolved: it is the sign of
the mass of the second species that will determine the sign of
$\sigma_{xy}$.

The second scenario is not new but is known from the work of Ludwig et
al \cite{lfsg} who studied a disordered electron model on a square
lattice with a magnetic flux of $\pi$ per plaquette.  The low-energy
limit of that model consists of two species of Dirac fermions, one
with a light mass and the other with a heavy mass.  The Hall
conductivity of the model does not vanish when the light species
becomes massless (and hence the low-energy sector parity-invariant),
and is determined by the sign of the Dirac mass of the heavy species.

\end{multicols}

\begin{thebibliography}{99}

\bibitem{Gorkov} L. P. Gorkov and P. A. Kalugin, Pis'ma Zh. Eksp.
  Teor. Fiz.~{\bf 41}, 208 (1985) [JETP Lett. {\bf 41}, 253 (1985)].

\bibitem{Lee} P. A. Lee, Phys. Rev. Lett. {\bf 71}, 1887 (1993).

\bibitem{hwe} P. J. Hirschfeld, P. W\"olfle and D. Einzel, Phys. Rev.
  B {\bf 37}, 83 (1988).

\bibitem{zhh} K. Ziegler, M. H. Hettler, and P. J. Hirschfeld, Phys.
  Rev. Lett. {\bf 77}, 3013 (1996); Phys. Rev. B {\bf 57}, 10825
  (1998).

\bibitem{ntw} A. A. Nersesyan, A. M. Tsvelik, and F. Wenger, Phys.
  Rev.  Lett.~{\bf 72}, 2628 (1994); Nucl. Phys. B {\bf 438}, 561
  (1995).

\bibitem{sfbn} T. Senthil, M. P. A. Fisher, L. Balents, and C. Nayak,
  Phys. Rev. Lett. {\bf 81}, 4704 (1998).

\bibitem{mp} P. Monthoux and D. Pines, Phys. Rev. B {\bf 49}, 4261
  (1994).

\bibitem{Balatsky94} A. V. Balatsky, A. Rosengren, and B. L.
  Altshuler, Phys. Rev. Lett. {\bf 73}, 720 (1994).

\bibitem{Balatsky96} A. V. Balatsky and M. I. Salkola, Phys. Rev.
  Lett.~{\bf 76}, 2386 (1996); A. V. Balatsky et al., Phys. Rev. B
  {\bf 51}, 1547 (1995).

\bibitem{Joynt} R. Joynt, J. Low Temp. Phys. {\bf 109}, 811 (1997).

\bibitem{Balents} L. Balents, M. P. A. Fisher, and C. Nayak, Int. J.
  Mod. Phys. B {\bf 12}, 1033 (1998).

\bibitem{smf} T. Senthil, J. B. Marston, and M. P. A. Fisher, Phys.
  Rev. B {\bf 60}, 4245 (1999).

\bibitem{sf} T. Senthil and M. P. A. Fisher, Phys. Rev. B {\bf 60},
  6893 (1999).

\bibitem{vsf} S. Vishveshwara, T. Senthil and M. P. A. Fisher,
  Phys. Rev. B {\bf 61}, 6966 (2000).

\bibitem{dl} A. C. Durst and P. A. Lee, Phys. Rev. B {\bf 62}, 1270 (2000).

\bibitem{mag} R. Moradian, J. F. Annett and B. L. Gy\"orffy, Phys.
  Rev. B {\bf 62}, 3508 (2000).

\bibitem{Pepin} C. P\'epin and P. A. Lee, Phys. Rev. Lett. {\bf 81},
  2779 (1998); Phys. Rev. B {\bf 63}, 4502 (2001).

\bibitem{Fukui} T. Fukui, Phys. Rev. B {\bf 62}, R9279 (2000).

\bibitem{fk} P. Fendley and R. M. Konik, Phys. Rev. B {\bf 62}, 9359 (2000).

\bibitem{ahm} W. A. Atkinson, P. J. Hirschfeld, A. H. MacDonald, Phys.
  Rev. Lett. {\bf 85}, 3922 (2000); W. A. Atkinson, P. J.  Hirschfeld,
  A. H. MacDonald, K. Ziegler, Phys. Rev. Lett. {\bf 85}, 3926 (2000).

\bibitem{zst} J.-X. Zhu, D.N. Sheng, and C.S. Ting, cond-mat/0005266.

\bibitem{fradkin} The first study of the $(2+1)-$dimensional Dirac
  equation with certain types of disorder, a problem closely related
  to both the dirty $d$-wave superconductor and the quantum Hall
  plateau transition, was made in: M. P. A. Fisher and E. Fradkin,
  Nucl. Phys.  B {\bf 241}, 457 (1985); E. Fradkin, Phys. Rev. B {\bf
    33}, 3257 (1986); {\it ibid} {\bf 33}, 3263 (1986).

\bibitem{kulic1} M. L. Kuli\'c and V. Oudovenko, Solid State Commun.
  {\bf 104}, 375 (1997).

\bibitem{kulic2} O. V. Dolgov, O. V. Danylenko, M. L. Kuli\'c, and V.
  Oudovenko, Int. J. Mod. Phys. B {\bf 12}, 3083 (1998); Eur. Phys. J.
  B {\bf 9}, 201 (1999).

\bibitem{kulic3} M. L. Kuli\'c and O. V. Dolgov, Phys. Rev. B {\bf
    60}, 13062 (1999).

\bibitem{gs} L. P. Gorkov and J. R. Schrieffer, Phys. Rev. Lett. {\bf
    80}, 3360 (1998).

\bibitem{anderson} P. W. Anderson, cond-mat/9812063.

\bibitem{ft} M. Franz and Z. Tesanovic, Phys. Rev. Lett. {\bf 84}, 554
  (2000). 

\bibitem{melnikov} A. S. Melnikov, J. Phys. Cond. Matt. {\bf 21}, 4219
  (1999).

\bibitem{mhs} L. Marinelli, B. I. Halperin, and S. H. Simon, Phys.
  Rev. {\bf 62}, 3488 (2000).

\bibitem{wm} H. Won and K. Maki, cond-mat/0004105.

\bibitem{gorkov2} L. P. Gorkov, Sov. Phys. JETP {\bf 7}, 505 (1959).
  
\bibitem{fn_gorkov} Central to this paper are the Green functions of
  quasi-particles subject to an effective, spatially varying pairing
  field.  These objects, and the underlying Hamiltonian, have first
  been formulated and investigated by Gorkov. For this reason we use
  the terminology ``Gorkov equations'' and ``Gorkov Hamiltonian''
  (deviating from the commonly used attribute
  ``Bogoliubov-deGennes'').

\bibitem{Altland} A. Altland and M. R. Zirnbauer, Phys. Rev. B {\bf
    55}, 1142 (1997).

\bibitem{Oppermann} R. Oppermann, Physica A {\bf 167}, 301 (1990); V.
  E. Kravtsov and R. Oppermann, Phys. Rev. B {\bf 43}, 10865 (1991).

\bibitem{fn_parity} In its standard definition, the parity operation
  denotes a point reflection at the origin, $(x,y) \mapsto (-x,-y)$.
  Within the context of anomalous superconductivity, however,
  ``parity'' mostly stands for a planar reflection, $(x,y) \mapsto
  (x,-y)$, or any other discrete transformation that reverses the
  orientation of 2d space.

\bibitem{class_warning} While the classification scheme is complete
  (in the sense that it does provide closure under renormalization),
  it is important to recognize that it does not uniquely determine the
  physical behaviour of a system. For example, inside the symmetry
  classes $D$ and $D$III in two dimensions, a number of different
  physical behaviours (localized, critical, or metallic) is possible.

\bibitem{Witten84} E. Witten, Comm. Math. Phys. {\bf 92}, 455 (1984).

\bibitem{mcw} C. Mudry, C. Chamon, and X.-G. Wen, Phys. Rev. B {\bf
    53}, 7638 (1996); Nucl. Phys. B {\bf 466}, 383 (1996).

\bibitem{Laughlin} R. B. Laughlin, Phys. Rev. Lett. {\bf 80}, 5188
  (1998).

\bibitem{lhw} M. R. Li, P. J. Hirschfeld, and P. W\"olfle, Phys. Rev.
  B {\bf 63}, 054504 (2001).

\bibitem{glr} I. Gruzberg, A. W. W. Ludwig, and N. Read, Phys. Rev.
  Lett. {\bf 82}, 4524 (1999).

\bibitem{Kagalovsky} V. Kagalovsky, B. Horovitz, Y. Avishai, and J. T.
  Chalker, Phys. Rev. Lett. {\bf 82}, 3516 (1999).

\bibitem{cardysqh} J. L. Cardy, Phys. Rev. Lett. {\bf 84}, 3507
  (2000).

\bibitem{bele} D. Bernard and A. LeClair, cond-mat/0003075.

\bibitem{ye} J. Ye, Phys. Rev. Lett. 86, 316 (2001).

\bibitem{sfD} T. Senthil and M. P. A. Fisher, Phys. Rev. B {\bf 61},
  9690 (2000).

\bibitem{rg} N. Read and D. Green, Phys. Rev. B {\bf 61}, 10267
  (2000).

\bibitem{bsz} M. Bocquet, D. Serban, and M. R. Zirnbauer, Nucl. Phys.
  B {\bf 578}, 628 (2000).

\bibitem{rss} M. R. Zirnbauer, J. Math. Phys. {\bf 37} 4986 (1996).

\bibitem{AndersonT} P. W. Anderson, Phys. Rev. Lett. {\bf 3}, 325
  (1959).

\bibitem{efetov} K. B. Efetov, Adv. Phys. {\bf 32}, 53 (1983).

\bibitem{Nagaosa95} N. Nagaosa and T. K. Ng, Phys. Rev. B {\bf 51},
  15588 (1995).

\bibitem{Krivenko95} S. Krivenko and G. Khaliullin, JETP Lett. {\bf
    62}, 723 (1995); G. Khaliullin et al., Phys. Rev. B {\bf 56},
  11882 (1997).

\bibitem{Kaminski99} A. Kaminski et al., Phys. Rev.  Lett. {\bf 84},
  1788 (2000).

\bibitem{Taillefer97} L. Taillefer et al., Phys. Rev. Lett. {\bf 79},
  482 (1997).

\bibitem{Chiao99} M. Chiao et al., Phys. Rev. B {\bf 62}, 3554 (2000).

\bibitem{Millis} J. Orenstein and A. Millis, Science {\bf 288}, 468
  (2000).

\bibitem{Hussey} N. E. Hussey et al., Phys. Rev. Lett. {\bf 85}, 4140
  (2000).

\bibitem{Ong} Y. Zhang et al., Phys. Rev. Lett. {\bf 84}, 2219 (2000).

\bibitem{volovik} G. E. Volovik, JETP Lett. {\bf 58}, 469 (1993).

\bibitem{dyson} F. J. Dyson, J. Math. Phys. {\bf 3}, 1199 (1962).
  
\bibitem{v1} J. J. M. Verbaarschot and I. Zahed, Phys. Rev. Lett. {\bf 70},
  3852 (1993).

\bibitem{v2} J. J. M. Verbaarschot, Phys. Rev. Lett. {\bf 72}, 2531
  (1994).

\bibitem{complex} Examples of a group and its complex extension are
  ${\rm U}(N)$ and ${\rm GL}(N,C)$.  Another example is $O(N)$ and
  $O(N,C)$, the latter being the orthogonal group over the complex
  number field.  More generally, we call a manifold $M_C$ the complex
  extension of a manifold $M$ if $M$ is a submanifold of $M_C$ with
  half the dimension and the complexification $T_p M + i T_p M$ of the
  tangent space of $M$ in every point $p \in M$ coincides with the
  tangent space of $M_C$ in $p$.

\bibitem{bcsz1} R. Bundschuh, C. Cassanello, D. Serban, and M. R.
  Zirnbauer, Nucl. Phys. B {\bf 532}, 689 (1998).

\bibitem{bcsz2} R. Bundschuh, C. Cassanello, D. Serban, and M. R.
  Zirnbauer, Phys. Rev. B {\bf 59}, 4382 (1999).

\bibitem{zamoRG} A. B. Zamolodchikov, Sov. J. Nucl. Phys. {\bf 46}
  1090 (1987).

\bibitem{cardysbook} J. L. Cardy, {\it Scaling and Renormalization in
    Statistical Physics} (Cambridge University Press, Cambridge,
  1996).

\bibitem{ast_sc} A. Altland, B. D. Simons, and D. Taras-Semchuk, Adv.
  Phys. {\bf 49}, 321 (2000).

\bibitem{friedan} D. H. Friedan, Ann. Phys. {\bf 163}, 318 (1985).

\bibitem{helgason} S. Helgason, {\it Differential geometry, Lie groups
    and symmetric spaces} (Academic Press, New York, 1978).

\bibitem{lfsg} A. W. W. Ludwig, M. P. A. Fisher, R. Shankar, and G.
  Grinstein, Phys. Rev. B {\bf 50}, 7526 (1994).

\bibitem{affleck} I. Affleck, Phys. Rev. Lett. {\bf 55}, 1355 (1985).

\bibitem{gll} S. Guruswamy, A. LeClair, and A. W. W. Ludwig, Nucl.
  Phys. B {\bf 583}, 475 (2000).

\bibitem{kz} V. G. Knizhnik and A. B. Zamolodchikov, Nucl. Phys. B
  {\bf 247}, 83 (1984).

\bibitem{llp} H. Levine, S. B. Libby, and A. M. M. Pruisken, Phys.
  Rev. Lett. {\bf 51}, 1915 (1983).

\bibitem{pruisken} A. M. M. Pruisken, Nucl. Phys. B {\bf 235}, 277
  (1984).

\bibitem{sl} S. H. Simon and P. A. Lee, Phys. Rev. Lett. 78, 5029
  (1997).

\bibitem{vmz} A. Vishwanath, cond-mat/0104213; Vafek, A. Melikyan, and Z. Tesanovic,
  cond-mat/0104516.

\bibitem{gawedzki} K. Gawedzki, hep-th/9904145.

\bibitem{zhh_reply} K. Ziegler, M. H. Hettler, and P. J. Hirschfeld,
  Phys. Rev. Lett. {\bf 78}, 3982 (1997).

\bibitem{nt_comment} A. A. Nersesyan and A. M. Tsvelik, Phys. Rev.
  Lett. {\bf 78}, 3981 (1997).
  

\bibitem{ast} A. V. Andreev, B. D. Simons, and N. Taniguchi, Nucl.
  Phys.~B {\bf 432}, 487 (1994).

\bibitem{huckalt} B. Huckestein and A. Altland, cond-mat/0102124.
  
\bibitem{Gornyi} A. G. Yashenkin, D. V. Khveshchenko, I. V. Gornyi,
  cond-mat/0011249; A. G. Yashenkin, W. A. Atkinson, I. V. Gornyi, P.
  J. Hirschfeld, D. V. Khveshchenko, cond-mat/0102310.

\bibitem{gurarie} V. Gurarie, cond-mat/9907502.

\bibitem{Alvarez} O. Alvarez, Nucl. Phys. B {\bf 238}, 61 (1984).

\bibitem{zinnjustin} J. Zinn-Justin, {\it Quantum field theory and
    critical phenomena} (Clarendon Press, Oxford, 1989).

\bibitem{bgv} N. Berline, E. Getzler, and M. Vergne, {\it Heat Kernels
    and Dirac Operators} (Springer, Berlin, 1992).

\bibitem{novikov} S. P. Novikov, Uspekhi Mat. Nauk {\bf 37}:5, 3
  (1982).

\bibitem{ctt} J. S. Caux, N. Taniguchi, and A. M. Tsvelik, Phys. Rev.
  Lett. {\bf 80}, 1276 (1998); Nucl. Phys. B {\bf 525}, 671 (1998).

\bibitem{berezin} F. A. Berezin, {\it Introduction to Superanalysis}
  (Reidel, Dordrecht, 1987).

\bibitem{cftiqhe} M.R. Zirnbauer, hep-th/9905054.

\bibitem{pwrel} A. M. Polyakov and P. B. Wiegmann, Phys. Lett. B {\bf
    141}, 223 (1984).
\end{thebibliography}
\end{document}